\documentclass[aps,prb,floatfix,amsmath,amssymb,preprint,eqsecnum,nofootinbib,superscriptaddress]{revtex4}
\usepackage{graphicx}
\usepackage{dcolumn}
\usepackage{bm}
\usepackage{epstopdf}
\usepackage{setspace}   
\usepackage[colorlinks=true,linktoc=page,citecolor=red,linkcolor=blue]{hyperref}

\newcommand{\beq}{\begin{equation}}
\newcommand{\eeq}{\end{equation}}
\newcommand{\ba}{\begin{array}{ccc}}
\newcommand{\ea}{\end{array}}
\newcommand{\nn}{\nonumber}
 \renewcommand{\d}{\partial}
\def\bZ{\mathbb{Z}}
\def\bR{\mathbb{R}}
\def\mRP{\mathbb{RP}^4}
\def\bea{\begin{eqnarray}}
\def\eea{\end{eqnarray}}

\def\<{\langle}
\def\>{\rangle}

\usepackage{amsmath}
\usepackage{amssymb}
\usepackage{slashed}
\usepackage{color}

\begin{document}

\title{$S$-duality of $u(1)$ gauge theory with $\theta =\pi$ on non-orientable manifolds:
Applications to topological insulators and superconductors}

\author{Max A. Metlitski}
\affiliation{Perimeter Institute for Theoretical Physics,  Waterloo, ON N2L 2Y5, Canada.}
\affiliation{Kavli Institute for Theoretical Physics, UC Santa Barbara, CA 93106, USA. }

\begin{abstract} 
Electric-magnetic duality ($S$-duality) is a well-known property of pure $u(1)$ gauge theory in 3+1 dimensions. In this paper, we investigate the compatibility of this duality with time-reversal symmetry. We consider two theories obtained by coupling a Dirac fermion with an ``inverted" sign of the mass $m$ to a $u(1)$ gauge field. Time-reversal in the two theories is implemented respectively via the $T$ and $CT$ symmetries of the Dirac fermion. It was recently conjectured  (C.~Wang and T.~Senthil (arXiv:1505.03520), and M.~Metlitski and A.~Vishwanath (arXiv:1505.05142)) that in the $|m| \to \infty$ limit these two theories are $S$-dual to each other. We provide support for this conjecture by studying partition functions of the two theories on non-orientable manifolds as a way to probe the realization of time-reversal. Upon integrating out the Dirac fermion, topological terms in the actions of the two theories are generated. While on an orientable manifold topological terms in both theories reduce to a $\theta$-term with $\theta = \pi$, on a non-orientable manifold they are distinct. 
We explicitly compute partition functions of the two theories on the manifold $\mathbb{RP}^4$ and show that they are equal; this result combined with certain physical arguments is sufficient to establish the duality. The two theories can be viewed as a gauged topological insulator in class AII and a gauged topological superconductor in class AIII, and the bulk duality allows us to derive previously conjectured non-trivial symmetric gapped surface states of these phases. 

\end{abstract}

\maketitle

\tableofcontents

\newpage

\section{Introduction}
Electric-magnetic duality, or $S$-duality, is a well-known property of pure $u(1)$ gauge theory in 3+1 dimensions.\cite{CardyRabinovici, Cardy, WittenS}
Let us consider a $u(1)$ gauge-theory with {\it bosonic} matter\footnote{By a $u(1)$ gauge theory with bosonic matter we will mean that all excitations with no magnetic charge (including excitations with finite electric charge) are bosons. By a $u(1)$ gauge theory with fermionic matter we will mean that all excitations with an odd electric charge (and no magnetic charge) are fermions, while all excitations with even electric charge (and no magnetic charge) are bosons.} and Lagrangian,
\beq L = -\frac{1}{4 e^2} f_{\mu \nu} f^{\mu \nu} + \frac{\theta}{32 \pi^2} \epsilon^{\mu \nu \lambda \sigma} f_{\mu \nu} f_{\lambda \sigma}  \label{eq:Ltheta}\eeq
and define a complex coupling constant $\tau$,
\beq  \tau = \frac{\theta}{2\pi} - \frac{2 \pi i}{e^2} \label{eq:introtaudef} \eeq
$S$ duality then relates theories with coupling constants $\tau$ and $-1/\tau$:
\beq S: \quad \tau \to -\frac{1}{\tau} \label{eq:Stau}\eeq
In the simplest case when $\theta  = 0$ this operation simply sends $e \to \frac{2 \pi}{e}$. A second duality of the $u(1)$ gauge theory is, ${\cal T}: \theta \to \theta + 4\pi$, i.e.
\beq {\cal T}: \tau \to \tau + 2\label{eq:Ttau} \eeq
The operations $S$ and ${\cal T}$ together generate a subgroup of $SL(2,\bZ)$. 

There are two complementary ways to understand $S$ and ${\cal T}$ duality. The first way is in terms of the dyon excitations of the $u(1)$ gauge-theory.\cite{CardyRabinovici,Cardy} One can show that the duality corresponds to a simple re-labeling of dyons, which preserves their statistics and interactions (see section \ref{sec:dyons} for a review). In particular, at $\theta = 0$ the $S$-operation simply exchanges the electric and magnetic charges of the theory. Crucially, at $\theta = 0$ both of these charges are bosons, so the re-labeling preserves the dyon statistics. One can also understand the periodicity $\theta \sim \theta + 4\pi$ in terms of the dyon quantum numbers: the electrically neutral monopole is a boson at $\theta = 0$, but a fermion at $\theta = 2\pi$, and returns to being a boson only at $\theta  =4\pi$.\cite{MetlitskibTI}

An alternative way to understand the  duality is by considering the partition function of the theory on an arbitrary 4-dimensional (Euclidean) oriented closed manifold. It was shown by Witten \cite{WittenS} that the partition function $Z(\tau)$ is, indeed, invariant (more strictly covariant) under the $SL(2,\bZ)$ transformations (\ref{eq:Stau}), (\ref{eq:Ttau}).  

We can also consider $u(1)$ gauge theory with fermionic charge matter. In this case, it will be convenient to think of the $\theta$ term as arising from integrating out a Dirac fermion $\psi$ with a complex mass,
\beq L = \bar{\psi} i \gamma^{\mu} (\d_{\mu} - i a_{\mu}) \psi  - |m| \cos \theta \bar{\psi} \psi - i |m| \sin \theta \bar{\psi} \gamma^5 \psi \label{eq:Lpsiintro}\eeq
and taking the limit $|m| \to \infty$. Indeed, on an  orientable Euclidean manifold $M$, the partition function of a Dirac fermion $Z_\psi$ takes the form,
\beq \frac{Z_{\psi}(\theta, |m|)}{Z_{\psi}(\theta = 0, |m|)} = \exp\left[i \theta \left(\frac{1}{32 \pi^2} \int_M d^4 x \, \epsilon^{\mu \nu \lambda \sigma} f_{\mu \nu} f_{\lambda \sigma} -\frac{1}{8} \sigma(M)\right)\right] \label{eq:Zpsio}\eeq
with $\sigma(M)$ - the signature of the manifold. Note that we have normalized the partition function of the Dirac fermion by its value at $\theta = 0$ to cancel out the non-topological contribution to the effective action.

In the case of fermionic matter, the monopole is always a boson and the $\theta$ angle is periodic modulo $2\pi$.  Furthermore, if we start at $\theta = 0$ and exchange the electric and magnetic charges, we get a theory with a bosonic charge and a fermionic monopole. Thus, $S$-duality maps a theory with fermionic charges and $\theta = 0$ to a theory with bosonic charges and $\theta = 2 \pi$. More generally, a fermionic theory with coupling constant $\tau$ gets mapped to a bosonic theory with coupling constant $-1/\tau + 1$:
\beq S_{bf}:\quad \tau \to -\frac{1}{\tau} +1, \quad\quad {\rm fermionic \,\, matter} \to {\rm bosonic \,\, matter}\eeq
 We will confirm this fact by comparing partition functions of fermionic and bosonic theories on an arbitrary oriented manifold. Note that we always take the mass of the matter fields to infinity, so naively the partition function  does not know whether the matter content is fermionic or bosonic. However, this is not correct: in a theory with bosonic matter the $u(1)$ gauge field is a connection on a complex line bundle and the partition function involves a sum over all line bundles. On the other hand, when we have fermionic charge matter, we need to equip the manifold with a Spin$_c$ structure and the partition function involves a sum over all such structures. Note that the duality is not restricted to Spin manifolds.

\begin{figure}[t!]
\begin{center}
\includegraphics[width = 5.5in]{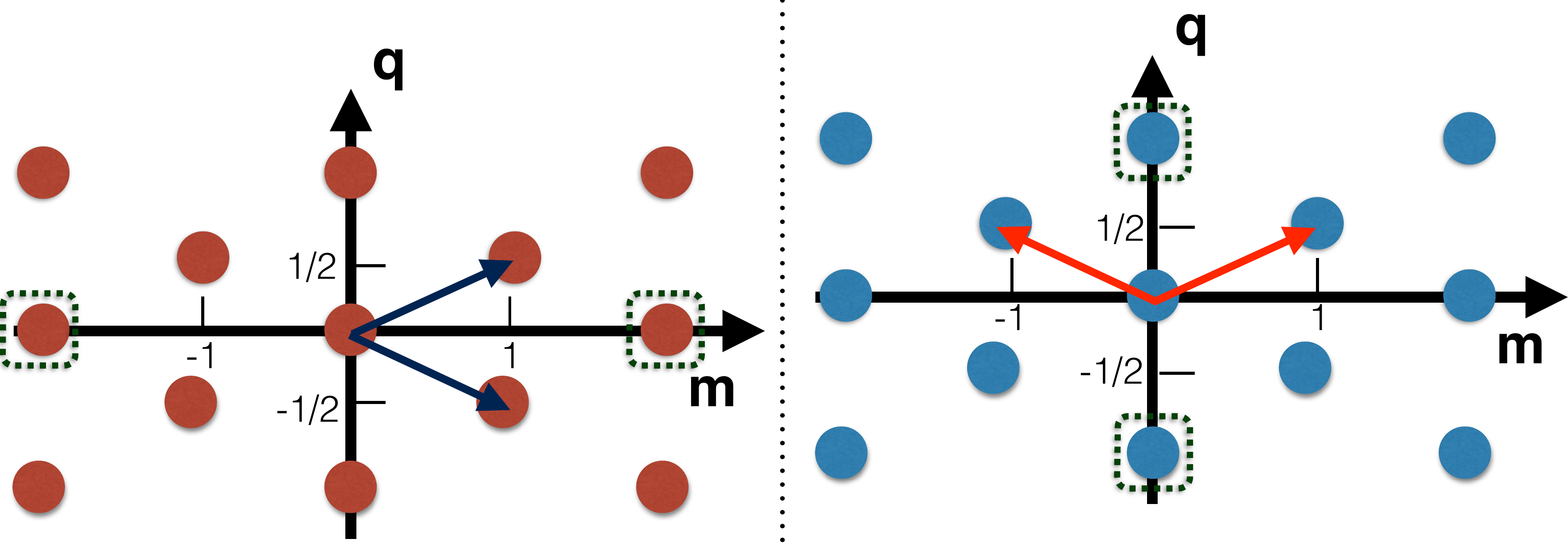}
\vspace{-.2in}
\end{center}
\caption{Lattice of dyon excitations in a $u(1)$ gauge-theory with fermionic matter and $\theta = \pi$. $q$ and $m$ denote the electric and magnetic charges. Left and right sides correspond to different implementations of time-reversal symmetry. Left side corresponds to the theory ${\cal L}_{CT}$, where time-reversal acts on the Dirac fermion via Eq.~(\ref{eq:CTintro}), so $CT: q \to - q, m \to m$. Blue arrows mark the dyons $d_{\pm}: (q = \pm 1/2, m = 1)$, which are partners under $CT$. Right side corresponds to the theory ${\cal L}_T$, where time-reversal acts on the Dirac fermion via Eq.~(\ref{eq:Tintro}), so $T: q \to q, m \to -m$. Red arrows mark the dyons $\tilde{d}_{\pm}: (q = 1/2, m = \mp 1)$, which are partners under $T$.   The duality (\ref{eq:Sfdyonsintro}) maps ${\cal L}_{CT}$ to ${\cal L}_T$, sending $d_{\pm} \to \tilde{d}_{\pm}$. In particular, the Kramers doublet fermion $d_+ d_-: (q = 0, m = 2)$ is mapped to the Kramers doublet fermion $\tilde{d}_+ \tilde{d}_-: (q = 1, m = 0)$. }
\label{fig:Dyons}
\end{figure}

We can also generate a duality which maps a theory with fermionic matter back to itself by acting with $S_f = S^{-1}_{bf} {\cal T}^{-1}_b S_{bf}$, with ${\cal T}_b$ being the bosonic ${\cal T}$ operation (\ref{eq:Ttau}). Under this combined operation, the coupling constant $\tau$ transforms as,
\beq S_f: \tau \to \frac{\tau}{2 \tau + 1} \label{eq:Sf}\eeq
Note that $S_f$ sends coupling constant $(e, \theta = -\pi) \to (\frac{4 \pi}{e}, \theta = \pi)$. Since  $\theta = \pi$ and $\theta = -\pi$ are identical in a fermionic theory, $S_f$ fixes the value $\theta = \pi$, where it sends $e \to \frac{4 \pi}{e}$ and transforms electric charges $q$ and magnetic charges $m$ as,
\beq S_f:\quad \left(\begin{array}{c} q \\ m\end{array}\right) \to  \left(\begin{array}{cc} 0& 1/2\\ -2 & 0 \end{array}\right)\left(\begin{array}{c} {q} \\ {m}\end{array}\right), \quad e \to \frac{4 \pi}{e},  \quad \quad\theta = \pi \label{eq:Sfdyonsintro}\eeq
This transformation of the dyon lattice is illustrated in Fig.~\ref{fig:Dyons}.

The new question that this paper addresses is the compatibility of $S$-duality with time-reversal symmetry. We will be mostly interested in the case of fermionic matter. We again take the matter to be a single Dirac fermion $\psi$,  Eq.~(\ref{eq:Lpsiintro}). We must now specify how time-reversal symmetry acts on $\psi$. There are two natural options: time-reversal can be implemented via standard $T$ symmetry or via $CT$ symmetry:
\bea  &&T:\quad \,\,\,\,\,\psi(\vec{x}, t) \to C^{\dagger} \gamma^5 \psi(\vec{x}, -t),  \quad i \to -i  \label{eq:Tintro}\\
  &&CT: \quad \psi(\vec{x}, t) \to  (\bar{\psi}(\vec{x}, -t) \gamma^5)^T, \quad i \to -i  \label{eq:CTintro}\eea
We have written the symmetry actions in Minkowski space and have stressed that $T$ and $CT$ are anti-unitary operators. $T$ and $CT$ are symmetries only when $\theta = 0$ or $\theta = \pi$. Here we concentrate on the $\theta = \pi$ case - i.e. inverted sign of $\bar{\psi} \psi$ mass term in Eq.~(\ref{eq:Lpsiintro}). $T$ and $CT$ acts on the electric and magnetic charges via,
\bea T: \, (q, m) \to (q, -m) \label{eq:Tqm}\\
CT: \, (q, m) \to (-q, m) \label{eq:CTqm}\eea
Naively, the two implementations of time-reversal: via $T$ and via $CT$ are distinct, giving rise to two time-reversal invariant $u(1)$ gauge theories, which we label ${\cal L}_T$ and ${\cal L}_{CT}$. However, it was conjectured in Refs.~\onlinecite{Wang2015, MVDuality} that at $\theta = \pi$ these two theories are, in fact, related by the duality $S_f$, Eq.~(\ref{eq:Sfdyonsintro}). Namely, $S_f$ maps $T$ symmetry to $CT$ symmetry. This conjecture was made based on the action of $T$ and $CT$ on the dyon lattice, see Fig.~\ref{fig:Dyons}. In particular, note that  $S_f$ at $\theta = \pi$ exchanges electric and magnetic charges, Eq.~(\ref{eq:Sfdyonsintro}), which is consistent with Eqs.~(\ref{eq:Tqm}), (\ref{eq:CTqm}).

While the preservation of dyon quantum numbers is very suggestive, as we will explain in section \ref{sec:ambiguity} it is not sufficient to establish a full duality between the two theories as time-reversal invariant theories. One may ask: given two theories with a global symmetry $G$, how do we tell if they are dual? In the case when $G$ is an internal symmetry, one may compute the partition function of the theory on a manifold with $G$-fluxes inserted through its cycles. If partition functions of two theories on an arbitrary oriented closed manifold in the presence of an arbitrary background $G$ gauge field agree then we say that the two $G$-invariant theories are dual.\footnote{In certain cases we may need to restrict to flat $G$ gauge fields.}

How do we generalize the above discussion to the case of time-reversal symmetry? In Refs.~\onlinecite{KapustinBos, KapustinBosTI, KapustinFerm} it was suggested that here the analogue of coupling the theory to a background $G$ gauge field is placing the system on a non-orientable (Euclidean) manifold. Recall that a manifold is a collection of patches with coordinate charts and transition functions specifying how the patches are glued together. If the manifold is non-orientable then some of the transitions between patches will reverse the chart orientation. If we place a field on such a manifold, in order to glue the fields in different patches together we must specify how the field transforms under orientation-reversing coordinate changes, i.e. under the discrete symmetries of time-reversal and spatial reflection. In particular, time-reversal symmetry tells us how to parallel transport a field around an orientation-reversing $1$-cycle. This is similar to the way the usual internal symmetry tells us how to parallel transport a field around a $1$-cycle with some non-trivial gauge flux.

We note that all the theories considered in this paper are Lorentz invariant. After continuation to Euclidean space, time-reversal and spatial reflections become related by Euclidean space-time rotations. Thus, in a Euclidean description time-reversal symmetry acts on par with reflection; in particular, it is an ordinary symmetry of the path-integral and does not involve complex conjugation. When we discuss a duality between two $T$-invariant theories, we implicitly mean that it also preserves the spatial reflection symmetry and Lorentz symmetry. 

Thus, to establish the duality between our two $u(1)$ gauge theories ${\cal L}_T$ and ${\cal L}_{CT}$ we need to compute their partition functions on non-orientable manifolds. Integrating out the Dirac fermion $\psi$ generates the following terms in the effective action,
\bea \frac{Z_{\psi, T}(\theta = \pi, |m|)}{Z_{\psi, T}(\theta = 0, |m|)} &=& (-1)^{N[a_{\mu}]/2}  \label{eq:ZTIintro}\\
\nn\\
\frac{Z_{\psi, CT}(\theta = \pi, |m|)}{Z_{\psi, CT}(\theta = 0, |m|)} &=& \exp\left(2 \pi i \eta[a_{\mu}]\right) \label{eq:ZTScintro}\eea
Here $Z_{\psi, T}$ and $Z_{\psi, CT}$ denote partition functions where orientation-reversing transformations on the manifold are implemented using the $T$ and $CT$ symmetries respectively.\footnote{More precisely, the Euclidean counterparts Eqs.~(\ref{eq:TEu}), (\ref{eq:CTEu}) of Eqs.~(\ref{eq:Tintro}), (\ref{eq:CTintro}).}  As before, we have normalized the partition function at $\theta = \pi$ by the partition function at $\theta = 0$ in order to cancel out the non-topological contribution to the effective action. The limit $|m| \to \infty$ is assumed in all expressions. In the case of $T$, the partition function (\ref{eq:ZTIintro}) is expressed in terms  of $N$ - the number of zero-modes of a a certain ``doubled" Dirac operator (see section \ref{sec:Zfrtimes}). The number $N$ is always even. In the case of $CT$, the partition function is expressed in terms of the $\eta$-invariant measuring the spectral asymmetry of the Dirac operator (see section \ref{sec:Zftimes}).\footnote{An analogous expression for the partition function of a massive Majorana fermion in terms of the $\eta$-invariant of the Dirac operator was recently obtained in Ref.~\onlinecite{WittenSPT}.}  In space-time dimension $D  = 4$, $\eta$ is always a multiple of $1/8$.\cite{GilkeyPinC} Both $(-1)^{N/2}$ and $e^{2 \pi i \eta}$ are ``topological terms," i.e. they are invariant under smooth deformations of the background gauge field $a_{\mu}$. On an orientable manifold  both terms reduce to a $\theta$ term for $a_{\mu}$ with $\theta = \pi$, Eq.~(\ref{eq:Zpsio}). %
However, on a non-orientable manifold $(-1)^{N/2}$ and $e^{2 \pi i \eta}$ are generally different. 
We then take the partition functions of our two gauge theories to be
\bea Z_T(e) &=& \int D a_{\mu} \,e^{-S[a_{\mu}]} (-1)^{N[a_{\mu}]/2} \quad \quad \label{eq:ZgT}\\
Z_{CT}(e) &=& \int D a_{\mu}\, e^{-S[a_{\mu}]}e^{2 \pi i \eta[a_{\mu}]} \quad \quad  \label{eq:ZgCT} \eea
with $S[a_{\mu}]$ given by the Maxwell action,
\beq S[a_{\mu}] = \frac{1}{4 e^2} \int_M d^4x \, \sqrt{g} f_{\mu \nu} f^{\mu \nu} \label{eq:introMaxwell}\eeq
The statement of duality between our two theories then becomes,
\beq Z_{T}(e) = Z_{CT}\left(\frac{4 \pi}{e}\right) \label{eq:introZduality}\eeq
We will show that Eq.~(\ref{eq:introZduality}) holds on an arbitrary orientable manifold. On general non-orientable manifolds we have been able to reduce Eq.~(\ref{eq:introZduality}) to the identity (\ref{eq:Finalf}) involving only a sum over topologically distinct saddle point gauge configurations. While we have not been able to prove this identity for arbitrary non-orientable manifolds, we will demonstrate that it holds on the manifold $\mathbb{RP}^4$. As we discuss in section \ref{sec:ambiguity}, physical arguments suggest that this is sufficient to establish the full duality. 

In addition to investigating the duality between theories with fermionic matter in Eqs.~(\ref{eq:ZgT}), (\ref{eq:ZgCT}), we also study $S$-duality in pure Maxwell theory with bosonic matter and with no topological terms ($\theta = 0$). Here we are able to prove $S$-duality on arbitrary non-orientable manifolds by showing that it reduces to the equality between Ray-Singer analytic torsion and Reidemeister torsion of a manifold.\footnote{Strictly speaking, the original proofs \onlinecite{Cheeger, Muller} of the equality between Ray-Singer analytic torsion and Reidemeister torsion assume an oriented manifold. However, we expect that the equality continues to hold on non-orientable manifolds.} We are grateful to E. Witten for pointing out this connection.



The formal duality between the two $u(1)$ gauge theories with fermionic matter ${\cal L}_T$ (\ref{eq:ZgT}) and ${\cal L}_{CT}$ (\ref{eq:ZgCT}) has direct implications for the physics of topological insulators and superconductors in 3+1D. In fact, ${\cal L}_T$ and ${\cal L}_{CT}$ can be respectively obtained from topological insulators in class AII and topological superconductors in class AIII by gauging the $u(1)$ symmetry (see section \ref{sec:TIreview} for a review). Refs.~\onlinecite{MVDuality,ChongSenthilDirac} have used the duality (\ref{eq:introZduality})  to obtain a dual surface theory of original (ungauged) topological insulators and superconductors. This dual surface theory allows one to deduce previously conjectured non-trivial symmetric gapped surface phases of topological insulators and superconductors, resolving certain previous ambiguities.\cite{Chen2014PRB, Bonderson2013, FidkowskiChenAV} We give a brief summary of these results in section \ref{sec:surface}.

This paper is organized as follows. In section \ref{sec:TIreview}, we review the relation of gauge theories ${\cal L}_T$ and ${\cal L}_{CT}$ to topological insulators and superconductors in 3+1D. In section \ref{sec:dyons}, we review the arguments of Refs.~\onlinecite{Wang2015,MVDuality} for duality between ${\cal L}_T$ and ${\cal L}_{CT}$ based on the quantum numbers of dyon excitations. We then describe in section \ref{sec:ambiguity} why the preservation of dyon quantum numbers by $S_f$ is insufficient to establish a full duality. We also present here physical arguments why checking the duality (\ref{eq:introZduality}) on manifolds $\mathbb{RP}^4$ and $\mathbb{CP}^2$ is, in fact, sufficient to establish the duality on all manifolds. Sections \ref{sec:Dirac} and \ref{sec:Gauge} are devoted to the main subject of this paper: the definition and study of partition functions (\ref{eq:ZgT}), (\ref{eq:ZgCT}) on non-trivial manifolds.  In section \ref{sec:Dirac}, we show that the partition function of a Dirac fermion on a non-orientable manifold is given by Eqs.~(\ref{eq:ZTIintro}), (\ref{eq:ZTScintro}).
Section \ref{sec:Gauge} demonstrates the equality of partition functions (\ref{eq:introZduality}) on arbitrary orientable manifolds and on the non-orientable manifold $\mathbb{RP}^4$. As a by-product, section \ref{sec:bosonsS} demonstrates $S$-duality in pure Maxwell theory with bosonic matter and $\theta = 0$ on arbitrary manifolds by relating it to the equality of analytic torsion and Reidemeister torsion. Section \ref{sec:surface} reviews some consequences of the bulk $S$-duality (\ref{eq:introZduality}) for the surface phases of topological insulators and superconductors.\cite{ChongSenthilDirac,MVDuality} Section \ref{sec:SW} has a different focus from the rest of the paper: it discusses a proposal for classifying topological superconductors in class AIII (topological insulators in class AII) in any dimension using Pin$_c$ (Pin$_{\tilde{c}}$) bordism. 

\subsection{Relation to topological insulators and superconductors}
\label{sec:TIreview}

In this section, we give a brief review of topological insulators and superconductors in 3+1D.

It is convenient to regard topological insulators and superconductors as examples of symmetry protected topological (SPT) phases.\cite{Chen2013} We recall that SPTs are gapped phases of matter possessing a global symmetry $G$. When the symmetry $G$ is broken, one can continuously connect the ground state of an SPT phase to a trivial product state. This implies that an SPT phase has no intrinsic topological order, i.e. no excitation with non-trivial statistics and no ground state degeneracy on an arbitrary spatial manifold. In cases where a field theory description of an SPT phase is known, the partition function on a closed space-time manifold  in the presence of a  background $G$ gauge field is a pure phase, $Z_{{\rm SPT}} = e^{i \omega}$, which is a topological invariant.\cite{LevinGu,KapustinBos} As already discussed, when the symmetry group $G$ contains time-reversal, one may consider the partition function $Z_{\rm SPT}$ on non-orientable manifolds. When one places an SPT on a manifold with a boundary, the boundary supports protected states with an ``anomalous" implementation of the symmetry $G$.


A topological insulator (TI) of fermions (class AII in condensed matter terminology) is an SPT with $u(1)$ and $T$ symmetry. The $u(1)$ symmetry acts on fermion fields $\psi$ via,
\beq u(1): \quad \psi(x) \to e^{i \alpha} \psi(x) \label{eq:U1intro} \eeq
while the time-reversal symmetry acts on $\psi$ via,
\beq T:\quad \psi(\vec{x}, t) \to U_T \psi(\vec{x}, -t), \quad i \to {-i} \quad \quad ({\rm TI})\label{eq:Tgintro}\eeq
with $U_T$ - a unitary matrix. 
In a topological insulator, the fermion is assumed to be a Kramers doublet under $T$: $T^2 = (-1)^F$, with $(-1)^F$ - the fermion parity, i.e. $U^*_T U_T = -1$. Due to the anti-unitary nature of $T$, the $u(1)$ and $T$ operations do not commute: $T u_\alpha T^{-1} = u_{-\alpha}$, where $u_\alpha$ is a rotation by $\alpha$ in the $u(1)$ group. For this reason, we will refer to the symmetry group of a TI as $u(1) \rtimes T$.\footnote{Strictly speaking, the group is not a semidirect product since $T^2 = (-1)^F$; we will abuse the notation slightly.}

If fermions are taken to be non-interacting, topological insulators in $3+1$D have a $\bZ_2$ classification, i.e. there is just a single non-trivial TI phase.\cite{FuKaneMele,KitaevNI,LudwigNI} The  interface of this phase with vacuum (i.e. with the trivial phase) supports an odd number of gapless $2+1$D Dirac fermions. A $3+1$D non-interacting TI has a convenient field theoretic representation as a massive (4-component) Dirac fermion,
\beq L = \bar{\psi} i \gamma^{\mu} (\d_{\mu} - i a_{\mu}) \psi - m \bar{\psi} \psi \label{eq:Diracintro}\eeq
$u(1)$ symmetry acts on $\psi$ via Eq.~(\ref{eq:U1intro}) and $T$ acts as the standard $T$ symmetry, Eq.~(\ref{eq:Tintro}). 
For future convenience, we have included the coupling of $\psi$ to a background $u(1)$ gauge field $a_{\mu}$. 
The non-trivial and trivial TI phases can be represented by the above Lagrangian with $m < 0$ an $m > 0$ respectively.\footnote{The choice of overall sign is a convention, but the relative sign is important.} One can easily check that if one lets $m$ vary in space from $m< 0$ to $m > 0$ then the interface supports a single (2-component) gapless $2+1$D Dirac fermion, as required. Thus, the $u(1)$ gauge theory ${\cal L}_T$ is obtained by starting with the non-trivial TI phase and promoting $a_{\mu}$ to a dynamical gauge field.

Now let's proceed to topological superconductors (TSc) of fermions (class AIII in condensed matter terminology). A TSc is an SPT phase of fermions with $u(1)$ and $T$ symmetry.\footnote{Physically, one typically thinks of the $u(1)$ symmetry as conservation of the $z$-component of electron spin in a superconductor.}  The $u(1)$ symmetry again acts on fermions $\psi$ via (\ref{eq:U1intro}), but the time-reversal symmetry acts as
\beq T: \quad \psi(\vec{x}, t) \to \tilde{U}_T \psi^{\dagger}(\vec{x}, -t), \quad i \to -i\quad \quad ({\rm TSc}) \label{eq:Tgintro2}\eeq
Note that $T$ and $u(1)$ symmetries now commute, so the symmetry group is $u(1) \times T$.  Canonically, one chooses $\tilde{U}^2_T = -1$, so that $T^2 = (-1)^F$ and $\psi$ is nominally a Kramers doublet; however, in the present case this is a matter of convention, since one can combine $T$ with a $\pi/2$ rotation in the $u(1)$ group, obtaining $\tilde{T} = u_{\pi/2} T$ with $\tilde{T}^2 = +1$. 

In the absence of interactions, topological superconductors in class AIII have an integer classification.\cite{KitaevNI, LudwigNI} The surface of a phase in class $\nu \in \bZ$ supports $|\nu|$ gapless $2+1$D Dirac fermions. In the presence of interactions, the phases $\nu$ and $\nu +8$ become continuously connected.\cite{FidkowskiChenAV, Wang2014, MetlitskiChenFidkowskiAV2014}

 Non-interacting topological superconductors again have a convenient field-theoretic representation in terms of a massive (4-component) Dirac fermion  (\ref{eq:Diracintro}). The $u(1)$ symmetry acts on the Dirac fermion $\psi$ via phase rotations, Eq.~(\ref{eq:U1intro}), and the time-reversal symmetry acts as $CT$ of Eq.~(\ref{eq:CTintro}). 
The trivial ($\nu = 0$) phase is represented by the Dirac fermion (\ref{eq:Diracintro}) with $m > 0$, and the $\nu = 1$ phase is represented by the Dirac fermion with $m < 0$. More generally, the phase $\nu$ is represented by $\nu$ Dirac fermions with $m < 0$ (and one uses $\nu$ Dirac fermions with $m > 0$ as a reference trivial phase). The $u(1)$ gauge theory  ${\cal L}_{CT}$ is obtained by starting with the $\nu = 1$ TSc and gauging the $u(1)$ symmetry. We will comment on $u(1)$ gauge-theories derived from TSc's with other $\nu$ in sections \ref{sec:ambiguity} and \ref{sec:Zftimes}.

\section{Dyon spectrum and $S$-duality}
\label{sec:dyons}
In this section we review dyon properties in a $u(1)$ gauge theory with topological angle $\theta$ and discuss the action of $S$-duality on the dyons. We also review the quantum numbers of dyons under time-reversal in the two $u(1)$ gauge theories ${\cal L}_T$ and ${\cal L}_{CT}$ and show that these are preserved by  $S_f$ duality, Eq.~(\ref{eq:Sfdyonsintro}).

Let us begin with a $u(1)$ gauge theory with bosonic matter and action (\ref{eq:Ltheta}). All dyons in this theory can be labeled by two quantum numbers: the electric charge $q$ and the magnetic charge $m$. The magnetic charge $m$ is an integer at any $\theta$. Due to the Witten effect\cite{WittenEffect}, the electric charge $q$ satisfies,
\beq q = n + \frac{\theta m}{2 \pi}, \quad n \in \bZ \label{eq:Witten} \eeq
where the integer $n$ reflects the freedom of adding an arbitrary number of electric charges to the monopole. Thus, the dyon excitations form a two-dimensional lattice. 
Two dyons $(q,m)$ and $(q',m')$ experience a (static) Coulomb interaction
\beq E = \frac{1}{4 \pi r} \left(e^2 q q' + \frac{(2\pi)^2}{e^2} m m'\right) \label{eq:Coulomb} \eeq
Dyons also experience the following statistical interaction: if we place $(q',m')$ at the origin and let $(q,m)$ move along a closed path $C$, the statistical Berry's phase is $e^{i (q m' - m q') \Omega/2}$, with $\Omega$ - the solid angle subtended by $C$. In addition to mutual statistics, the excitations have the self-statistics $(-1)^{n m} = (-1)^{(q - \theta m/(2\pi)) m}$, with $+1$ - corresponding to bosonic statistics and $-1$ to fermionic.\cite{Goldhaber,MetlitskibTI} Note that while electric charges of dyons are periodic under $\theta \to \theta + 2\pi$, the statistics is only periodic under $\theta \to \theta + 4\pi$. In particular, the neutral monopole is a boson at $\theta = 0$, a fermion at $\theta = 2\pi$ and a boson again at $\theta = 4\pi$. 

Under $SL(2,\bZ)$ duality the dyons get relabeled.\cite{Cardy} The operation $S$ corresponds to a relabeling $\tilde{n} = m, \tilde{m} = -n$ accompanied by a change of the coupling constant $\tilde{\tau} = -1/\tau$, with $\tau$ given by Eq.~(\ref{eq:introtaudef}). Here $n$ is defined via Eq.~(\ref{eq:Witten}). Equivalently, in terms of the total electric and magnetic charge,
\beq \left(\begin{array}{c} \tilde{q} \\ \tilde{m}\end{array}\right) = \left(\begin{array}{cc} \frac{\frac{\theta}{2\pi}}{\left(\frac{\theta}{2\pi}\right)^2 + \left(\frac{2\pi}{e^2}\right)^2} & \frac{\left(\frac{2\pi}{e^2}\right)^2}{\left(\frac{\theta}{2\pi}\right)^2 + \left(\frac{2\pi}{e^2}\right)^2}\\ -1 & \frac{\theta}{2\pi}\end{array}\right)\left(\begin{array}{c} q \\ m\end{array}\right) \label{eq:qm}\eeq
  We see that the Coulomb interaction (\ref{eq:Coulomb}) between the dyons is preserved by this relabeling, as is the statistical interaction and the dyon self-statistics. 

The ${\cal T}$ operation corresponds to a relabeling $\tilde{n} = n+ 2m$, $\tilde{m} = m$, accompanied by $\tilde{\tau} = \tau+2$, or equivalently $\tilde{q} = q$, $\tilde{m} = m$. Thus, this operation is just a shift $\tilde{\theta} = \theta + 4\pi$, $\tilde{e} = e$.  

We next proceed to a $u(1)$ gauge theory with fermionic matter and action (\ref{eq:Ltheta}). The only difference in the dyon spectrum compared to the bosonic case is that now dyon statistics is given by $(-1)^{n (m+1)} = (-1)^{(q-\theta m/(2\pi))(m+1)}$, reflecting the fermionic nature of the charge excitation. As a result, the $\theta$ angle is periodic modulo $2\pi$. In particular, a single monopole (of any charge) is a boson at any $\theta$. Let us discuss $S$-duality in the present case. For simplicity, first imagine that $\theta = 0$. The single charge is a fermion and a single monopole is a boson. If we interchange these excitations we get a theory with a bosonic charge and a fermionic monopole, i.e. a theory with bosonic matter at $\theta = 2 \pi$. More generally, if we start with a theory with fermionic charge matter and  relabel charges and monopoles according to Eq.~(\ref{eq:qm}), we get a theory with bosonic matter and coupling constant $\tilde{\tau} =- \tau^{-1} + 1$. 

As already noted in the introduction, one can also find an operation $S_f$ which maps the theory with fermionic matter back to itself. Indeed, let $S_f = S^{-1}_{bf} {\cal T}^{-1}_b S_{bf}$, where $S_{bf}: \tau \to -\tau^{-1} + 1$ maps a fermionic theory to a bosonic theory, and ${\cal T}_b: \tau \to \tau+2$, shifts $\theta$ in the bosonic theory by $4\pi$. Then $S_f: \tau \to \tau/(2 \tau +1)$, is a duality of the fermionic theory. Observe that $S_f$ fixes the value $\theta  = \pi$; $S_f: (\theta = -\pi, e) \to (\theta = \pi, \frac{4 \pi}{e})$. The corresponding action on electric and magnetic charges at $\theta = \pi$ is:
\beq S_f:\quad \left(\begin{array}{c} \tilde{q} \\ \tilde{m}\end{array}\right) =  \left(\begin{array}{cc} 0& 1/2\\ -2 & 0 \end{array}\right)\left(\begin{array}{c} {q} \\ {m}\end{array}\right) \label{eq:Sfdyons}\eeq

Next, we discuss the dyon lattice in our two $u(1)$ gauge-theories ${\cal L}_T$ and ${\cal L}_{CT}$. Recall that these are obtained by starting with the Dirac theory (\ref{eq:Lpsiintro}) with $\theta = \pi$ and implementing time-reversal via $T$ and $CT$ respectively. We begin with the case ${\cal L}_{CT}$. After integrating the gapped fermions out (in flat space), one obtains an effective action for the gauge field $a_{\mu}$ in Eq.~(\ref{eq:Ltheta}) with $\theta = \pi$. Thus, the electric charge $q$ and the magnetic charge $m$ of a dyon satisfy $q = n + \frac{m}{2}$ with $n \in \bZ$.  The electric charge is inverted by $CT$ and the magnetic charge is preserved by $CT$, Eq.~(\ref{eq:CTqm}). We can also ask about the Kramers parity $(CT)^2$ of the dyons. In general, only excitations whose topological sector is preserved by $CT$ can be assigned a value of $(CT)^2$. Thus, the only dyons which carry a definite value of $(CT)^2$ are pure monopoles $(q = 0, m)$, with $m$ - an even integer. The minimal such excitation - the neutral double monopole $(q = 0, m = 2)$ - is a Kramers doublet ($(CT)^2 = -1$) fermion, as we review below.\cite{Wang2014, MetlitskiChenFidkowskiAV2014}  
A convenient basis for the lattice of dyon excitations is provided by dyons $d_+: (q = 1/2, m = 1)$ and $d_-: (q = -1/2, m = 1)$. These excitations are exchanged by the time-reversal symmetry, $CT: d_+\leftrightarrow d_-$. Furthermore,  $d_+$ and $d_-$ fuse to the double monopole $(q = 0, m = 2)$. The Kramers parity $(CT)^2 = -1$ of this fusion product is fixed by the fact that $d_+$ and $d_-$ are $CT$-partners and possesses non-trivial mutual statistics.\cite{Metlitski2013,Wang2013a} 

Next, consider the theory ${\cal L}_T$. Integrating the gapped fermions out, we again get the effective action (\ref{eq:Ltheta}) with $\theta = \pi$, so the electric and magnetic charges of dyons are related by $q = n + \frac{m}{2}$ with $n \in \bZ$. However, the transformation properties of dyons under time-reversal differ from those in ${\cal L}_{CT}$: the electric charge is preserved by $T$ and the magnetic charge is inverted, Eq.~(\ref{eq:Tqm}). The only dyons which can be assigned a definite Kramers parity are pure charges $(q, m = 0)$ with $q \in \bZ$. The minimal such excitation is the single charge $(q = 1, m = 0)$, which is the Kramers doublet ($T^2 = -1$) fermion $\psi$. We can choose the following basis for the dyon lattice: $\tilde{d}_+: (q = 1/2, m = -1)$ and $\tilde{d}_-: (q = 1/2, m = 1)$. Time-reversal exchanges these two dyons, $T: \tilde{d}_+ \leftrightarrow \tilde{d}_-$. The fusion product $\tilde{d}_+ \tilde{d}_-$ is the single charge $(q = 1, m = 0)$, which as we noted, is a Kramers doublet fermion in accordance with the non-trivial mutual statistics between $\tilde{d}_+$ and $\tilde{d}_-$.

Now, if we consider the theory ${\cal L}_{CT}$ with coupling constant $e$ and the theory ${\cal L}_T$ with coupling constant $4 \pi/e$, in the absence of time-reversal symmetry these theories are clearly related by $S_f$ duality (\ref{eq:Sfdyons}). What is more surprising is that the duality preserves the action of time-reversal on the dyon lattice, mapping $CT$ symmetry to $T$ symmetry (see Fig.~\ref{fig:Dyons}). Indeed, under (\ref{eq:Sfdyons}), $S_f: d_+ \to \tilde{d}_+, \,\, d_- \to \tilde{d}_-$. Thus, the pair of $CT$ partners $d_{+}$, $d_-$ is mapped to the pair of $T$ partners $\tilde{d}_+$, $\tilde{d}_-$. As these dyons generate the entire dyon lattice, we conclude that the action of time-reversal on the entire lattice is preserved. In particular, the $(CT)^2 = -1$ double  monopole of ${\cal L}_{CT}$, $d_+ d_-$, is mapped to the $T^2 = -1$ single charge of ${\cal L}_T$, $\tilde{d}_+ \tilde{d}_-$. 

\subsection{Bosonic SPT ambiguity}
\label{sec:ambiguity}

As already remarked in the introduction, while the preservation of the dyon interactions, statistics and quantum numbers under time-reversal by $S_f$ suggests that ${\cal L}_T$ and ${\cal L}_{CT}$ are dual to each other, it is not sufficient to establish a full duality. 
Indeed, these two $u(1)$ gauge theories could potentially differ by an  SPT phase of neutral bosons with time-reversal symmetry. 

We recall that SPT phases of bosons with $T$-symmetry in $3+1$D are believed to have a $\bZ_2^2$ classification.\cite{AVTS,Burnell2013,KapustinBos,KapustinBosTI} The two $\bZ_2$ root phases are most conveniently labeled by their gapped $T$-preserving topologically ordered $2+1$D surface states. The surface of one of the phases supports a toric code topological order ($\bZ_2$ gauge theory) where electric and magnetic charges are Kramers doublet bosons under $T$. This surface topological order and its corresponding bulk SPT phase is labeled eTmT. The surface of the second $\bZ_2$ phase supports a topological order with anyon content $\{1, f_1, f_2, f_3\}$, where $f_1$, $f_2$, $f_3$  are Kramers singlet fermions. The fusion rules are $f_i \times f_i = 1$ and $f_i \times f_j = f_k$ for $i \neq j \neq k$.  This topological order and its bulk SPT phase is labeled FFF. 

The bulk of an SPT phase is fully gapped and possesses no topologically non-trivial excitations. Therefore, adding an SPT phase to one of our $u(1)$ gauge theories will not affect the bulk dyon properties. It will, however, affect the surface spectrum of the $u(1)$ gauge theory. The difficulty is that ${\cal L}_T$ and ${\cal L}_{CT}$ can each support many different surface phases and establishing a direct correspondence between them is challenging. 




We can, however, detect an SPT phase in the bulk by computing its partition function on a non-trivial closed manifold. For SPT phases protected by time-reversal symmetry, certain phases can only be detected by studying their partition function on non-orientable manifolds. Expressions for the partition functions of the eTmT and FFF phases on an arbitrary manifold $M$ in terms of the Stiefel-Whitney classes of the manifold have been deduced by Kapustin in Refs.~\onlinecite{KapustinBos, KapustinBosTI}. The partition function of the eTmT phase is
\beq Z_{{\rm eTmT}} = \exp\left(\pi i \int_M w^4_1\right) \label{eq:eTmTa}\eeq
and the partition function of the FFF phase is 
\beq Z_{{\rm FFF}} = \exp\left(\pi i \int_M w_4\right) = (-1)^{\chi} \label{eq:FFFa} \eeq
Here $w_1 \in H^1(M, \bZ_2)$ is the first Stiefel-Whitney class. Given a closed 1-cycle $C$, $\int_C w_1$ measures whether $C$ is orientation-reversing. $w_4 \in H^4(M, \bZ_2)$ is the top Stiefel-Whitney class, whose integral over the manifold is equal to the Euler number $\chi$ modulo $2$. Products of Stiefel-Whitney classes are taken using the cup product. In Ref.~\onlinecite{KapustinBos} it was shown that when the bulk actions (\ref{eq:eTmTa}), (\ref{eq:FFFa}) are placed on a manifold with a boundary, gauge-invariance $w_i \to w_i + d \alpha_{i-1}$ is spoiled. However, if one places the action of the eTmT/FFF TQFT on the boundary, its anomaly precisely compensates the bulk anomaly. 

From Eq.~(\ref{eq:eTmTa}), the partition function of the eTmT phase on an arbitrary orientable manifold is equal to $1$, but is equal to $-1$ on the non-orientable manifold $\mathbb{RP}^4$. The FFF phase can be ``identified" already on orientable manifolds: for instance, the orientable manifold $\mathbb{CP}^2$ has Euler number $\chi  = 3$, so $Z_{{\rm FFF}}(\mathbb{CP}^2) = -1$. Therefore, we can rule out the possibility that $u(1)$ gauge theories ${\cal L}_T$ and ${\cal L}_{CT}$ differ by the eTmT or FFF phase if we establish the equality of their partition functions (\ref{eq:introZduality})  on manifolds $\mathbb{RP}^4$ and $\mathbb{CP}^2$.  The rest of this paper is devoted to this task (in fact, we will be able to establish the duality on $\mathbb{RP}^4$ and on arbitrary orientable manifolds.) While a full mathematical classification of time-reversal invariant $u(1)$ gauge theories has not been developed to date, 
one might expect that if two such $u(1)$ gauge theories have the same interactions, statistics and quantum numbers of dyon excitations then they differ at most by a bosonic SPT phase with $T$ symmetry.\cite{Wang2015} We will rule out the latter possibility so one may argue that ${\cal L}_T$ and ${\cal L}_{CT}$ are fully dual.

The reader may wonder why we are only considering $T$-symmetric SPT phases of ${\it bosons}$ as a potential ambiguity in the duality and excluding SPT phases of fermions. The reason is that we view ${\cal L}_T$ and ${\cal L}_{CT}$ as gauge theories emergent from a microscopic Hilbert space made of bosons. Indeed, all gauge-invariant local operators in ${\cal L}_T$ and ${\cal L}_{CT}$ are bosons. The fermionic matter fields are not gauge-invariant and so are not local operators. Moreover, the partition function of these theories can be defined on an arbitrary 4D manifold and does not take a Spin structure as an input (or Pin structure in the non-orientable case).  

We conclude this section by noting that there is a simple way to generate a phase differing from ${\cal L}_{CT}$ by the eTmT SPT. Consider a variation of our $u(1)$ gauge theories, where instead of a single Dirac fermion with mass $m < 0$,  Eq.~(\ref{eq:Diracintro}), we take $\nu$ such Dirac fermions with $\nu$ - odd. In flat space (and, in fact, on an arbitrary oriented manifold), the resulting gauge theory will still have a $\theta$-term (\ref{eq:Zpsio}) with $\theta = \pi$, so the quantum numbers of dyon excitations will be the same as for $\nu = 1$. In the case when time-reversal is implemented with $T$, changing $\nu \to \nu + 2$, indeed, does not alter the theory (in the $|m| \to \infty$ limit), as is evident from the partition function (\ref{eq:ZTIintro}) for $\nu  =1$ being equal to $\pm 1$ on an arbitrary manifold. 
 Recall that a single Dirac fermion with $m < 0$ has the interpretation of a topological insulator, so this is consistent with non-interacting TIs having a $\bZ_2$ classification. 
Turning to the case when time-reversal is implemented with $CT$, a single Dirac fermion with $m < 0$ represents a $\nu = 1$ topological superconductor in class AIII. Such TSc's have an integer non-interacting classification $\nu \in \bZ$. In the presence of interactions,  phases $\nu$ and $\nu + 8$ become continuously connected. \cite{Wang2014, MetlitskiChenFidkowskiAV2014}  This is reflected in the $\nu  =1$ partition function (\ref{eq:ZTScintro}) being a power of $e^{2\pi i/8}$ on an arbitrary manifold.  Furthermore, the $\nu = 4$ TSc is precisely the eTmT phase of neutral bosons,\cite{FidkowskiChenAV, Wang2014, MetlitskiChenFidkowskiAV2014} as we will verify using the bulk partition functions (\ref{eq:ZTScintro}), (\ref{eq:eTmTa}) in section \ref{sec:SW}. Thus, $\nu = 1$ and $\nu  = 5$ TSc phases differ by eTmT phase, and so do $\nu = 3$ and $\nu = 7$. Furthermore, as we will show in section \ref{sec:Zftimes}, once the $u(1)$ gauge field $a_{\mu}$ is made dynamical, phases $\nu$ and $-\nu$ coalesce, so $\nu = 1 \sim -1 \sim 7$ and $\nu = 3 \sim  -3 \sim 5$. Thus, among ${\cal L}_{CT}$ theories based on odd $\nu$, only $\nu  =1$ and $\nu = 5$ are distinct, differing precisely by the eTmT phase.

\section{Dirac fermion in curved space}
\label{sec:Dirac}

We now proceed with our main program of establishing the equality of partition functions (\ref{eq:introZduality}) of theories ${\cal L}_T$ and ${\cal L}_{CT}$ on an arbitrary closed (Euclidean) manifold.  As a first step, we need to deduce the partition functions of the Dirac fermion with $T$ and $CT$ symmetry,  Eqs.~(\ref{eq:ZTIintro}) and (\ref{eq:ZTScintro}). 
In the introduction, our expressions for the Lagrangian of the Dirac fermion and the action of time-reversal symmetry were presented in flat Minkowski space. We begin this section by transitioning to flat Euclidean space. We then discuss the generalization to curved Euclidean space and the computation of partition functions.

\subsection{Flat Euclidean space}

We start with a Dirac fermion in flat Euclidean space:
\beq L = \bar{\psi} \gamma^{\mu}(\d_{\mu} - i a_{\mu})\psi + m \bar{\psi} \psi \eeq
The 4x4 $\gamma$ matrices satisfy,  $\{\gamma^{\mu}, \gamma^{\nu}\} = 2 \delta^{\mu \nu}$, $(\gamma^{\mu})^{\dagger} = \gamma^{\mu}$, and we define $\gamma^5 = \gamma^0 \gamma^1 \gamma^2 \gamma^3$, so that $(\gamma^5)^{\dagger} = \gamma^5$ and $\{\gamma^{\mu}, \gamma^5\} = 0$. 

The theory possesses the following discrete symmetries:

\bea &C:& \quad \psi(x) \to C \bar{\psi}^T(x), \quad  \bar{\psi}(x) \to - \psi^T(x) C^{\dagger}, \quad a_{\mu}(x) \to - a_{\mu} (x),\\
\nn\\
&T:& \quad \psi(\vec{x}, \tau) \to i \gamma^0 \gamma^5 C \bar{\psi}^T(\vec{x}, -\tau), \quad \bar{\psi}(\vec{x}, \tau) \to i\psi^T(\vec{x}, -\tau) C^{\dagger} \gamma^5 \gamma^0, \nn\\
&& \quad a_{\tau}(\vec{x}, \tau) \to a_\tau(\vec{x}, -\tau), \,\,
a_{i}(\vec{x}, \tau) \to -a_i(\vec{x}, -\tau), \quad\quad\quad ({\rm TI}) \label{eq:TEu}\\
\nn\\
&CT:& \quad \psi(\vec{x}, \tau) \to i \gamma^0 \gamma^5 \psi(\vec{x}, -\tau), \quad \bar{\psi}(\vec{x}, \tau) \to -i\bar{\psi}(\vec{x}, -\tau) \gamma^5 \gamma^0, \nn\\
&& \quad a_{\tau}(\vec{x}, \tau) \to -a_\tau(\vec{x}, -\tau), \,\,
a_{i}(\vec{x}, \tau) \to a_i(\vec{x}, -\tau) \quad\quad\quad ({\rm TSc}) \label{eq:CTEu}\eea
where $C$ is a unitary matrix satisfying, $C (\gamma^{\mu})^T C^{\dagger} = -\gamma^{\mu}$, $C (\gamma^5)^T C^{\dagger} = \gamma^5$ and $C^* C = - 1$. The physical time-reversal symmetry of the TI is represented by the $T$ symmetry (\ref{eq:TEu}) and the physical time-reversal symmetry of the TSc is represented by the $CT$ symmetry (\ref{eq:CTEu}). Note that $C$, $T$ and $CT$ should be thought of as symmetries of the Grassmann path integral. In particular, they do not involve any complex conjugation. Thus, time-reversal symmetry, which is an anti-unitary symmetry in real-time Hamiltonian formulation becomes an ordinary (unitary-like) symmetry of the Grassmann path integral. In fact, in Euclidean space it is just related by a space-time rotation to a spatial reflection symmetry. The latter is a unitary symmetry already in the real-time formulation. The form of Eqs.~(\ref{eq:TEu}) and (\ref{eq:CTEu}) appears counter-intuitive with $\psi \to \bar{\psi}$ for $T$, and $\psi \to \psi$ for $CT$, which is opposite of what happens in the real-time formulation, Eqs.~(\ref{eq:Tintro}) and (\ref{eq:CTintro}). This is again a consequence of the time-reversal symmetry involving no complex conjugation in the path-integral formulation. Note that in the path integral formulation $T$ and a $u(1)$ rotation $u_\alpha: \psi \to e^{i \alpha} \psi$  satisfy $T u_\alpha T^{-1} = u_{-\alpha}$, while $CT$ commutes with $u(1)$ rotations. These relationships also hold for the real-time implementations (\ref{eq:Tintro}) and (\ref{eq:CTintro}). However, note that in Euclidean space both $T^2:\, \psi \to + \psi$ and $(CT)^2:\, \psi \to +\psi$. While the latter relationship can be changed by combining $CT$ with a $u(1)$ rotation, the former relationship cannot. The change of $T^2 = (-1)^F$ to $T^2 = +1$ upon continuation to Euclidean space is in accordance with the discussion in Ref.~\onlinecite{KapustinFerm}.

\subsection{Curved Euclidean space}
\label{sec:curved}
Next, we generalize the above discussion to curved Euclidean space. Now the action of the theory is,
\beq S = \int d^4 x \sqrt{g} \,L \eeq
where 
\beq L = \bar{\psi} e^{\mu}_a \gamma^a (\d_{\mu} + i \omega_{\mu} - i a_{\mu}) \psi + m \bar{\psi} \psi \label{eq:LDcurved}\eeq
Let us summarize the notation. $g_{\mu \nu}(x)$ denotes the metric of the manifold, $g^{\mu \nu}$ is the inverse of the metric, $\sqrt{g} = (\det g_{\mu \nu})^{1/2}$. $e^{\mu}_a$ is the vielbein, satisfying
\beq e^{\mu}_a e^{\nu}_b g_{\mu \nu} = \delta_{ab} \eeq Latin letters $a, b, c, \ldots$ run over local frame indices, while $\mu, \nu, \lambda$ still run over the coordinate indices. We also define, $e^{a}_{\mu} = g_{\mu \nu} e^{\nu}_a$, which satisfy,
\beq e^{a}_{\mu} e^a_{\nu} = g_{\mu \nu},\quad e^{\mu}_a e^{a}_{\nu} = \delta^{\mu}_{\nu}, \quad e^{a}_{\mu} e^{\mu}_{b}  = \delta_{ab}\eeq
Generally, Greek indices are raised and lowered using the metric $g$. Raising/lowering of frame indices has no effect. The spin connection $\omega_{\mu}$ is given by
\beq \omega_{\mu} = \frac{1}{2} \omega^{ab}_{\mu} \Sigma_{ab}, \quad \Sigma_{ab} = \frac{-i}{4} \left[ \gamma^a, \gamma^b  \right] \label{eq:spincon}\eeq
where $\Sigma_{ab}$ are the generators of local $SO(4)$ transformations and
\beq \omega^{ab}_{\mu} = e^a_{\nu} (\d_{\mu} e^{\nu}_b + \Gamma^{\nu}_{\mu \lambda} e^{\lambda}_b) = e^a_{\nu} (\nabla_{\mu} e_b)^\nu \label{eq:spincon2}\eeq
Here $\Gamma^{\mu}_{\nu \lambda}$ is the Christoffel symbol and $\nabla$ denotes the covariant derivative.

Generally, we need several coordinate patches to cover our manifold. Suppose we have a patch 1 with local coordinates $x$ and a patch 2 with local coordinates $y$. We will denote fields on patch $2$ with a tilde. On the intersection of these patches we have the following transformation rules:

\bea \tilde{g}_{\mu \nu}(y) &=& \frac{\d x^{\alpha}}{\d y^{\mu}} \frac{\d x^{\beta}}{\d y^{\nu}} g_{\alpha \beta}(x(y)) \\
\tilde{e}^a_{\mu}(y) &=& \frac{\d x^{\alpha}}{\d y^{\mu}} R^{ab} (y) e^b_\alpha(x(y)) \label{eq:vieltransf}\eea
Here $R(y) \in O(4)$ is a coordinate dependent matrix describing the transformation between the frames in the two patches. If the manifold is oriented then $R(y) \in SO(4)$. 

Let us discuss the transition functions on an oriented manifold first. Since $R$ is an $SO(4)$ matrix in this case, we may write it as,
\beq R(y) = \exp\left(\frac{i}{2} \theta_{ab}(y) L_{ab}\right) \eeq
where $L_{ab}$ is a generator of $SO(4)$: $(L_{ab})_{cd} = -i (\delta^{ac} \delta^{bd} - \delta^{bc} \delta^{ad})$. Now, we can lift $R$ to the group $\mathrm{Spin}(4)$, i.e. the double-cover of $SO(4)$, as
\beq U(y) = \exp\left(\frac{i}{2} \theta_{ab}(y) \Sigma_{ab}\right) \label{eq:SO4lift}\eeq
(The lift is $1$ to $2$, as both $U(y)$ and $-U(y)$ project back onto $R$). We glue the fermion fields on the two patches using,
\beq \widetilde{\psi}(y) = e^{i \alpha(y)} U(y) \psi(x(y)), \quad \widetilde{\bar{\psi}}(y) =  \bar{\psi}(x(y)) U^{\dagger}(y) e^{-i \alpha(y)} \label{eq:psior} \eeq
 The gauge field needs to transform accordingly:
\beq \tilde{a}_{\mu}(y) = \frac{\d x^{\beta}}{\d y^{\mu}} a_{\beta}(x(y)) + \d_{\mu} \alpha(y) \label{eq:atransf}\eeq
Note that the transition functions $g_{21}(y) = e^{i \alpha(y)} U(y)$ belongs to the group $\mathrm{Spin}(4)_c = \left(u(1) \times \mathrm{Spin}(4)\right)/\bZ_2$, with $\bZ_2$ generated by $(-1,-1)$. If we have an intersection of 3 patches, the transition functions must be consistent on this intersection, i.e. $g_{31} = g_{32} g_{21}$ (this is known as the cocycle condition). Such a set of transition functions on a manifold is known as a $\mathrm{Spin}_c$ structure. All oriented four-dimensional manifolds admit a $\mathrm{Spin}_c$ structure.\cite{HH} For a given manifold, two different $\mathrm{Spin}_c$ structures differ only by the $u(1)$ factors, $e^{i \alpha'} = e^{i \beta} e^{i \alpha}$, and $e^{i \beta}$ must themselves satisfy the cocycle condition. Thus, two different $\mathrm{Spin}_c$ structures differ by a complex line bundle. 

Using the transformation law for the vielbein (\ref{eq:vieltransf}) one can show that the spin-connection (\ref{eq:spincon}), (\ref{eq:spincon2}) transforms as,
\beq \tilde{\omega}_{\mu}(y) = \frac{\d x^{\beta}}{\d y^{\mu}} U \omega_{\beta} U^{\dagger} + i \d_{\mu} U(y) U^{\dagger}(y) \label{eq:omegatransf}\eeq
so that the Lagrangian (\ref{eq:LDcurved}) is invariant under the change of patch.

We now proceed to the case of a non-orientable manifold. For orientation-preserving patch changes we still have transformation properties (\ref{eq:psior}), (\ref{eq:atransf}). However, some of patch changes are now orientation-reversing, in which case the matrix $R$ in Eq.~(\ref{eq:vieltransf}) has $\det R(y) = -1$. We may then write 
\beq R(y) = R_0(y) R_\tau\eeq
where $R_\tau = \mathrm{diag}(-1,1,1,1)$ and $\mathrm{det} R_0 = 1$. We will need to use the physical time-reversal symmetry to glue the fermion fields on the two patches. In the case of ${\cal L}_{CT}$ (TSc) the physical time reversal symmetry is represented by $CT$, so we have
\beq {\cal L}_{CT}: \quad \widetilde{\psi}(y) = e^{i \alpha(y)} U(y)  \psi(x(y)), \quad \widetilde{\bar{\psi}} = \bar{\psi}(x(y)) U^{\dagger}(y) e^{-i \alpha(y)} \label{eq:AIIIglue}\eeq
where
\beq U(y) = U_0(y) i \gamma^{0} \gamma^5 \label{eq:Upin}\eeq
and $U_0(y)$ is the lift (\ref{eq:SO4lift})  of $R_0(y) \in SO(4)$ to $\mathrm{Spin}(4)$. $U(y)$ is now an element of the group $\mathrm{Pin}(4)_+$. The overall transition functions $e^{i \alpha} U$ are elements of group ${\rm Pin}(4)_c = \left(u(1) \times \mathrm{Pin}(4)_+\right)/\bZ_2$, with $\bZ_2$ generated by $(-1,-1)$. The transition functions must again satisfy the cocycle condition. Such a structure on a manifold is known as $\mathrm{Pin}_c$ structure. Not all 4-dimensional manifolds admit a Pin$_c$ structure (e.g. $\mathbb{RP}^2 \times \mathbb{RP}^2$ does not).\cite{BahriGilkey, RP2RP2}  When one Pin$_c$ structure on a manifold does exists, other Pin$_c$ structures  again differ only by the $u(1)$ phase factors, i.e. by complex line bundles. 

The transformation law of the gauge field in ${\cal L}_{CT}$  under orientation-reversing patch changes is still given by Eq.~(\ref{eq:atransf}). The spin connection $\omega_{\mu}$ still transforms according to Eq.~(\ref{eq:omegatransf}), with $U$ given by the full expression (\ref{eq:Upin}). 

Proceeding to ${\cal L}_T$ (TI) on a non-orientable manifold, the physical time-reversal symmetry is represented by $T$, so under orientation-reversing patch transformations we should glue the spinor fields as,
\beq {\cal L}_T: \quad \widetilde{\psi}(y) = e^{i \alpha(y)} U(y)  C \bar{\psi}^T(x(y)), \quad \widetilde{\bar{\psi}}(y) = - \psi^T(x(y)) C^{\dagger} U^{\dagger}(y) e^{-i \alpha(y)} \label{eq:AIIglue}\eeq
with $U(y)$ still given by Eq.~(\ref{eq:Upin}). The transformation of the gauge field under an orientation-reversing patch change is now given by,
\beq {\cal L_{T}}: \quad \tilde{a}_{\mu}(y) = - \frac{\d x^{\beta}}{\d y^{\mu}} a_{\beta}(x(y)) + \d_{\mu} \alpha(y) \label{eq:atransfAII}\eeq
As the time-reversal transformations now mix $\psi$ and $\bar{\psi}$, it will be more convenient to go to a ``Majorana" basis of Grassmann fields, $\chi = (\chi_1, \chi_2)$, 
\beq \chi_1 = \frac{\psi + \psi_c}{2}, \quad \chi_2 = \frac{\psi - \psi_c}{2 i} \eeq
with $\psi_c = C \bar{\psi}^T$.\footnote{In Minkowski space and in a basis of $\gamma$ matrices where Lorentz transformations are purely real, we have $\chi^{\dagger}_1 = \chi_1$ and $\chi^{\dagger}_2 = \chi_2$, so these are, indeed, Majorana fields.} The Lagrangian can be re-written as 
\beq L = \chi^T C^* \left[ e^{\mu}_a \gamma^a \left(\d_{\mu} + i \omega_{\mu} + i a_{\mu} \rho^2\right) + m\right]\chi \label{eq:LMajorana}\eeq
Here and below the Pauli matrices $\rho^{1,2,3}$ act on the two components of $\chi$. For orientation-reversing transformations, the transition functions (\ref{eq:AIIglue}) act on $\chi$ as,
\beq {\cal L}_T: \quad\tilde{\chi}(y) = U(y) e^{-i \alpha(y) \rho^2} \rho^3 \chi(x(y)) \label{eq:transchiT}\eeq
with $U(y)$ given by Eq.~(\ref{eq:Upin}), while for orientation-preserving transformations,
\beq  \tilde{\chi}(y) = U(y) e^{-i \alpha(y) \rho^2} \chi(x(y)) \label{eq:transchi}\eeq
with $U(y)$ given by Eq.~(\ref{eq:SO4lift}). The transition functions must again satisfy the cocycle condition; we call the resulting structure on the manifold a Pin$_{\tilde{c}}$ structure. Distinct Pin$_{\tilde{c}}$ structures on a manifold differ by ``twisted" complex line-bundles. Here and below we define a ``twisted" complex line-bundle  as a set of transition functions between patches $h_{ij} (y) \in O(2)$ satisfying $\det h  = +1$ if the transition preserves chart orientation and $\det h = -1$ if the transition reverses chart orientation, together with the cocycle condition $h_{31 } = h_{32} h_{21}$. 

\subsection{Partition function of a Dirac fermion: oriented manifold}

Having discussed in the previous section how to place a Dirac fermion on an arbitrary closed manifold, we now proceed to compute its partition function in the presence of a background $u(1)$ gauge field $a_{\mu}$. We begin with the case of an orientable manifold. Here the partition function of a Dirac fermion of mass $m$ is given by,

\beq Z[a_{\mu}] =  \mathrm{det}({\slashed{D}} + m)  \label{eq:Zoriented} \eeq
where 
\beq {\slashed D} =  e^{\mu}_a \gamma^a (\d_{\mu} + i \omega_{\mu} - i a_{\mu}) \eeq
The transition functions (\ref{eq:psior}) are part of the definition of ${\slashed D}$. 

Note that on an orientable manifold the partition function is independent of whether the physical time-reversal symmetry is $T$ or $CT$: as we saw in the previous section, the action of time-reversal symmetry only enters when orientation reversing transitions between patches are present. Thus, the partition functions on an orientable manifold are the same for ${\cal L}_T$ (TI) and ${\cal L}_{CT}$ ($\nu = 1$ TSc). Recall that we are representing a non-trivial TI/TSc by a Dirac fermion with negative mass $-|m|$ and the trivial TI/TSc by a Dirac fermion with positive mass $|m|$. Since we are only interested in the topological contribution to the partition function, which should be absent for a trivial TI/TSc, we will take the partition function of a $\nu = 1$ TI/TSc to be

\beq Z^o_{\psi, T/CT}[a_{\mu}] = \frac{\mathrm{det}({\slashed{D}} - |m|)}{\mathrm{det}({\slashed{D}} + |m|)} \label{eq:Zfodef} \eeq
The superscript $o$ serves to remind that we are working on an oriented manifold. Note that $- i {\slashed D}$ is a Hermitian operator with respect to the natural inner product $\langle \chi, \psi \rangle = \int d^4x \, \sqrt{g} \, \chi^{\dagger} \psi$. Denoting the eigenvalues of $- i {\slashed D}$ as $\lambda$, 

\beq  Z^o_{\psi, T/CT}[a_{\mu}]= \prod_{\lambda} \frac{ i \lambda - |m|}{i \lambda + |m|} \label{eq:Zfo}\eeq
Locally $\{ {\slashed D}, \gamma^5\} = 0$. Moreover, on an oriented manifold $\gamma^5$ commutes with the patch transition functions (\ref{eq:psior}) (as we will see, this is no longer true on a non-orientable manifold). Therefore, on an oriented manifold all eigenvalues of $- i {\slashed D}$ with $\lambda \neq 0$ come in pairs $\lambda$, $-\lambda$, with the corresponding eigenvectors $\psi_\lambda$, $\psi_{-\lambda} = \gamma^5 \psi_\lambda$. Moreover, all eigenvectors with $\lambda = 0$ can be chosen to be simultaneous eigenvectors of $\gamma^5$. Let the number of zero-modes of ${\slashed D}$ with $\gamma^5 = + 1$ be $N_+$ and the number of zero-modes with $\gamma^5 = -1$ be $N_-$. Then,

\beq  Z^o_{\psi, T/CT}[a_{\mu}] = (-1)^{N_+ + N_-} = (-1)^{N_+ - N_-}  \label{eq:Zfofinal}\eeq
An index theorem\cite{Atiyah} states that
\beq N_+ - N_- = \frac{1}{32 \pi^2} \int d^4 x \, \epsilon^{\mu \nu \lambda \sigma} f_{\mu \nu} f_{\lambda \sigma} - \frac{1}{768 \pi^2} \int d^4x \, \epsilon^{\mu \nu \lambda \sigma} R^{\alpha}_{\,\beta \mu \nu} R^{\beta}_{\,\alpha \lambda \sigma} \label{eq:index}\eeq
where $f_{\mu \nu} = \d_{\mu} a_{\nu} - \d_{\nu} a_{\mu}$ and $R^{\alpha}_{\, \beta \mu \nu}$ is the Riemann curvature tensor. The second term in Eq.~(\ref{eq:index}) is related to the signature of the manifold,
\beq \sigma(M) = \frac{1}{96 \pi^2}  \int d^4x \, \epsilon^{\mu \nu \lambda \sigma} R^{\alpha}_{\,\beta \mu \nu} R^{\beta}_{\,\alpha \lambda \sigma} \label{eq:signature}\eeq
so we can write,
\beq N_+ - N_- = \frac{1}{2 (2\pi)^2} \int_M \, f \wedge f - \frac{1}{8} \sigma(M) \label{eq:indexsigma}\eeq
with $f = \frac{1}{2} f_{\mu \nu} dx^{\mu} \wedge dx^{\nu}$. Thus, on an orientable manifold the partition function $Z^o_\psi$ has a local expression in terms of the gauge field strength and the curvature. Note also that $Z^o_\psi$ is invariant under smooth changes of the gauge field/metric. Indeed, as we change $a_{\mu}$, some eigenvalues of $-i {\slashed D}$ with $\lambda \neq 0$ might descend to $\lambda = 0$. However, they always descend in pairs $\psi_\lambda$, $\gamma^5 \psi_\lambda$, so that $\Delta N_+ = \Delta N_-$ and the partition function is unchanged.

While our primary focus is on the time-reversal invariant massive Dirac fermion, it will be convenient to consider it as a special case of a Dirac fermion with a complex mass,
\beq  L_m = |m| (\cos \theta \bar{\psi} \psi + i \sin \theta \bar{\psi} \gamma^5\psi) \label{eq:Diractheta}\eeq  
Such an action breaks the time-reversal symmetry as long as $\theta/\pi$ is not an integer. Thus, such a theory cannot generally be placed on a non-orientable manifold. However, on an oriented manifold the theory with an arbitrary $\theta$ is well-defined and the corresponding partition function is,
\beq Z^o_\psi[a_{\mu}, \theta] = \frac{\mathrm{det}({\slashed{D}} + |m| e^{i \theta \gamma^5})}{\mathrm{det}({\slashed{D}} + |m|)} \label{eq:Zftheta} \eeq
Here, we have again normalized the parition function so that $Z^o_\psi = 1$ for $\theta = 0$. The previously discussed partition function, Eq.~(\ref{eq:Zfodef}), corresponds to $\theta = \pi$. It is easy to see that the contributions of eigenstates of $- i {\slashed D}$ with $\lambda \neq 0$ to Eq.~(\ref{eq:Zftheta}) cancel out and
\beq Z^o_\psi[a_{\mu}, \theta] = e^{i \theta (N_+ - N_-)} \label{eq:Zfthetafinal}\eeq
with $N_+ - N_-$ given by the index theorem (\ref{eq:indexsigma}).

\subsection{Partition function of a Dirac fermion: non-orientable manifold, $CT$}
\label{sec:Zftimes}
We now proceed to the partition function of ${\cal L}_{CT}$ ($\nu = 1$ TSc) on an non-orientable manifold. Recall that we are representing a TSc by a massive Dirac fermion, where orientation reversing transformations between patches are implemented by Eq.~(\ref{eq:AIIIglue}). The Grassmann path integral is again given by 
\beq Z[a_{\mu}] = \det ({\slashed D} + m) \label{eq:Zdettimes}\eeq
To isolate the purely topological contribution we again divide the partition function of the $\nu = 1$ phase ($m < 0$) by the partition function of the trivial $\nu = 0$ phase ($m > 0$), obtaining,
\beq Z_{\psi,CT}[a_{\mu}] = \frac{\det ({\slashed D} - |m|)}{\det ({\slashed D} +|m|)} = \prod_\lambda \frac{i \lambda - |m|}{i \lambda + |m|} \label{eq:Zftimes} \eeq
The transition functions for orientation preserving (\ref{eq:psior}) and orientation reversing transformations (\ref{eq:AIIIglue}) are part of the definition of the operator ${\slashed D}$, i.e. the operator depends on the choice of a Pin$_c$ structure.  On an orientable manifold, (\ref{eq:Zftimes}) coincides with our previous definition (\ref{eq:Zfodef}). 

Crucially, the form of the spectrum of $-i {\slashed D}$ on a non-orientable manifold is generally different from that on an orientable manifold. While $\gamma^5$ again anticommutes with ${\slashed D}$ in each patch, $\gamma^5$ does not commute with the orientation reversing transition functions (\ref{eq:Upin}). Acting with $\gamma^5$ in each patch changes all orientation reversing transition functions by a factor of $(-1)$, i.e. the holonomy of the Dirac fermion around every orientation reversing loop is changed by a factor $-1$. Thus, acting with $\gamma^5$ maps a $\mathrm{Pin}_c$ structure to a different $\mathrm{Pin}_c$ structure: one with an inverted spectrum of the Dirac operator. For a given $\mathrm{Pin}_c$ structure eigenvalues of $- i {\slashed D}$ do not generally come in $\lambda, - \lambda$ pairs. 

Now, from Eq.~(\ref{eq:Zftimes}),
\beq \log Z_{\psi, CT}[a_{\mu}] = 2 i \sum_{\lambda} \tan^{-1}\left(\frac{|m|}{\lambda}\right) \eeq 
Here, we are only interested in $\log Z$ modulo $2 \pi i$. We  only consider energy scales below the mass gap of the Dirac fermion. Thus, we take the limit $m \to \infty$, obtaining,
\beq \log Z_{\psi, CT}[a_{\mu}] = \pi i \left(\sum_{\lambda \neq 0} \mathrm{sgn} (\lambda) + N_0\right) \label{eq:Zfsgn} \eeq
where $N_0$ is the number of zero-modes of ${\slashed D}$. Recall that the $\eta$-spectral function of a Dirac operator is defined as 
\beq \eta(s) = \sum_{\lambda \neq 0} \mathrm{sgn}(\lambda) |\lambda|^{-s}  \label{eq:etasdef}\eeq
The sum over $\lambda$ is convergent for large $Re(s)$ and can be analytically continued to $s = 0$. One then defines the $\eta$ invariant as,
\beq \eta = \frac{1}{2} (\eta(0) + N_0) \label{eq:etadef}\eeq
We have not been careful about regularization in deducing Eq.~(\ref{eq:Zfsgn}). In appendix \ref{sec:Eta}, we use $\zeta$-function regularization on the partition function (\ref{eq:Zftimes}) from the outset and obtain,
\beq \log Z_{\psi, CT}[a_{\mu}] = 2 \pi i \eta \label{eq:Zftimesfinal} \eeq
a result already suggested by the formal expression (\ref{eq:Zfsgn}). Note that $e^{2 \pi i \eta}$ is invariant under smooth deformations of the gauge field, i.e. it only depends on the Pin$_c$ structure.\cite{GilkeyPinC} In fact, a stronger statement holds: it is a bordism invariant of the $\mathrm{Pin}_c$ structure.\cite{GilkeyPinC} Further note that for an orientable manifold, $\eta(s) = 0$, so $\eta = \frac12 N_0$ in agreement with our discussion in the previous section. However, unlike for the case of an orientable manifold, on a non-orientable manifold $e^{2 \pi i \eta}$ generally does not have a local expression.

Eq.~(\ref{eq:Zftimesfinal}) gives the partition function of a $\nu =1$ TSc on an arbitrary non-orientable manifold. Note that this partition function depends not only on the manifold itself, but also on the Pin$_c$ structure, which is necessary to put a charged fermion on the manifold. The partition function of a non-interacting TSc phase with general $\nu$ is $e^{2 \pi i \nu \eta}$. It is known that the $\eta$-invariant on a Pin$_c$ manifold in four dimensions is always a multiple of $1/8$.\cite{GilkeyPinC} From this we conclude that non-interacting TSc phases $\nu$ and $\nu + 8$ have the same partition function on an arbitrary Pin$_c$ manifold, so they must, in fact, belong to the same phase in agreement with previous physical arguments.\cite{FidkowskiChenAV, Wang2014, MetlitskiChenFidkowskiAV2014} Furthermore, for $\nu = 4$ the partition function is $e^{8 \pi i \eta}$, which on all Pin$_c$ manifolds  coincides with the partition function of the eTmT SPT phase of bosons deduced in Refs.~\onlinecite{KapustinBos, KapustinBosTI},
\beq Z_{{\rm eTmT}} = \exp\left(\pi i \int_M w^4_1\right) = \exp(8 \pi i \eta) \label{eq:ZeTmT}\eeq
(see section \ref{sec:SW} for details).  Importantly, the first Stiefel-Whitney class $w_1$ depends only on the topology of the manifold, but not on  the Pin$_c$ structure. (Indeed, the partition function of a SPT phase of neutral bosons does not require a Pin$_c$ structure as an input). For instance, the manifold $\mathbb{RP}^4$ admits two Pin$_c$ structures with correspoding values of $\eta = \pm 1/8$ (see appendix \ref{app:DRP4}). Hence, the partition function of the $\nu = 4$ phase on $\mRP$ is independent of the Pin$_c$ structure and is equal to $-1$, which  precisely coincides with $\exp(\pi i \int_{\mRP} w^4_1)$. The identification of the $\nu =4$ TSc with the bosonic eTmT phase is again in agreement with previous physical arguments based on surface topological order.\cite{Wang2014,MetlitskiChenFidkowskiAV2014}

Another conclusion that follows from Eq.~(\ref{eq:Zftimesfinal}) is that TSc phases $\nu$ and $-\nu$ coalesce once the $u(1)$ symmetry is gauged. Indeed, the partition function of the theory obtained by gauging a TSc in class $\nu$ is,
\beq Z^{\nu}_{CT}(e) = \int D a_{\mu}\, e^{-S[a_{\mu}]} e^{2 \pi i \nu \eta[a_{\mu}]} \label{eq:Znu} \eeq
with $S[a_{\mu}]$ given by the classical Maxwell action (\ref{eq:introMaxwell}). The measure $D a_{\mu}$ includes a sum over all Pin$_c$ structures. As a result, the partition function of the gauge theory does not take a Pin$_c$ structure as an input, allowing us to interpret it as a partition function of a system built out of bosons. As we already remarked, given a Pin$_c$ structure on the manifold, we can locally act with the $\gamma^5$ operator in each patch obtaining a generally different Pin$_c$ structure with the same value of $f_{\mu \nu}$ but an inverted spectrum of the Dirac operator, i.e. with the opposite value of $\eta$. From this, we immediately conclude that 
\beq Z^{\nu}_{CT}(e) = Z^{-\nu}_{CT}(e) \label{eq:Zpmnu}\eeq
 This is again in agreement with physical arguments presented in Refs.~\onlinecite{Wang2014,MVDuality}. Thus, as claimed in section \ref{sec:dyons}, gauging of odd $\nu$ non-interacting topological superconductors gives rise to only two distinct phases: one descendant from $\nu =1$ and one descendant from $\nu = 5$. Moreover, from Eq.~(\ref{eq:ZeTmT}) we have
\beq Z^{\nu = 5}_{CT}(e) = Z^{\nu = 1}_{CT}(e) Z_{\rm eTmT}\eeq
since $e^{8 \pi i \eta}$ is independent of the Pin$_c$ structure and field-strength $f_{\mu \nu}$. This confirms our statement that the gauged $\nu =1$ TSc and the gauged $\nu = 5$ TSc differ by the bosonic eTmT phase.

A final curious observation, which is not directly related to the main subject of this paper, concerns the gauged $\nu = 2$ TSc. From Eqs.~(\ref{eq:ZeTmT}), (\ref{eq:Zpmnu}), we obtain
\beq Z^{\nu = 2}_{CT}(e) = Z^{\nu = -2}_{CT}(e) = Z^{\nu = 2}_{CT}(e) Z_{\rm eTmT} \eeq
This means that if we begin with the gauged $\nu = 2$ TSc and add a bosonic eTmT phase, we obtain a state in the same gauged $\nu = 2$ phase that we started with. Mathematically, this means that the partition function $Z^{\nu = 2}_{CT}$ vanishes whenever $Z_{\rm eTmT} \neq 1$. Physical arguments for this result were given in Ref.~\onlinecite{Wang2015}. The fact that an SPT phase can sometimes be ``absorbed" by a long-range-entangled phase has previously been known in the context of 2+1D symmetry enriched topological phases.\cite{Maissam} Note that the gauged $\nu = 1$ phase does not absorb the eTmT phase, i.e. gauged $\nu = 1$ and $\nu= 5$ phases are different. Indeed, we show in section \ref{sec:fermionsRP4} that $Z^{\nu =1}_{CT} \neq 0$ on the manifold $\mRP$, where $Z_{{\rm eTmT}} = -1$. 

\subsection{Partition function of a Dirac fermion: non-orientable manifold, $T$}
\label{sec:Zfrtimes}
We now discuss the partition function of ${\cal L}_T$ (TI)  on an non-orientable manifold. Here, orientation reversing transformations between patches are implemented by Eq.~(\ref{eq:AIIglue}). In order to perform the Grassmann path-integral over $\psi$, $\bar{\psi}$ it is more convenient to use the Lagrangian in the Majorana basis, Eq.~(\ref{eq:LMajorana}), where the orientation-reversing/orientation-preserving transitions between patches are given by Eqs.~(\ref{eq:transchiT})/(\ref{eq:transchi}). 
The Grassmann path-integral over $\chi_1$, $\chi_2$, then gives,
\beq Z[a_{\mu}]  = (\mathrm{det} ({\slashed D}^{\chi} + m))^{1/2} \label{eq:ZPf}\eeq
where 
\beq {\slashed D}^{\chi} = e^{\mu}_a \gamma^a \left(\d_{\mu} + i \omega_{\mu} + i a_{\mu} \rho^2\right) \label{eq:Dchi}\eeq
Note that $-i {\slashed D}^{\chi}$ is a Hermitian operator and the transition functions (\ref{eq:transchiT}), (\ref{eq:transchi}) are part of its definition. Further note that $-i {\slashed D}^{\chi}$ possesses an anti-unitary symmetry,
\beq \left[-i {\slashed D}^{\chi}, C K\right] = 0\eeq
where $K$ denotes complex conjugation. As $(CK)^2 = C C^* = -1$, every eigenvalue of $-i {\slashed D}^{\chi}$ is doubly degenerate. As a result, no non-analyticities are introduced by the square root in Eq.~(\ref{eq:ZPf}). Furthermore, the ambiguity in the sign of the square root cancels out when one divides the partition function at $m < 0$ by the partition function at $m > 0$ to obtain the topological partition function of the non-trivial TI:
\beq Z_{\psi, T}[a_{\mu}] = \prod_{\langle \lambda \rangle} \frac{(i \lambda  - |m|)^{d(\lambda)/2}}{(i \lambda + |m|)^{d(\lambda)/2}} \label{eq:Zfrtimes}\eeq
Here, the product is over non-repeated eigenvalues $\lambda$ of $- i {\slashed D}^{\chi}$ only, and we've denoted the degeneracies of eigenvalues by $d(\lambda)$ (with $d(\lambda)$ - even). Furthermore, we now have a matrix $\gamma^5 \rho^2$, which anticommutes locally with $  {\slashed D}^{\chi}$, and {\it commutes} with the transition functions in both the orientation reversing (\ref{eq:transchiT}) and orientation-preserving (\ref{eq:transchi}) cases. Thus, all non-zero eigenvalues of $-i {\slashed D}^{\chi}$ come in pairs $\lambda$, $-\lambda$, with corresponding eigenvectors $\chi_\lambda$, $\chi_{-\lambda} = \gamma^5 \rho^2 \chi_\lambda$. Thus, the contribution of all non-zero eigenvalues to Eq.~(\ref{eq:Zfrtimes}) cancels and
\beq  Z_{\psi, T}[a_{\mu}] = (-1)^{N^{\chi}_0/2} \label{eq:Zfrtimesfinal}\eeq
with $N^{\chi}_0$ - the number of zero-modes of ${\slashed D}^{\chi}$.  Note that $(-1)^{N^{\chi}_0/2}$ is invariant under smooth deformations of $a_{\mu}$, i.e. it only depends on the Pin$_{\tilde{c}}$ structure. Indeed, finite eigenvalues $\lambda, -\lambda$ can descend to $\lambda = 0$. However, due to the symmetry $CK$, $d(\lambda) = d(-\lambda)$ must be even. Thus, $\Delta N^{\chi}_0$ can only change in multiples of $4$ and $(-1)^{N^{\chi}_0/2}$ is invariant under smooth deformations. 
Again, an even stronger statement holds: $(-1)^{N^{\chi}_0/2}$ is a bordism invariant of the  Pin$_{\tilde{c}}$ structure.\footnote{This follows from an application of Atiyah, Patodi, Singer theorem,\cite{APS} which completely parallels that used to show that $e^{\pi i \eta}$ is a bordism invariant of a Pin$_+$ structure, with $\eta$ - the spectral asymmetry of a single (non-doubled) Dirac operator with no $u(1)$ gauge field.\cite{Stolz}}
 The fact that the partition function of a $\nu = 1$ TI on an arbitrary manifold is equal to $\pm 1$ is consistent with $\nu$ being an integer modulo $2$ for non-interacting TIs.


\section{$S$-duality from bulk partition function}
\label{sec:Gauge}

Having discussed the partition function of a Dirac fermion in the background of a gauge field $a_{\mu}$, we now proceed to make the gauge field fluctuating, thus studying the partition functions of the gauge theories ${\cal L}_T$, Eq.~(\ref{eq:ZgT}), and ${\cal L}_{CT}$, Eq.~(\ref{eq:ZgCT}). We write out the measure of the path integeral a bit more explicitly:
\bea Z_{T}(e) &=& \frac{1}{\mathrm{Vol}({\cal G})} \sum_{{\rm Pin}_{\tilde{c}}} \int D a_{\mu} \, e^{- S[a]}  (-1)^{N^{\chi}_0/2} \quad\quad ({\rm gauged \,\,TI}), \label{eq:Zrtimesm}\\
 Z_{CT}(e) &=& \frac{1}{\mathrm{Vol}({\cal G})} \sum_{{\rm Pin}_{c}} \int D a_{\mu} \, e^{- S[a]} e^{2 \pi i \eta} \quad\quad\quad\quad ({\rm gauged \,\,} \nu = 1 {\,\,\rm TSc}) \label{eq:Ztimesm} \eea
with $S[a]$ - the classical Maxwell action (\ref{eq:introMaxwell}). 
We've normalized our partition functions by the volume of the gauge group $\mathrm{Vol}({\cal G})$ following Ref.~\onlinecite{KlebanovF}. $\sum_{{\rm Pin}_{c}} $ denotes a sum over gauge-inequivalent Pin$_c$ structures  and $\sum_{{\rm Pin}_{\tilde{c}}}$ denotes a sum over gauge inequivalent Pin$_{\tilde{c}}$ structures. If no Pin$_c$ (Pin$_{\tilde{c}}$) structure on the manifold exists, we set $Z_{CT}$ ($Z_T$) to $0$. As already remarked, when a Pin$_c$ structure on a manifold does exist, inequivalent Pin$_c$ structures on a manifold differ by complex line bundles, which are in one-to-one correspondence with elements of the second cohomology group of the manifold $H^2(M, \bZ)$. Gauge fields $a_{\mu}$ compatible with a given Pin$_c$ structure differ by one-forms (a one form transforms according to Eq.~(\ref{eq:atransf}) with $\alpha = 0$). Similarly, inequivalent Pin$_{\tilde{c}}$ structures on a manifold differ by twisted complex line bundles, which are in one-to-one correspondence with elements of the second cohomology group with local coefficients $H^2(M, \tilde{\bZ})$, which by Poincare duality is isomorphic to the second homology group $H_2(M, \bZ)$. Gauge fields $a_{\mu}$ compatible with a given Pin$_{\tilde{c}}$ structure differ by pseudo-one-forms (a pseudo-one-form transforms according to Eq.~(\ref{eq:atransf}) with $\alpha = 0$ under orientation-preserving patch changes, and according to  Eq.~(\ref{eq:atransfAII}) with $\alpha = 0$ under orientation-reversing patch changes).

As we discussed, on an orientable manifold  $(-1)^{N^{\chi}_0/2}$ and $e^{2 \pi i \eta}$ coincide and are given by Eqs.~(\ref{eq:Zfofinal}), (\ref{eq:indexsigma}). Thus, on an orientable manifold, $Z_{CT}(e) = Z_{T}(e)$. On the other hand, on a non-orientable manifold not only are $(-1)^{N^{\chi}_0/2}$ and $e^{2 \pi i \eta}$ generally different, but also the  Pin$_c$/Pin$_{\tilde{c}}$ structures and corresponding admissible gauge fields $a_{\mu}$ in the two path integrals (\ref{eq:Zrtimesm}), (\ref{eq:Ztimesm}) are different.

We would like to show,
\beq Z_{CT}(e) = Z_{T}(4 \pi/e)  \label{eq:ZCTTpr}\eeq
We will be able to demonstrate this statement on an arbitrary orientable manifold and on the single (but important) non-orientable manifold $\mRP$. 

This section is organized as follows. We begin in section \ref{sec:bosonsoriented} by reviewing Witten's proof of $S$-duality in a theory with bosonic matter on orientable manifolds. In section \ref{sec:Sfor} we extend this to a proof of $S_f$ duality (\ref{eq:Sf}) in a theory with fermionic matter on  orientable manifolds, in particular demonstrating that Eq.~(\ref{eq:ZCTTpr}) holds on orientable manifolds. We then proceed to the more difficult case of non-orientable manifolds. We start with $S$-duality in a theory with bosonic matter on non-orientable manifolds. As a warm up, in section \ref{sec:RP4bos} we explicitly compute the partition function of the bosonic theory with $\theta = 0$ on the manifold $\mathbb{R P}^4$ and show that it obeys $S$-duality. We then prove in section \ref{sec:bosonsS} $S$-duality in the bosonic theory with $\theta = 0$ on an arbitrary non-orientable manifold by reducing it to the equality of Ray-Singer analytic torsion and Reidemeister torsion. Finally, in section \ref{sec:fermionsRP4} we discuss the duality (\ref{eq:ZCTTpr}) in a theory with fermionic matter on non-orientable manifolds. We show that Eq.~(\ref{eq:ZCTTpr}) reduces to the identity (\ref{eq:Finalf}) involving only a sum over classical saddle point mimima for the gauge field. We explicitly check this identity on the manifold $\mathbb{RP}^4$. Physical arguments presented in section  \ref{sec:ambiguity} then strongly suggest that Eq.~(\ref{eq:ZCTTpr}) holds on all manifolds.

\subsection{$S$-duality: bosonic matter, oriented manifold}
\label{sec:bosonsoriented}

As a warm up, we begin by reviewing Witten's demonstration of $S$-duality in a $u(1)$ gauge theory with bosonic matter on an oriented manifold.\cite{WittenS} Here, the partition function is given by 
\beq Z^{b}[e, \theta] =  \frac{1}{\mathrm{Vol}({\cal G})} \sum_{L} \int D a_{\mu}  e^{- S[a]} \label{eq:Zboson}\eeq
with $\sum_L$ denoting the sum over complex line bundles. While we are mostly interested in the time-reversal invariant situation, on an oriented manifold we will allow for a $T$-breaking $\theta$ term, taking
\beq S[a] = \frac{1}{4 e^2} \int_M d^4 x \, \sqrt{g} \, f_{\mu \nu} f^{\mu \nu} + \frac{i \theta}{32 \pi^2} \int_M d^4 x \, \epsilon^{\mu \nu \lambda \sigma} f_{\mu \nu} f_{\lambda \sigma} \label{eq:Satheta} \eeq 
which can also be written in form notation (see appendix \ref{app:forms}) as
\beq S[a] = \frac{1}{2 e^2} \int_M \, f \wedge * f + \frac{i \theta}{2 (2\pi)^2} \int_M \, f \wedge f \eeq
We will show,
\beq Z^b(\tau) = Z^b(-1/\tau) \label{eq:Sboson}\eeq
with $\tau$ given by Eq.~(\ref{eq:introtaudef}). 

Let's begin by unpacking the definition (\ref{eq:Zboson}). For each line-bundle we may write the admissible gauge fields $a_{\mu} = a^{cl}_{\mu} + a^{q}_{\mu}$. $a^{cl}$ is one representative gauge field compatible with the line-bundle (satisfying Eq.~(\ref{eq:atransf})), while $a^q$ is a one-form. The integral $D a_{\mu}$ in Eq.~(\ref{eq:Zboson}) is then an integral over $D a^{q}_{\mu}$. Gauge-inequivalent line-bundles on an oriented manifold $M$ are in one-to-one correspondence with elements of the cohomology group $H^2(M, \bZ)$.  This cohomology group generally has a free part (sum of $\bZ$'s) and a torsion part (finite group). $a^{cl}$ representing the torsion part can be chosen to be flat, $f^{cl}_{\mu \nu} = 0$. The contribution of this torsion part thus just gives an overall multiplicative constant to $Z^b(\tau)$ equal to the order of torsion subgroup $|T| = |\mathrm{Tor}(H^2(M,\bZ))|$. Since this constant is $\tau$ independent, it drops out in the proof of the $S$-duality (\ref{eq:Sboson}). Turning our attention to the free part of $H^2(M, \bZ)$, we can promote its coefficients to $\mathbb{R}$ obtaining the cohomology group $H^2(M, \mathbb{R})$, which is isomorphic to the de-Rham cohomology $H^2_{dR}(M)$. In turn, $H^2_{dR}(M)$ is in one-to-one correspondence with harmonic 2-forms on $M$, i.e. ones satisfying $\Delta f^{cl} = 0$.

Thus, for each line bundle, we can choose $a^{cl}$ such that $\Delta f^{cl} = 0$ (This in turn implies $\delta f^{cl} = 0$ and $d f^{cl} = 0$, the latter condition being trivially satisfied as locally $f^{cl} = d a^{cl}$). We must also require that $\frac{1}{2\pi} \int_U \, f^{cl} \in \bZ$ for each closed $2$-cycle $U$ in $M$. This ``Dirac condition" projects one back from $H^2(M, \mathbb{R})$ to $H^2(M, \bZ)$. 
Concretely, we may choose a basis of harmonic two-forms $f_p$ for the free part of $H^2(M, \bZ)$, where $p = 1, 2, \ldots b_2$ and $b_2 = \mathrm{dim}\left[H^2(M, \mathbb{R})\right]$ is the second Betti number, and write $f^{cl} =  2 \pi c_p f_p$, with $c_p$ - integers. The sum over line bundles is then given by $\sum_L = |T| \sum_{c_p \in \bZ}$. Note that since $f_p$ is harmonic so is $* f_p$, although $* f_p$ need not satisfy $\int_U (* f_p) \in \bZ$. Therefore, we may write,
\beq * f_p = S_{r p} f_r \eeq
with $S$ - a (not necessarily integer-valued) matrix. Since $* * f_p = f_p$, $S^2 = 1$. 

Since $\delta f^{cl} = 0$, the action (\ref{eq:Satheta}) splits into,
\beq S[a^{cl} + a^{q}] = S_{cl} + S_q\eeq
where 
\bea S_{cl} &=& \frac{1}{2 e^2} \int \,f^{cl} \wedge * f^{cl}+ \frac{i \theta}{2 (2\pi)^2} \int \, f^{cl} \wedge f^{cl}\\
 S_q &=& \frac{1}{2 e^2} \int \, d a^q \wedge * d a^q\eea
and so we may write
\beq Z^{b}(\tau) = Z^b_{cl}(\tau) Z^b_{q}(\tau)\eeq
where
\beq Z^b_{cl}(\tau) =  |T| \sum_{c_p \in Z} e^{-S_{cl}} \label{eq:Zcl}\eeq
and 
\beq Z^b_q(\tau) = \frac{1}{\mathrm{Vol}({\cal G})} \int da^q_{\mu} \, e^{-S_q} \label{eq:Zq}\eeq

Let us begin with the ``classical" part $Z^b_{cl}$. For $\eta, \xi \in \mathrm{Free}(H^2(M, \bZ))$, we can define the intersection form $(\eta, \xi)$,
\beq (\eta, \xi) = \int_M \eta \wedge \xi \eeq
Using our basis $f_p$, we let $Q_{pr} =  (f_p, f_r)$. An important fact is that $Q_{pr}$ is symmetric, integer-valued and by Poincare duality $|\mathrm{det} \, Q| = 1$.\cite{Hatcher} Thus, we can rewrite
\beq S_{cl} = \frac{(2 \pi)^2}{2 e^2} c^T Q S c + \frac{i \theta}{2} c^T Q c \label{eq:Sclc}\eeq
We recall the inner product on two-forms on a manifold,
\beq \langle \eta, \xi \rangle = \frac12 \int_M d^4x \, \sqrt{g} \, \eta_{\mu \nu}\xi^{\mu \nu} = \int_M \eta \wedge * \xi \eeq
is symmetric and positive definite, so $Q S = S^T Q$ and $Q S$ is positive definite. Thus, the number $b^+_2$ ($b^-_2$) of positive (negative) eigenvalues of $Q$ is equal to the number of $+1$ ($-1$) eigenvalues of $S$, i.e the number of self-dual (anti-self-dual) harmonic two-forms. The signature of the manifold $\sigma(M) = b^+_2 - b^-_2$. 

From the fact that $Q$ is integer-valued, the action (\ref{eq:Sclc}) is invariant under the ${\cal T}$ operation $\theta \to \theta + 4 \pi$ ($\tau \to \tau + 2$). As discussed in section \ref{sec:dyons}, this is the correct periodicity of $\theta$ in a theory with bosonic matter. Note that $Q$ is not an even form on a general manifold. For instance, on the manifold $\mathbb{CP}^2$ there is just one non-trivial $2$-cycle and $Q$ is a $1\times1$ matrix: $Q = 1$. Thus, the partition function of a theory with bosonic matter on a general manifold is not invariant under $\theta \to \theta + 2 \pi$. On a Spin manifold $Q$ is, indeed, an even form, however, we are assuming that our $u(1)$ gauge theory emerges in a system of bosons with short-range interactions, so there is no reason to restrict one's attention to Spin manifolds.

To understand how $Z^b_{cl}(\tau)$ transforms under $S$-duality, we perform Poisson re-summation on Eq.~(\ref{eq:Zcl})
\beq Z^b_{cl}(\tau) = |T| \sum_{d_p \in \bZ} \int d c \exp\left(- \frac{1}{2} c^T Q \left[\frac{(2\pi)^2}{e^2} S + i \theta\right] c + 2 \pi i d^T c\right)\eeq
Actually, since $\mathrm{det} Q = \pm 1$, we can write $d  = Q d'$, with $d' \in \bZ^{b_2}$, so that
\beq Z^b_{cl}(\tau) = |T| \sum_{d'_p \in \bZ} \int d c \exp\left(- \frac{1}{2} c^T Q \left[\frac{(2\pi)^2}{e^2} S + i \theta\right] c + 2 \pi i (d')^T Q c\right)\eeq
Performing the integral over $c$ one obtains,
\beq Z^b_{cl}(\tau) = (i \tau)^{-b^+_2/2} (- i \bar{\tau})^{-b^-_2/2} Z^b_{cl}(-1/\tau) \label{eq:classbdual}\eeq

Next, let's proceed to the ``quantum" part, Eq.~(\ref{eq:Zq}). Note that this part only depends on $e$ but not on $\theta$. We will closely follow the discussion of Ref.~\onlinecite{KlebanovF}. We write (see appendix \ref{app:forms} for notation),
\beq S_q = \frac{1}{2e^2}\langle da^q, da^q \rangle = \frac{1}{2e^2}\langle a^q, \delta d a^q \rangle \eeq
Any $a^q$ may be decomposed as 
\beq a^q_{\mu} = b_{\mu} + \d_{\mu} \alpha' \label{eq:adecomp} \eeq
where $\delta b  = 0$ and $\alpha'$ is a (non-constant) function. Then,
\beq S_q = \frac{1}{2e^2}\langle b, \delta d b \rangle = \frac{1}{2e^2} \langle b, \Delta b \rangle \eeq
We take the measure
\beq D a^q = \prod_n d u_n \eeq
where $a^q(x) = \sum_n u_n \phi_{n} (x)$, and $\phi_{n}(x)$ may be taken as eigenstates of $\Delta^1$ with the normalization $\langle \phi_n, \phi_m \rangle = \delta_{nm}$. 
The superscript $1$  on the Laplacian $\Delta$ underscores that it acts on 1-forms. All eigenstates of the Laplacian $\Delta^1$ can be divided into two groups: i) eigenstates $\phi^{\perp}_n$ satisfying $\delta \phi^{\perp}_n = 0$, ii) eigenstates of the form $\phi^{\parallel}_n = {\cal N}_n   d \alpha_n$, with $\alpha_n$ - a function satisfying $\Delta^0 \alpha_n = \lambda^0_n \alpha_n$ and with $\alpha_0 = const$ excluded (the superscript $0$ on the Laplacian $\Delta$ underscores that it acts on zero-forms, i.e. functions). If $\alpha_n$ are normalized so that $\langle \alpha_n, \alpha_m \rangle = \delta_{nm}$ then we must choose the factor ${\cal N}_n = (\lambda^0_n)^{-1/2}$. Let us write $b$ in Eq.~(\ref{eq:adecomp}) as $b_{\mu}(x) = \sum_n u^{\perp}_n \phi^{\perp}_{n, \mu} (x)$, and $\alpha'$ in Eq.~(\ref{eq:adecomp}) as $\alpha'(x) = \sum_{n \neq 0} v_n \alpha_n(x)$. Then with the natural measures 
\beq D b = \prod_n du^{\perp}_n, \quad D \alpha' = \prod_{n \neq 0} d v_n \label{eq:bmeasure}\eeq
we have
\beq D a^q = D b D \alpha' (\mathrm{det}' \Delta^0)^{1/2} \eeq
where the prime on $\alpha$ indicates that the constant mode of $\alpha$ is removed and similarly for the determinant of the Laplacian $\Delta^0$. 

Now the volume of the gauge group in Eq.~(\ref{eq:Zq}) corresponds precisely to an integral over gauge transformations $\alpha$,
\beq \mathrm{Vol}({\cal G}) = \int D \alpha\eeq
where again writing  $\alpha(x) = \sum_n  v_n \alpha_n(x)$, we have $D \alpha = \prod_n d v_n$. Note that the constant gauge-transformation $\alpha_0(x) = \frac{1}{\sqrt{V(M)}}$ is ${\it not}$ excluded from the gauge group (here $V(M)$ denotes the volume of the manifold). Since gauge transformations differing by $2\pi$ are identified, $\alpha(x) \sim \alpha(x) + 2 \pi$, the integral over the coefficient of the zero-mode $v_0$ gives $2 \pi \sqrt{V(M)}$, so
\beq \mathrm{Vol}({\cal G}) = 2 \pi \sqrt{V(M)} \int D \alpha' \label{eq:VolG}\eeq
The integral $\int D \alpha'$ cancels in the numerator and denominator of Eq.~(\ref{eq:Zq}) giving,
\beq Z^b_q(\tau) = \frac{(\mathrm{det}' \Delta^0)^{1/2}}{2 \pi \sqrt{V(M)}}  \int d b \exp\left(-\frac{1}{2e^2} \langle b, \Delta b \rangle\right) \label{eq:Zqb} \eeq
The integral over $b$ consists of two parts: zero-modes of $\Delta^1$ and non-zero-modes. The zero-modes are in one-to-one correspondence with elements of the de-Rham cohomology group $H^1_{dR}(M) = H^1(M, \mathbb{R})$ and their number is given by $b_1 = \mathrm{dim}[H^1(M, \mathbb{R})]$.  An integral over them gives a $\tau$ independent constant ${\cal C}$ (see section \ref{sec:bosonsS} for more details). With the measure (\ref{eq:bmeasure}), the integral over the non-zero-modes gives $\left(\mathrm{det}'_{\perp} \frac{\Delta^1}{2 \pi e^2}\right)^{1/2}$, where the prime indicates that the zero-modes are removed and the subscript $\perp$ reminds us that only transverse modes $\phi^{\perp}_n$ are taken into account. We, thus, arrive at the final expression:

\beq Z^b_q(\tau) = \frac{{\cal C}}{2 \pi \sqrt{V(M)}} \left(\mathrm{det}' \Delta^0\right)^{1/2} \left(\mathrm{det}'_{\perp} \frac{\Delta^1}{2 \pi e^2}\right)^{-1/2} \label{eq:Zqfinal}\eeq

To prove $S$-duality on an oriented manifold we only need to know the dependence of $Z_q$ on the coupling $e$. From Eq.~(\ref{eq:Zqfinal}), $Z^b_q(\tau) \propto e^{\tilde{N}^{1}_{\perp}}$, where $\tilde{N}^1_{\perp}$ is the number of transverse non-zero eigenvalues of the Laplacian $\Delta^1$. Since the number of zero-modes of $\Delta^1$ is $b_1$, we have $\tilde{N}^1_{\perp} = N^1_{\perp} - b_1$, with $N^1_{\perp}$ the total number of transverse eigenmodes. Subtracting the number of longitudinal eigenmodes, $N^1_{\perp} = N^1 - (N^0 - 1)$, where $N^1$ is the total number of eigenvalues of $\Delta^1$, and $N^0$ - the total number of eigenvalues of $\Delta^0$, so,
\beq Z^b_q(\tau) \propto e^{N^1 - N^0 - b_1 + 1} \label{eq:Zbe}\eeq
Eq.~(\ref{eq:Zqfinal}), and hence $N^1$, $N^0$ require regularization. However, in any sensible regularization $N^1$ and $N^0$ will have an expression as an integral of a local geometric quantity (analytic in the local curvature) over the manifold. For instance, one could use a heat-kernel regularization,
\beq N^1 = \mathrm{Tr} \, \exp(-t \Delta^1), \quad\quad N^0 =   \mathrm{Tr} \,\exp(-t \Delta^0)\eeq
with $t \to 0$. The heat-kernel expansion then gives,
\beq N = \sum_{n = 0}^{\infty} A_n t^{(n-D)/2} \label{eq:Nheatk} \eeq
with $D = 4$ - the space-time dimension, and $A_n$ - integrals of local quantities analytic in curvature. (Here, we have suppressed the superscripts $0$, $1$). In this paper it will be more convenient to use $\zeta$-function regularization, which is implicitly assumed in all expressions below,
\beq N = N_0 + \lim_{s \to 0} \sum_{\lambda \neq 0} \lambda^{-s}\eeq
with $N_0$ - the number of  zero-modes and $\lambda$ - eigenvalues of the Laplacian. The $\lim_{s \to 0}$ on the right-hand-side is understood to be performed by analytically continuing the sum from large $Re(s)$. With this regularization $N$ is simply given by the coefficient $A_{4}$ in the heat-kernel expansion (\ref{eq:Nheatk}).\cite{Vassilevich} Explicitly,\cite{Vassilevich}
\beq A_4 = \frac{1}{90 (64 \pi^2)}  \int_M d^4 x \sqrt{g} \left(-a E_4 + c C^2 + d {\cal R}^2\right) \label{eq:A4}\eeq
where $E_4$ is the Euler density,
\beq E_4 = R_{\mu \nu \lambda \sigma} R^{\mu \nu \lambda \sigma} - 4 R_{\mu \nu} R^{\mu \nu} + {\cal R}^2 \label{eq:E4}\eeq
$C^2$ is the square of the Weyl tensor,
\beq C^2 = C_{\mu \nu \lambda \sigma} C^{\mu \nu \lambda \sigma} =  R_{\mu \nu \lambda \sigma} R^{\mu \nu \lambda \sigma} - 2 R_{\mu \nu} R^{\mu \nu} + {\cal R}^2\eeq
and $R^{\mu}_{\nu \lambda \sigma}$ is the Riemann curvature tensor, $R_{\mu \nu} = R^{\lambda}_{\mu \lambda \nu}$ is the Ricci tensor, and ${\cal R} = R^{\mu}_{\mu}$ is the scalar curvature. We remind the reader that,
\beq \chi = \frac{1}{32 \pi^2} \int_M d^4x \sqrt{g} E_4 \label{eq:chi}\eeq
with $\chi$ - the Euler characteristic of the manifold. The coefficients $a$, $c$ and $d$ for Laplacian acting on zero and one-forms are given in table \ref{tbl:charges}.
\begin{table}[t]
\beq
\begin{array}{|c|c|c|c|}
\hline
   q  & a & c & d \\ \hline
0    & 1&  3 & 5   \\ \hline
1  &  64 & 42 &  10 \\ \hline
\end{array}\nn\eeq
\caption{Coefficients $a$, $c$, $d$ in the heat-kernel expansion (\ref{eq:A4}). The rows correspond to  Laplacian $\Delta^q$ for zero-forms ($q = 0$) and one-forms ($q = 1$).}
\label{tbl:charges}
\end{table}
Therefore, in $\zeta$-function regularization we have,
\beq N^1 - N^0 = \frac{1}{90 (64 \pi^2)}  \int_M d^4 x \sqrt{g} \left(-63 E_4 + 39 C^2 + 5 {\cal R}^2\right) \label{eq:NvmNs}\eeq
Now, let us define a new partition function $Z^b_{r}(\tau)$ as
\beq Z^b_{r}(\tau) = (\sqrt{2 \pi} e)^{N^0 - N^1} Z^b(\tau) \label{eq:Zr}\eeq
This simply amounts to adding a term to the action of the theory,
\beq S_r = S + \log(\sqrt{2 \pi} e) (N^1 - N^0) \eeq
The added term is local, furthermore, one can turn on this term smoothly starting with the original action $S$. $N^1 - N^0$ does not break the time-reversal symmetry (the integral (\ref{eq:NvmNs}) does not require specifying the orientation of the manifold), thus, as we turn this term on, time-reversal is preserved. Therefore, if we start with the time-reversal invariant theory ($\theta = 0$), the  theory with action $S_r$ is smoothly connected to our original theory - i.e. they describe the same $T$-respecting phase. We will, therefore, work with $Z_r$ below instead of $Z$. Note that the factor $\sqrt{2 \pi}$ in Eq.~(\ref{eq:Zr}) is inserted purely for future convenience. 

From Eq.~(\ref{eq:Zbe}),
\beq Z^b_{q,r}(\tau) \equiv (\sqrt{2 \pi} e)^{N^0 - N^1} Z^b_q(\tau)\sim e^{1-b_1} \eeq
Therefore, 
\beq Z^b_{q,r}(-1/\tau) = (\tau \bar{\tau})^{(1-b_1)/2} Z^b_{q,r}(\tau) \label{eq:Zbqrdual}\eeq 
Combining this with Eq.~(\ref{eq:classbdual}) we obtain,\cite{WittenS}
\beq Z^b_r(-1/\tau) = (i \tau)^{(\chi+\sigma)/4} (-i \bar{\tau})^{(\chi - \sigma)/4} Z^b_r(\tau)\eeq
where $\chi = 2- 2 b_1 + b^+_2 + b^-_2$ is the Euler characteristic and $\sigma = b^+_2 - b^-_2$ is the signature. On an oriented manifold (i.e. ignoring time-reversal symmetry), we can further re-define,
\beq \bar{Z}^b(\tau) = (i \tau)^{(\chi + \sigma)/8}(-i \tau)^{(\chi-\sigma)/8} Z^b_r(\tau) \label{eq:barb}\eeq
Again, since $\chi$ and $\sigma$ have local expressions (\ref{eq:chi}), (\ref{eq:signature}), $\bar{Z}^b(\tau)$ defines a sensible local theory, which describes a state continuously connected to our original state. Then,
\beq \bar{Z}^b(-1/\tau) = \bar{Z}^b(\tau) \label{eq:bdualityfinal}\eeq
which concludes the proof of $S$-duality for a pure gauge theory with bosonic matter on an oriented manifold. Note that after the re-definition, (\ref{eq:barb}), ${\bar Z}(\tau + 2) \neq {\bar Z}(\tau)$, i.e. the partition function ${\bar{Z}}$ now transforms trivially under $S$, but covariantly under ${\cal T}$. So, we have to contend with the partition function $Z^b_r(\tau)$ transforming covariantly under the subgroup of $SL(2, \bZ)$ generated by $S$ and ${\cal T}$ as a modular form of dimensions $((\chi+\sigma)/4, (\chi - \sigma)/4)$.\cite{WittenS} 

In the time-reversal symmetric case, which is of most interest to us here, $\theta = 0$ so
\beq Z^{b}_r\left(\frac{2 \pi}{e}\right) = \left(\frac{e^2}{2 \pi}\right)^{-\chi/2} Z^b_r(e) \eeq
Now,
\beq \bar{Z}^b(e) =  \left(\frac{e^2}{2 \pi}\right)^{-\chi/4} Z^b_r(e) \label{eq:redefbtheta0} \eeq
and
\beq \bar{Z}^{b}\left(\frac{2 \pi}{e}\right)  = \bar{Z}^b(e)\eeq Note that the redefinition (\ref{eq:redefbtheta0}) now does not involve the signature $\sigma$. Thus, this redefinition is meaningful on an arbitrary manifold, including non-orientable manifolds, i.e. it preserves time-reversal symmetry. Therefore, $\bar{Z}^b(e)$ defines a theory describing a state in the same time-reversal invariant phase as our original theory. This demonstrates that at least on an orientable manifold, $S$-duality in a theory with bosonic matter and $\theta = 0$ is compatible with time-reversal symmetry. In section \ref{sec:bosonsS}, we will show that in this theory $S$-duality, in fact, also holds on non-orientable manifolds.

\subsection{$S$-duality: fermionic matter, oriented manifold}
\label{sec:Sfor}

We next proceed to discuss $S$-duality in a gauge theory with fermionic matter on an oriented manifold. While we are mostly interested in the time-reversal invariant case of $\theta =\pi$, for further generality let us allow a $T$-breaking $\theta$ term, defining the partition function of the theory as,
 \beq Z^{f}[e, \theta] =  \frac{1}{\mathrm{Vol}({\cal G})} \sum_{{\rm Spin}_c} \int D a_{\mu}  e^{- S[a]} \label{eq:Zf}\eeq
with 
\beq S[a] = \frac{1}{2 e^2} \int \, f \wedge * f + i \theta \left(\frac{1 }{2 (2\pi)^2} \int \, f \wedge f  - \frac{\sigma}{8}\right)\eeq
We think of the $\theta$ term as coming from integrating out a gapped Dirac fermion with a complex mass (\ref{eq:Diractheta}), which gives Eq.~(\ref{eq:Zfthetafinal}). The partition functions of ${\cal L}_T$ and ${\cal L}_{CT}$ on an orientable manifold coincide and are given by Eq.~(\ref{eq:Zf}) with $\theta = \pi$. Note that the partition function now involves a sum over all Spin$_c$ structures.

The proof of $S$-duality is almost identical to one  in section \ref{sec:bosonsoriented}. Again, even though the charge matter in our theory is fermionic, we assume that the gauge theory emerges in a system of bosons with short-range interactions. Therefore, we should be able to place the theory on  arbitrary manifolds $M$, including ones which do not admit a Spin structure. Luckily, all four-dimensional manifolds admit a Spin$_c$ structure.\cite{HH} Let $w_2 \in H^2(M, \bZ_2)$ be the second Stiefel-Whitney class describing obstruction to Spin. In four dimensions $w_2$ always has a lift to $H^2(M, \bZ)$. In order to get a  Spin$_c$ structure, on every two-cycle $U$ of $M$ with $\int_U w_2$ - odd, we must put a gauge flux $\int_U \, f = 2 \pi (n + 1/2)$, with $n \in \bZ$. Note that this choice of gauge fluxes ensures that the quantity on the right-hand-side of Eq.~(\ref{eq:indexsigma}) is an integer. This implies that the partition function is invariant under $\theta \to \theta + 2\pi$. For instance, as already noted, for the non-spin manifold $\mathbb{CP}^2$,  we have just a single non-trivial 2-cycle. This two-cycle has $w_2 \neq 0$. The intersection form is given by $Q = 1$, so $\sigma = 1$. If we place flux $2 \pi (n+1/2)$ on the 2-cycle then 
\beq \frac{1 }{2 (2\pi)^2} \int \, f \wedge f  - \frac{\sigma}{8} = \frac{1}{2} (n + \frac12)^2 - \frac{1}{8} = \frac12 n(n+1) \in \bZ \label{eq:CP2}\eeq

For a given Spin$_c$ structure, we again write $a = a^{cl} + a^q$ with $\Delta a^{cl} = 0$, and $a^q$ - a one-form. The partition function again splits into a product of a classical piece and a quantum piece. 
The classical part of the partition function becomes,
\beq Z^f_{cl}(\tau) = |T| e^{i \theta \sigma/8}\sum_{c_p \in \bZ} \exp\left(- \frac{1}{2} \left(c+ \frac{w_2}{2}\right)^T Q \left[\frac{(2\pi)^2}{e^2} S + i \theta\right] \left(c+ \frac{w_2}{2}\right)  \right)\eeq
Here, $w_2$ denotes some integer representation of $w_2(M)$ in our basis on $\mathrm{Free}(H^2(M,\bZ))$. As before, we Poisson resum the above expression,
\beq Z^f_{cl}(\tau) = |T| e^{i \theta \sigma/8} \sum_{d_p \in \bZ} \int d c \exp\left(- \frac{1}{2} c^T Q \left[\frac{(2\pi)^2}{e^2} S + i \theta\right] c + 2 \pi i d^T Q \left(c- \frac{w_2}{2}\right)\right)\eeq
A crucial observation is that for $\eta \in \mathrm{Free}(H^2(M,\bZ))$, $\int \eta \wedge w_2 = \int \eta \wedge \eta \,\, (\mathrm{mod} \,\,2)$. Therefore, $d^T Q w_2 = d^T Q d\,\, (\mathrm{mod} \,\,2)$. Performing the $c$ integral, we obtain
\beq Z^f_{cl}(\tau) = e^{i \pi (\tau + \bar{\tau}) \sigma/8}  (-i \tau)^{-b^+_2/2} (-i \bar{\tau})^{-b^-_2/2} Z^{b}_{cl}\left(-1/\tau + 1\right) \label{eq:Zfcldual}\eeq
where $Z^b_{cl}$ on the right-hand-side is the classical partition function in a theory with bosonic matter, Eqs.~(\ref{eq:Zcl}), (\ref{eq:Sclc}).

The quantum part of the partition function is the same as in a theory with bosonic matter $Z^f_q(e) = Z^b_q(e)$.  As before, we define,
\beq Z^f_{r}(\tau) = (\sqrt{2 \pi} e)^{N^0 - N^1} Z^f(\tau) \label{eq:Zfr}\eeq
with $N^1 - N^0$ given by Eq.~(\ref{eq:NvmNs}), and similarly for $Z^f_{q,r}(\tau)$. Then combining Eqs.~(\ref{eq:Zbqrdual}), (\ref{eq:Zfcldual}),
\beq Z^f_r(\tau) = e^{i \pi (\tau + \bar{\tau}) \sigma/8} (i \tau)^{-(\chi + \sigma)/4} (-i \bar{\tau})^{-(\chi - \sigma)/4} Z^b_r(-1/\tau + 1) \label{eq:ftobdual}\eeq
Thus, we find that the partition function of a theory with fermionic matter and coupling constant $\tau$ is related to the partition function of a theory with ${\it bosonic}$ matter and coupling constant $-1/\tau + 1$. This is in perfect agreement with the considerations of the dyon quantum numbers discussed in section \ref{sec:dyons}.

If we want to relate the theory with fermionic matter back to itself, we can use the invariance of $Z^b$ under $\tau \to \tau + 2$ to write $Z^b_r(-1/\tau+1) = Z^b_r(-1/\tau - 1) = Z^b_r(-z^{-1} + 1)$ with $z = \tau/(2\tau + 1)$, and apply the duality (\ref{eq:ftobdual}) backwards. It proves convenient to define a variable $\rho$,
\beq \rho = 2 \tau + 1 = \left(1 + \frac{\theta}{\pi}\right) - \frac{4 \pi i}{e^2} \label{eq:rhodef}\eeq 
Then,
\beq Z^f_r(-1/\rho) = \exp \left(- \frac{\pi i \sigma}{16}(\rho + \bar{\rho} + \rho^{-1} +\bar{\rho}^{-1})\right) (i \rho)^{(\chi+\sigma)/4} (-i \bar{\rho})^{(\chi-\sigma)/4} Z^f_r(\rho) \label{eq:dualffinal}\eeq
Note that with the definition of $\rho$ (\ref{eq:rhodef}), we have $Z^f_r(\rho+ 2) = Z^f_r(\rho)$. 

We are most interested in the time-reversal symmetric case $\theta = \pi \sim - \pi$. Plugging $\theta = -\pi$ into Eq.~(\ref{eq:dualffinal}), we obtain,
\beq Z^f_r\left(\frac{4 \pi}{e}, \theta = \pi\right) = \left(\frac{e^2}{4 \pi}\right)^{-\chi/2} Z^f_r(e, \theta =\pi)\eeq
Finally, re-defining,
\beq \bar{Z}^f(e, \theta = \pi) =  \left(\frac{e^2}{4 \pi}\right)^{-\chi/4} Z^f_r(e, \theta = \pi) \eeq
we obtain
\beq \bar{Z}^f\left(\frac{4 \pi}{e}, \theta = \pi\right) = \bar{Z}^f(e, \theta = \pi) \eeq
This proves that $S$-duality between ${\cal L}_T$ (gauged TI) and ${\cal L}_{CT}$ (gauged $\nu = 1$ TSc) holds  on an orientable manifold. We now proceed to the case of non-orientable manifolds.

\subsection{$S$-duality: bosonic matter, $\mRP$}
\label{sec:RP4bos}
We will now explicitly demonstrate $S$-duality in a $u(1)$ gauge-theory with bosonic matter and $\theta = 0$ on the non-orientable manifold $\mRP$. This will be a warm up for the proof of $S$-duality in this theory on arbitrary manifolds in the next section.

In a theory with bosonic matter, once we take time-reversal symmetry into account, the duality is really between two different $\theta = 0$ theories. In the first theory, time-reversal commutes with the $u(1)$ gauge-symmetry, so we denote it by $CT$ in analogy with the fermionic case, so the the full symmetry group is $u(1) \times CT$. In the second theory, time-reversal, which we label as $T$, does not commute with $u(1)$ rotations, and the full symmetry group is $u(1) \rtimes T$.  In both theories the single charge and the single monopole are bosons. However, in the $u(1) \times {CT}$ theory time-reversal acts as, $CT:\, (q, m) \to (-q, m)$ and the single monopole is a Kramers singlet, while in the $u(1) \rtimes T$ theory, $T:\, (q, m) \to (q, -m)$ and the single charge is a Kramers singlet. The Euclidean partition functions of the two theories are given by,
\bea Z^b_{CT}(e) &=& \frac{1}{\mathrm{Vol}({\cal G})} \sum_{L} \int D a_{\mu} \, e^{- S[a]} \label{eq:Zbos1} \\
Z^b_{T}(e) &=& \frac{1}{\mathrm{Vol}({\cal G})} \sum_{\tilde{L}} \int D a_{\mu} \, e^{- S[a]} \label{eq:Zbos2}\eea 
with the Maxwell-action (\ref{eq:introMaxwell}).
The two theories differ in the allowed set of line-bundles and gauge fields $a_{\mu}$. In the $u(1) \times CT$ case the (gapped) charged boson $B$ transforms under patch change as,
\beq u(1) \times CT: \, \quad \tilde{B}(y)  = e^{i \alpha(y)} B(x(y)), \quad \tilde{a}_{\mu}(y) = \frac{\d x^{\beta}}{\d y^{\mu}} a_{\beta}(x(y)) + \d_{\mu} \alpha(y) \label{eq:btransftimes} \eeq
The transition functions $e^{i \alpha}$ define a complex line-bundle. The sum in Eq.~(\ref{eq:Zbos1}) is over all line-bundles $L$, which are in one-to-one correspondence with elements of $H^2(M, \bZ)$. Gauge fields in each line bundle differ by one-forms.

On the other hand, in the $u(1) \rtimes T$ case, for orientation reversing patch changes,
\beq u(1) \rtimes T: \, \quad \tilde{B}(y)  = e^{i \alpha(y)} B^{\dagger}(x(y)), \quad \tilde{a}_{\mu}(y) = -\frac{\d x^{\beta}}{\d y^{\mu}} a_{\beta}(x(y)) + \d_{\mu} \alpha(y)  \eeq
while for orientation preserving patch changes Eq.~(\ref{eq:btransftimes}) still holds. The transition functions $e^{i \alpha}$ define a twisted complex line-bundle. The sum in Eq.~(\ref{eq:Zbos2}) is over all twisted line-bundles $\tilde{L}$, which are in one-to-one correspondence with elements of $H^2(M, \tilde{\bZ}) \sim H_2(M, \bZ)$.  Gauge fields in each twisted line-bundle differ by pseudo-one-forms.

Let us start with the $u(1) \times CT$ theory on $\mRP$. The non-equivalent line-bundles are given by $H^2(\mRP, \bZ) = \bZ_2$. It will be convenient to work on the double cover of $\mRP$, $S^4$. We denote the antipodal map on $S^4$ by $\imath$. The two distinct line-bundles then correspond to the following conditions on the matter field $B$ on $S^4$:
\beq B(\imath(x)) = \pm B(x) \label{eq:Bantip}\eeq
As in section \ref{sec:bosonsoriented}, we write $a_{\mu} = a^{cl}_{\mu} + a^q_{\mu}$. Both line-bundles admit a flat connection $f^{cl} = 0$. Thus, the sum over line-bundles just gives a factor $|T| = 2$ in the present case, and we are left with an integral over the ``quantum" field $a^q_{\mu}$, which is simply a $1$-form. Lifting $a^q$ to $S^4$, we have the condition
\beq \imath^* (a) = + a \label{eq:ia}\eeq
where $\imath^*$ is the pull-back of $a$ under the anti-podal map, i.e. the path integral is over $1$-forms with ``even" parity. We also need to discuss the factor of the gauge-group volume $\mathrm{Vol}({\cal G})$ in Eq.~(\ref{eq:Zbos1}). Consider a gauge transformation $B(x) \to e^{i \alpha(x)} B(x)$. Lifting the gauge-transformation to $S^4$, in order to preserve the condition (\ref{eq:Bantip}), we must have $\alpha(\imath(x)) = + \alpha(x)$, i.e. $\alpha$ must have even parity. $d \alpha$ then satisfies (\ref{eq:ia}), as required. Then, repeating the analysis in section \ref{sec:bosonsoriented}, we obtain
\beq Z^b_{CT}(e, \mRP) = \frac{2}{2 \pi \sqrt{V(\mRP)}} \left(\mathrm{det}'_+ \Delta^0\right)^{1/2} \left(\mathrm{det}_{\perp, +} \frac{\Delta^1}{2 \pi e^2}\right)^{-1/2} \label{eq:ZbRP4}\eeq
where $\Delta^1$ and $\Delta^0$ are Laplacians for $1$ and $0$ forms on $S^4$ repspectively. The subscript `$+$' on the determinant serves to remind us that the product is only over even parity modes. As before, the subscript $\perp$ on $\mathrm{det}_{\perp, +} \Delta^1$ serves to remind that only transverse modes $\delta a  =0$ are taken into account. All determinants here and below are regularized using $\zeta$-function regularization.

Next we proceed to the $u(1) \rtimes T$ theory on $\mRP$. We must now sum over inequivalent twisted line bundles, which are classified by $H^2(\mRP, \tilde{\bZ}) = H_2(\mRP, \bZ) = \bZ_1$. Thus, there is only a single gauge-inequivalent twisted line-bundle, which  corresponds to the constraint on $S^4$,
\beq B(\imath(x)) = B^{\dagger}(x) \label{eq:Bdantip}\eeq
Indeed, by a global phase rotation $B(x) \to e^{i \alpha} B(x)$ we may change the phase in Eq.~(\ref{eq:Bdantip}) to $B(\imath(x)) = e^{-2 i \alpha} B^{\dagger}(x)$, i.e. all such choices are gauge-equivalent. So, we have  just a single twisted line bundle, which again admits $f^{cl} = 0$. The gauge-transformations on $S^4$ preserving Eq.~(\ref{eq:Bdantip}) are $B(x) \to e^{i \alpha(x)} B(x) $ with $e^{i \alpha(\imath(x))} = e^{- i \alpha(x)}$. Hence, the gauge group ${\cal G}$ decomposes into two subgroups: i) a subgroup consisting of $\alpha(x)$ satisfying $\alpha(\imath(x)) = - \alpha(x)$, i.e. $\alpha(x)$ - an arbitrary odd parity function on $S^4$; ii) a $\bZ_2$ subgroup generated by $e^{i \alpha(x)} = -1$. Thus, we may write,
\beq \mathrm{Vol}({\cal G}) = 2 \int D \alpha_- \eeq
with the integral being over all odd parity functions $\alpha_-$ on $S^4$. Comparing to Eq.~(\ref{eq:VolG}), we see that the factor $2 \pi \sqrt{V(M)}$ has been replaced by a factor of $2$, due to the fact that the only constant gauge transformation allowed is $e^{i \alpha} = -1$.

The $a_{\mu}$ integral again reduces to the quantum part. Lifting the pseudo-one-form $a^q$ to $S^4$, we have 
\beq \imath^* (a) = - a \label{eq:iartimes}\eeq
i.e. the integral is over all $1$-forms of odd parity. Repeating the analysis in section \ref{sec:bosonsoriented} we obtain,
\beq Z^b_{T}(e, \mRP) = \frac12\left(\mathrm{det}_- \Delta^0\right)^{1/2} \left(\mathrm{det}_{\perp, -} \frac{\Delta^1}{2 \pi e^2}\right)^{-1/2} \label{eq:ZbrRP4}\eeq
with the `$-$' subscripts reminding us to take the product only over odd parity modes.

We observe that the two partition functions (\ref{eq:ZbRP4}) and (\ref{eq:ZbrRP4}) are, seemingly, very different. However, an explicit calculation performed in Appendix \ref{app:RP4} shows that they are, in fact, related. In particular, if we as in the case of oriented manifolds make the redefinition
\beq \bar{Z}^b(e) = \left(\frac{e^2}{2 \pi}\right)^{-\chi/4} (\sqrt{2 \pi} e)^{N^0 - N^1} Z^b(e) \label{eq:barZbred}\eeq
with $N^0 - N^1$ given as before by Eq.~(\ref{eq:NvmNs}) for both the $u(1)\times CT$ and $u(1)\rtimes T$ theories then
\beq \bar{Z}^b_{CT}(e, \mRP) = \bar{Z}^b_T(2\pi/e,\mRP) \label{eq:ZbRP4final}\eeq
This concludes the proof of $S$-duality in a theory with bosonic matter on $\mRP$.

\subsection{$S$-duality:  bosonic matter, non-orientable manifold. Relation to analytic torsion and Reidemeister torsion.}
\label{sec:bosonsS}
In this section we give a proof of $S$-duality in a $u(1)$ gauge theory with bosonic matter and $\theta = 0$ on an arbitrary non-orientable manifold. In particular, we find that $S$-duality in this theory reduces to the well-known equality of Ray-Singer analytic torsion and Reidemeister torsion.\cite{RaySinger, Cheeger,Muller}

First we write the partition functions $Z^b_{CT}$ and $Z^b_T$ on an arbitrary non-orientable manifold $M$.
\bea Z^b_{CT}(e) &=& \frac{{\cal C}_+}{2 \pi \sqrt{V(M)}} \left(\mathrm{det}'_+ \Delta^0\right)^{1/2} \left(\mathrm{det'}_{\perp, +} \frac{\Delta^1}{2 \pi e^2}\right)^{-1/2} Z^b_{cl,CT}(e) \label{eq:Zbgen1}\\
 Z^b_T(e) &=& \frac{{\cal C}_-}{2}\left(\mathrm{det}_- \Delta^0\right)^{1/2} \left(\mathrm{det'}_{\perp, -} \frac{\Delta^1}{2 \pi e^2}\right)^{-1/2} Z^b_{cl,T}(e) \label{eq:Zbgen2} \eea
Let us unpack the expressions above. Again, it is convenient to work on the orientation double cover of $M$, $\tilde{M}$, with the projection map being $p: \tilde{M} \to {M}$. Let $\imath$ denote the map on $\tilde{M}$ exchanging the two pre-images $p^{-1}(x)$ of a point $x \in M$. $n$-forms $\omega$ on $\tilde{M}$ can be divided into even forms $C^+_n$ and odd forms $C^-_n$ satisfying $\imath^* \omega = \pm \omega$ respectively. The even forms on $\tilde{M}$ correspond to forms on $M$, while odd forms on $\tilde{M}$ correspond to pseudo-forms on $M$. The Laplacian operator on $\tilde{M}$ preserves the parity of the form. The Hodge star operator on $\tilde{M}$ commutes with $\Delta$ and gives a one-to-one correspondence  $*: C^+_n \to C^{-}_{4-n}$. The subscripts $+$ ($-$) in Eqs.~(\ref{eq:Zbgen1}), (\ref{eq:Zbgen2}) denote a product over even (odd) eigenstates of $\Delta$ on $\tilde{M}$. The subscript $\perp$ on $\det \Delta^1$ again indicates that only transverse eigenmodes $\delta a_\perp = 0$ are considered. $\zeta$-function regularization is assumed throughout.

The constants ${\cal C_+}$ and ${\cal C_-}$ in Eqs.~(\ref{eq:Zbgen1}), (\ref{eq:Zbgen2}) denote contributions of zero-modes of $\Delta^1$ acting on the space of forms and pseudo-forms respectively. More explicitly, let $a^+_i$ ($a^-_i$) be zero-modes of $\Delta^1$ acting on forms (pseudo-forms). We need to perform an integral over these zero-modes modulo ``large gauge transformations." It is convenient to choose $2 \pi a^+_i$ and $2 \pi a^-_i$ to be precisely pure gauge fields corresponding to these large gauge transformations. The modes $a^+_i$ are then in one to one correspondence with elements of $H^1(M, \bZ)$, i.e. we have $b_1$ such modes, and they satisfy $\int_C a^+_i \in \bZ$ for any curve $C \in H_1(M, \bZ)$.  Likewise, the modes $a^-_i$ are in one to one correspondence with elements of $\mathrm{Free}(H^1(M, \tilde{\bZ}))$. 
By Poincare duality $H^1(M, \tilde{\bZ}) \sim H_3(M, \bZ)$, so there are $b_3$ zero-modes $a^-_i$, and these satisfy $\int_{\tilde{C}} a^-_i \in \bZ$ for any $\tilde{C} \in H_1(M, \tilde{\bZ})$. Writing the zero-mode part of the gauge field $a$ as $a^{\pm} = c^{\pm}_i a^{\pm}_i$, the coefficient $c^{\pm}_i$ now range from $0$ to $2 \pi$. The modes $a^{\pm}_i$ are generally not orthogonal; let us define:
\beq {\cal N}^{1,+}_{ij} = \langle a^+_i, a^+_j\rangle_M, \quad {\cal N}^{1,-}_{ij} = \langle a^-_i, a^-_j\rangle_M \label{eq:N1def}\eeq
Then, the zero-mode contributions are simply,
\beq {\cal C}_{+} = (2\pi)^{b_1} (\det {\cal N}^{1,+})^{1/2}, \quad {\cal C}_{-} = (2 \pi)^{b_3} (\det {\cal N}^{1,-})^{1/2} \eeq

We note that the factor of $2$ in the denominator of Eq.~(\ref{eq:Zbgen2}) has the same origin as in our explicit calculation on $\mRP$: it represents the large gauge transformation $e^{i \alpha} = -1$. 

We finally define the classical partition functions $Z^b_{cl, CT}(e)$ , $Z^b_{cl, T} (e)$. These correspond to the sum over non-trivial line-bundles and twisted line-bundles. In the $u(1)\times CT$ case these are given by elements of $H^2(M, \bZ)$, and in the $u(1) \rtimes T$ case by elements of $H^2(M, \tilde{\bZ})$, so
\bea Z^b_{cl, CT}(e)  &=& |{\rm Tor}(H^2(M, \bZ))| \sum_{f^+ \in {\rm Free}(H^2(M, \bZ))} \exp\left(-\frac{(2\pi)^2}{2e^2} \langle f^+, f^+\rangle_M\right)  \nn\\
Z^b_{cl, T}(e) &=& |{\rm Tor}(H^2(M, \tilde{\bZ}))| \sum_{f^- \in {\rm Free}(H^2(M, {\tilde{\bZ}}))} \exp\left(-\frac{(2\pi)^2}{2e^2} \langle f^-, f^-\rangle_M\right)\eea
$f^{\pm}$ are chosen to satisfy $\Delta f^{\pm} = 0$, moreover $\int_U f^+ \in \bZ$ for any 2-cycle $U \in H_2(M, \bZ)$ and $\int_{\tilde{U} }f^- \in \bZ$ for any $\tilde{U} \in H_2(M, \tilde{\bZ})$. We will alternately think of $f^{+}$ ($f^-$) as a 2-form (pseudo 2-form) on $M$ or as an even (odd) form on $\tilde{M}$. Note that by Poincare duality $H^2(M, \tilde{\bZ}) \sim H_2(M, \bZ)$. We write
\beq f^+ = c^+_i f^+_i, \quad f^- = c^-_i f^-_i\eeq
where $f^+_i$ ($f^-_i$), $i = 1\ldots b_2$, form a basis for  ${\rm Free}(H^2(M, \bZ))$ (${\rm Free}(H^2(M, \tilde{\bZ}))$) and $c^{\pm}_i$ are integer coefficients. Let us define, 
\beq Q_{i j} = \int_M f^+_i \wedge f^-_j = \frac{1}{2 } \int_{\tilde M} f^+_i \wedge f^-_j \eeq
$Q$ is an integer-valued (not necessarily symmetric) matrix and Poincare duality implies $|\det Q| = 1$. Recalling that the Hodge star operator sends even forms on $\tilde{M}$ to odd forms, we can write
\beq * f^+_i = f^-_j S_{j i}\eeq
where $S$ is an invertible (not necessarily integer-valued) matrix. We then have,
\bea \langle f^+, f^+\rangle_M &=& \frac{1}{2} \int_{\tilde{M}} f^+ \wedge * f^+ =  (c^+)^T Q S c^+ \nn\\
\langle f^-, f^- \rangle_M &=& \frac{1}{2} \int_{\tilde{M}} f^- \wedge * f^- =  (c^-)^T Q^T S^{-1} c^- \eea
Thus,
\bea Z^b_{cl, CT}(e)  &=& |{\rm Tor}(H^2(M, \bZ))| \sum_{c^+ \in \bZ^{b_2}} \exp\left(-\frac{(2\pi)^2}{2e^2} (c^+)^T Q S c^+\right)  \nn\\
Z^b_{cl, T}(e) &=& |{\rm Tor}(H_2(M, \bZ))| \sum_{c^- \in \bZ^{b_2}} \exp\left(-\frac{ (2 \pi)^2}{2e^2} (c^-)^T Q^T S^{-1} c^- \right)\eea
Using Poisson resummation we obtain,
\beq Z^b_{cl, CT}(e) = \frac{|{\rm Tor}(H^2(M, \bZ)|}{|{\rm Tor} (H_2(M, \bZ))|} \left(\frac{e^2}{2\pi}\right)^{b_2/2} \det (Q S)^{-1/2} Z^b_{cl, T}(2\pi/e)\eeq
It will prove convenient to define
\bea {\cal N}^{2, +}_{ij} &\equiv&\langle f^+_i, f^+_j\rangle_M =  (Q S)_{i j}   \\ 
{\cal N}^{2, -}_{ij} &\equiv&  \langle f^-_i, f^-_j\rangle_M = (Q^T  S^{-1})_{ij} \eea

Returning to the full partition-functions (\ref{eq:Zbgen1}), (\ref{eq:Zbgen2}), we seek to factor out the $e$ dependence,
 \bea Z^b_{CT}(e) &=& \frac{{\cal C}_+ (2 \pi e^2)^{(N^1_+ - N^0_+ - b_1 + 1)/2}}{2 \pi \sqrt{V(M)}} \mathrm{det}'_+ \Delta^0\left(\mathrm{det'}_{ +} \Delta^1\right)^{-1/2} Z^b_{cl,CT}(e) \label{eq:Zbnop1}\\
 Z^b_T(e) &=& \frac{{\cal C}_- (2 \pi e^2)^{(N^1_- - N^0_- - b_3)/2}}{2} \mathrm{det}_- \Delta^0 \left(\mathrm{det'}_{-} \Delta^1\right)^{-1/2} Z^b_{cl,T}(e) \label{eq:Zbnop2} \eea
Notice that now the determinants ${\rm det}'_{\pm} \Delta^1$ are taken over all modes (both transverse and longitudinal). Also note that  $N^1_+ = N^1_-$ and $N^0_+ = N^0_-$, and $N^1_+ - N^0_+ = N^1_- - N^0_-$ is still given by Eq.~(\ref{eq:NvmNs}). Indeed, $N^1_{\pm}$, $N^0_{\pm}$ can be obtained through heat-kernel expansion on $M$ and are, therefore, insensititve to non-local properties (i.e. whether we are dealing with forms or pseudo-forms). Considering the renormalized partition functions (\ref{eq:barZbred}), we have
\bea \frac{\bar{Z}^b_{CT}(e)}{\bar{Z}^b_T(2\pi/e)} =\frac{2  \left(\det {\cal N}^{2,+}\right)^{-1/2}  \left(\det {\cal N}^{1,+}\right)^{1/2} }{\sqrt{V(M)}  \left(\det {\cal N}^{1,-}\right)^{1/2}} \frac{ |{\rm Tor}(H^2(M, \bZ))|}{|{\rm Tor}(H_2(M, \bZ))|} \frac{\mathrm{det}'_+ \Delta^0}{\mathrm{det}_- \Delta^0} \left(\frac{\mathrm{det'}_{ +} \Delta^1}{\mathrm{det'}_{ -} \Delta^1}\right)^{-1/2}  \nn\\ \label{eq:detid} \eea
We see that all factors of $e$ have cancelled out. It will be convenient to write Eq.~(\ref{eq:detid}) in a more suggestive form. First, let us rewrite all expressions in terms of spectrum of Laplacian acting on forms only. Using the Hodge star operation we have, 
$\det_- \Delta^0 = \det_+ \Delta^4$, $\det'_- \Delta^1 = \det'_+ \Delta^3$. Furthermore, we can relate the inner product ${\cal N}^{1,-}$ on ${\rm Free}(H^1(M, \tilde{\bZ}))$ to the inner product on ${\rm Free}(H^3(M, \bZ))$. Indeed, let $\eta^+_i$, $i = 1\ldots b_3$, be a basis for ${\rm Free}(H^3(M, \bZ))$. Consider the pairing between ${\rm Free}(H^3(M, \bZ))$ and ${\rm Free}(H^1(M, {\tilde{\bZ}}))$,
\beq P_{i j} = \int_M \eta^+_i \wedge a^-_j \eeq 
By Poincare duality $P$ is an integer-valued matrix and $|\det P| = 1$. The Hodge star operator gives a mapping,
\beq *\eta^+_i = a^-_j S^{31}_{ji} \eeq
where $S^{31}$ is an invertible (not necessarily integer valued) matrix. We have an inner product on ${\rm Free}(H^3(M, {\bZ}))$,
\beq {\cal N}^{3, +}_{ij} \equiv \langle \eta_i, \eta_j \rangle = \int_M \eta_i \wedge * \eta_j = (P S^{31})_{ij}\eeq 
Similarly, the inner product on ${\rm Free}(H^1(M, \tilde{{\bZ}}))$ is
\beq {\cal N}^{1, -}_{ij} = \langle a^-_i, a^-_j \rangle = \int_M a^-_i \wedge * a^-_j = \left(P^T (S^{31})^{-1}\right)_{ij}\eeq
Thus, we conclude,
\beq \det {\cal N}^{1,-} = (\det {\cal N}^{3, +})^{-1}\eeq
Finally, viewing the constant function $\alpha_0 = 1$ as the generator of $H^0(M, \bZ)$ we can think of ${\cal N}^{0, +} \equiv V(M) = \langle \alpha_0, \alpha_0 \rangle$ as the inner product on $H^0(M, \bZ)$. Therefore, we may rewrite Eq.~(\ref{eq:detid}) as,
\beq \frac{\bar{Z}^b_{CT}(e)}{\bar{Z}^b_T(2\pi/e)} = \frac{T^{RS}(M)}{T^c(M)}\eeq
where
\beq T^{RS}(M) = (\mathrm{det}_+ \Delta^4)^{-1}  \left(\mathrm{det}'_+ \Delta^3\right)^{1/2} (\mathrm{det'}_{+} \Delta^1)^{-1/2}  \,\mathrm{det}'_+ \Delta^0 \label{eq:TRS} \eeq
and
\bea T^c(M) &=& \frac{|{\rm Tor} (H^3(M, \bZ))|}{|{\rm Tor} (H^4(M, \bZ))| |{\rm Tor} (H^2(M, \bZ))| }  \times\nn\\
 &\times& \left(\det {\cal N}^{3, +}\right)^{-1/2}\left(\det {\cal N}^{2, +}\right)^{1/2} \left(\det {\cal N}^{1, +}\right)^{-1/2} \left(\det {\cal N}^{0, +}\right)^{1/2} \label{eq:Tc}\eea
Here we've used the fact that ${\rm Tor}(H^3(M, {\bZ})) = {\rm Tor}(H_2(M, {\bZ}))$, and $H^4(M, \bZ) = \bZ_2$ for a non-orientable four-dimensional manifold. As shown below, $T^{RS}(M)$ in Eq.~(\ref{eq:TRS}) is the Ray-Singer analytic torsion of the manifold $M$.\cite{RaySinger}   $T^c(M)$ in Eq.~(\ref{eq:Tc}) is the Reidemeister torsion of the manifold $M$ (see appendix \ref{app:Reidemeister} for details).\footnote{The superscript $c$ stands for ``combinatorial," which is yet another name associated with the Reidemeister torsion.} We thank E. Witten for pointing out this connection to us. A well-known theorem\cite{Cheeger,Muller} states that the two torsions coincide, $T^{RS}(M) = T^c(M)$. The original proofs of this theorem in Refs.~\onlinecite{Cheeger,Muller} assume an oriented manifold\footnote{In fact, in even dimensions on oriented manifolds both $T^{RS}$ and $T^c$ are trivially equal to $1$.}; however, we expect that the theorem continues to hold on non-orientable manifolds. Then,
\beq \bar{Z}^b_{CT}(e) = \bar{Z}^b_T(2\pi/e) \label{eq:Sbosonsf}\eeq
completing our proof of $S$-duality in $u(1)$ gauge theory with bosonic matter on non-orientable manifolds.

We conclude this section by showing how Eq.~(\ref{eq:TRS}) is related to the standard definition of Ray-Singer analytic torsion,
\beq \log T^{RS}_{\rm standard}(M) = \frac12 \sum_{q = 0}^{D} (-1)^{q+1} q \log {\rm det}^{'}_+ \Delta^q \label{eq:TRSgenD}\eeq
i.e. 
\beq T^{RS}_{\rm standard}(M) = ({\rm det}_+ \Delta^4)^{-2} ({\rm det}'_+ \Delta^3)^{3/2} ({\rm det}'_+ \Delta^2)^{-1} ({\rm det}'_+ \Delta^1)^{1/2} \label{eq:TRSstand}\eeq
For an arbitrary manifold,
\beq \prod_{q = 0}^{D} ({\rm \det}'_+ \Delta^q)^{(-1)^q} = 1 \label{eq:telescope}\eeq
Indeed, the space of $q$-forms can be decomposed as ${\cal F}^q = {\cal F}^q_{\parallel} \oplus {\cal F}^q_0  \oplus {\cal F}^q_{\perp}$. Here ${\cal F}^q_{\parallel}$ is spanned by eigenvectors $\phi_{\parallel}$ of $\Delta^q$ with eigenvalue $\lambda \neq 0$ and $d \phi_{\parallel} = 0$. ${\cal F}^q_{\perp}$ is spanned by eigenvectors $\phi_{\perp}$ of $\Delta^q$ with eigenvalue $\lambda \neq 0$ and $\delta \phi_{\perp} = 0$. ${\cal F}^q_0$ is spanned by zero-modes of $\Delta^q$. There is a one to one mapping ${\cal F}^q_{\perp} \stackrel{d}{\to} {\cal F}^{q+1}_{\parallel}$ where an eigenvector of $\Delta^q$, $\phi^q_{\perp} \in {\cal F}^q_{\perp}$, with eigenvalue $\lambda$ is mapped to an eigenvector of $\Delta^{q+1}$, $d \phi^q_{\perp} \in {\cal F}^{q+1}_{\parallel}$, with the same eigenvalue. Eq.~(\ref{eq:telescope}) follows immediately. Now multiplying Eq.~(\ref{eq:TRSstand}) by Eq.~(\ref{eq:telescope}) we obtain Eq.~(\ref{eq:TRS}).

\subsection{$S$-duality: fermionic matter, non-orientable manifold}
\label{sec:fermionsRP4}
We are finally ready to discuss $S$-duality of ${\cal L}_T$ (gauged TI) (\ref{eq:Zrtimesm}) and ${\cal L}_{CT}$ (gauged $\nu  =1$ TSc)  (\ref{eq:Ztimesm}) on a non-orientable manifold. The only difference compared to the bosonic case (\ref{eq:Zbgen1}), (\ref{eq:Zbgen2}) comes in the ``classical" part of the partition function:
\bea Z^f_{CT}(e) &=& \frac{{\cal C}_+ (2 \pi e^2)^{(N^1_+ - N^0_+ - b_1 + 1)/2}}{2 \pi \sqrt{V(M)}} \mathrm{det}'_+ \Delta^0\left(\mathrm{det'}_{ +} \Delta^1\right)^{-1/2} Z^f_{cl,{CT}}(e) \label{eq:Zfgen1}\\
 Z^f_T(e) &=& \frac{{\cal C}_- (2 \pi e^2)^{(N^1_- - N^0_- - b_3)/2}}{2} \mathrm{det}_- \Delta^0 \left(\mathrm{det'}_{-} \Delta^1\right)^{-1/2} Z^f_{cl,T}(e) \label{eq:Zfgen2} \eea
 with
\bea Z^f_{cl, CT}(e) &=& \sum_{{\rm Pin}_{c}} \exp\left(-\frac{1}{2e^2} \langle f, f\rangle_M + 2 \pi i \eta \right) \label{eq:Zfcl1}\\
Z^f_{cl, T}(e) &=& \sum_{{\rm Pin}_{\tilde{c}}} \exp \left(-\frac{1}{2e^2} \langle f, f \rangle_M + \frac{\pi i N^\chi_0}{2}\right) \label{eq:Zfcl2}\eea
The sum in Eq.~(\ref{eq:Zfcl1}) is over Pin$_c$ structures on the manifold $M$. $f$ is the field strength for each such structure chosen to satisfy $\Delta f = 0$. $\eta$ is the $\eta$-invariant of the Pin$_c$ structure, see section \ref{sec:Zftimes}. We also recall here that different Pin$_c$ structures on a manifold differ by complex line bundles, i.e. by elements of $H^2(M, \bZ)$. 

The sum in Eq.~(\ref{eq:Zfcl2}) is over Pin$_{\tilde{c}}$ structures on the manifold $M$, discussed in section \ref{sec:Zfrtimes}. Again for each such structure we choose a representative gauge field satisfying $\Delta f  = 0$. $N^{\chi}_0$ is the number of zero-modes of the associated twisted doubled Dirac operator (see section \ref{sec:Zfrtimes}). Different Pin$_{\tilde{c}}$ structures on a manifold differ by twisted complex line bundles, i.e. by elements of $H^2(M, \tilde{\bZ})$.  

Now, consider the renormalized partition functions,
\beq \bar{Z}^f(e) = \left(\frac{e^2}{4\pi}\right)^{-\chi/4} (\sqrt{2 \pi} e)^{N^0 - N^1} Z^f(e)\eeq
defined the same way for $\bar{Z}^f_{CT}$ and $\bar{Z}^f_T$ with the same expression for $N^0 - N^1$, Eq.~(\ref{eq:NvmNs}). The statement of $S$-duality is $\bar{Z}^f_T\left(\frac{4 \pi}{e}\right) = \bar{Z}^f_{CT}(e)$. We can now use $S$-duality in a theory with bosonic matter (\ref{eq:Sbosonsf}) proved in the previous section. Then from Eq.~(\ref{eq:detid}), the statement of $S$-duality for fermions reduces to
\beq Z^f_{cl, T}\left(\frac{4 \pi}{e}\right) = 2^{(1-b_1 + b_3)/2} ({\rm det} {\cal N}^{2,+} )^{1/2} \frac{|{\rm Tor}(H_2(M, \bZ))|}{|{\rm Tor}(H^2(M, \bZ))|} \left(\frac{e^2}{4\pi}\right)^{-b_2/2} Z^f_{cl, CT}(e) \label{eq:Finalf} \eeq

We leave the general proof of Eq.~(\ref{eq:Finalf}) to future work. Here, we will check that this identity holds for the manifold $\mRP$. Let us compute the classical partition functions (\ref{eq:Zfcl1}), (\ref{eq:Zfcl2}) on $\mRP$. Starting with the case of ${\cal L}_{CT}$, $\mRP$ admits two Pin$_c$ structures, differing by the non-trivial line bundle (\ref{eq:Bantip}) corresponding to the non-trivial element of $H^2(\mRP, \bZ) = \bZ_2$. These Pin$_c$ structures are discussed explicitly in appendix \ref{app:DRP4}. Both structures admit a flat connection $f  = 0$. The $\eta$ invariants corresponding to the two structures are $\eta = \pm \frac{1}{8}$. (Ref.~\onlinecite{GilkeyPinC} gives an elegant indirect way to compute $\eta$ of $\mRP$; we also give a direct calculation in appendix \ref{app:DRP4}.) Thus, we obtain 
\beq Z^f_{cl, CT}(e, \mRP) = \sqrt{2} \eeq

Proceeding to ${\cal L}_T$, $\mRP$ admits just a single Pin$_{\tilde{c}}$ structure (since $H^2(\mRP, \tilde{\bZ}) = H_2(\mRP, \bZ) = \bZ_1$). Again, the corresponding gauge field can be chosen to be flat $f  =0$. As we show in appendix \ref{app:DRP4}, there are no zero-modes of the doubled Dirac operator on $\mRP$, $N^{\chi}_0 = 0$. Therefore,
\beq Z^f_{cl, T}(e, \mRP) = 1 \eeq
Now, on $\mRP$, $b_1 = b_2 = b_3 = 0$. Furthermore, since $b_2  =0$ the factor $({\rm det} {\cal N}^{2,+})^{1/2}$ in Eq.~(\ref{eq:Finalf}) should be set to $1$. As $|{\rm Tor}(H^2(\mRP,\bZ))| = 2$ and $|{\rm Tor}(H_2(\mRP, \bZ))| = 1$, we see that the identity (\ref{eq:Finalf}) holds!

\begin{table}[t]

\beq
\begin{array}{|c|c|c|c|c|c|c|c|c|}
\hline
k\rightarrow     & 0 & 1 & 2 & 3 & 4 & 5 & 6 & 7 \\ \hline
I    & 1&   & \,-i&   &1\ &   &\,-i&   \\ \hline
\sigma  &   & \,\,1\,\, &   & -1 &   &  -1 &   &  \,\,1\,\,  \\ \hline
\psi & -1 &   & i&   & -1 &   & i &  \\ \hline\hline 
T^2 & 1& \eta & & -\eta & -1 & -\eta& & \eta\\ \hline
\end{array}\nn\eeq
\caption{T-Pfaffian$_\eta$ topological orders with $\eta = \pm 1$.  T-Pfaffian topological order can be regarded as a restriction of the Ising$\times U(1)_{-8}$ topological order. The top table lists the topological spins of anyons; the column and row indices denote the $U(1)_{-8}$ charge and the Ising charge respectively. The missing entries do not correspond to anyons in the T-Pfaffian theory. The physical $U(1)$ charge of anyons $Q_{EM} = k/4$, with $k$ - the $U(1)_{-8}$ charge. Time-reversal maps $k$ to itself.  The bottom row lists the $T^2$ assignment of anyons (where defined). The $T^2$ assignment is independent of the Ising charge.  $\psi_4$ is the physical electron. T-Pfaffian$_+$ can be realized on the surface of an ordinary TI.
}
\label{tbl:TPfaffian}

\end{table}

\section{Surface states of topological insulators and superconductors}
\label{sec:surface}
In this paper, we have provided evidence for a recent conjecture that a gauged non-interacting topological insulator and a gauged non-interacting $\nu = 1$ topological superconductor  describe the same $T$-invariant phase of matter. In particular, we have demonstrated that these two phases do not differ by an SPT phase of bosons with $T$ symmetry. Besides being interesting in its own right, this duality was used in Refs.~\onlinecite{ChongSenthilDirac, MVDuality} to derive a dual description of the surface of a (non-gauged) topological insulator. Recall that in the absence of interactions the TI surface supports a single $2+1$D gapless Dirac cone,
\beq L_{\rm TI, boundary} = \bar{\psi} i \gamma^{\mu} (\d_{\mu} - i A_{\mu}) \psi \label{eq:Diracs}\eeq
where we've introduced an external non-dynamical $U(1)$ gauge field $A_{\mu}$ for future convenience. Ref.~\onlinecite{MVDuality} started with a gauged bulk topological insulator and a gauged bulk $\nu  = 1$ topological superconductor, and considered a Higgs transition on the TI side, which by $S_f$-duality corresponds to a confinement transition on the TSc side, deducing the following surface theory of a (non-gauged) TI 
\beq L_{{\rm QED}_3} = \bar{\psi}_{d} i \gamma^{\mu} (\d_{\mu} - i a _{\mu}) \psi_{d} - \frac{1}{4 e^2} f_{\mu \nu} f^{\mu \nu} + \frac{1}{4 \pi} A_{\mu} \epsilon^{\mu \nu \lambda} \d_{\nu} a_{\lambda} \label{eq:Diracdual}\eeq
This dual surface theory is just QED$_3$ with a single flavor of Dirac fermions $\psi_{d}$.\footnote{Conventional lore would require adding a Chern-Simons term with a half-odd-integer level for the dynamical gauge field $a_{\mu}$ to Eq.~(\ref{eq:Diracdual}). We stress that such a term is absent in the dual theory (\ref{eq:Diracdual}) and the parity anomaly is avoided instead by changing the compactification conditions on gauge field $a_{\mu}$ (see Ref.~\onlinecite{MVDuality} for details).} The theory (\ref{eq:Diracdual}) is a description of {\it some} interacting surface of the bulk TI. The bulk duality of our two $u(1)$ gauge theories guarantees that the two theories (\ref{eq:Diracs}), (\ref{eq:Diracdual}) have the same anomaly under the global $U(1) \rtimes T$ symmetry of the TI. It, however, does not gaurantee that the theories (\ref{eq:Diracs}) and (\ref{eq:Diracdual}) are identical in the infra-red. The fate of QED$_3$ with a single fermion flavor in the infra-red is an interesting open problem, and the possibility that it flows to the free Dirac theory (\ref{eq:Diracs}) remains open. A more mundane possibility is that it spontaneously breaks time-reversal symmetry in the infra-red, generating a mass term $m \bar{\psi}_{d} \psi_{d}$. A third possibility is that it flows to a CFT distinct from a free Dirac theory (\ref{eq:Diracs}).

Independent of its fate in the infra-red,  the dual surface theory (\ref{eq:Diracdual}) has the following useful application. Imagine introducing a charge $2$ Higgs field into the theory (\ref{eq:Diracdual}) and driving a surface Higgs transition with a condensate $\langle \psi^T_{d} C \psi_{d} \rangle$. One then obtains a symmetry-respecting gapped topologically ordered surface phase, which is known in the literature as $T$-Pfaffian, see table \ref{tbl:TPfaffian}.\cite{ChongSenthilDirac, MVDuality} While the possibility of such a surface termination of a TI was conjectured before,\cite{Bonderson2013,Chen2014PRB} the dual theory (\ref{eq:Diracdual}) gives one direct access to it. Furthermore, the dual description fixes a previous ambiguity in the quantum-numbers of this surface phase under $T$: indeed, the $T$-Pfaffian topological order comes in two varieties $T$-Pfaffian$_+$ and $T$-Pfaffian$_-$, which differ by the quantum numbers of the anyons under $T$. It was previously known that one of these varieties corresponds to a surface phase of a TI, and the other - to a surface of a bulk phase differing from the TI by an eTmT SPT phase of bosons. However, it was not known which variety corresponds to which phase. The results of this paper allow one to fix $T$-Pfaffian$_+$ as the surface of a TI.

The $T$-Pfaffian phase also makes an appearance when one studies surface topological orders of fermion topological superconductors (class DIII) in 3+1D. These are SPT phases of fermions with time-reversal symmetry only and $T^2 = (-1)^F$. In the absence of interactions such topological superconductors have an integer classification. The non-interacting surface of a phase $\nu \in \bZ$ hosts $|\nu|$ gapless Majorana cones. Interaction effects break this classification down to $\bZ_{16}$.\cite{Kitaev_pc,FidkowskiChenAV, Wang2014, MetlitskiChenFidkowskiAV2014, KapustinFerm}  Symmetry respecting surface topological orders have been deduced for all even $\nu$ topological superconductors in class DIII.\cite{FidkowskiChenAV, Wang2014, MetlitskiChenFidkowskiAV2014} In particular, the surface of the $\nu = 2$ phase admits the topological order $T$-Pfaffian$_+\times \{1, s\}$, where $s$ is a semion and $s \times s = 1$. We now know that $T$-Pfaffian$_+$ is a surface state of a non-interacting TI.  The surface of a TI can be driven into a trivial $T$-invariant gapped phase by explicitly breaking the $U(1)$ symmetry (e.g. by turning on a pairing term $\psi^T C \psi$ in Eq.~(\ref{eq:Diracs})).  Therefore, we can reduce the topological order of the $\nu = 2$ TSc in class DIII from $T$-Pfaffian$_+ \times \{1, s\}$ to $\{1, s\} \times \{1, \psi\}$, with $\psi$ - the electron, without breaking $T$-symmetry. As $T$ is the only symmetry protecting a TSc in class DIII, we conclude that the $\nu = 2$ TSc admits a very simple symmetry-preserving topological order $\{1, s\} \times \{1, \psi\}$, where the $T$-symmetry acts via $T:\, s \leftrightarrow s \psi$. 

\section{Classification of topological insulators and superconductors via bordism}
\label{sec:SW}
In this section, we shift our focus from duality between twisted $u(1)$ gauge theories to a general classification of interacting (non-gauged) topological superconductors in class AIII and topological insulators in class AII.

It was recently proposed that bordism theory may provide a way to classify symmetry protected topological phases in any dimension.\cite{KapustinBos,KapustinBosTI,KapustinFerm}  For instance, for an internal discrete symmetry group $G$, SPT phases of bosons in space-time dimension $D$  were proposed to be classified by ${\rm Hom}({\rm Tor}(\Omega_{SO,D}(BG)), U(1))$. Here, $\Omega_{SO, D}(X)$ is the  $D$-dimensional oriented bordism group of a space $X$. $BG$ is the classifying space of the discrete group $G$. ${\rm Tor}$ denotes the torsion subgroup, and ${\rm Hom}$ is the group of homomorphisms. The map from $\Omega_{SO,D}(BG)$ to $U(1)$ is interpreted as the partition function of the SPT phase on a manifold in the background of a flat $G$-connection, and this partition function is assumed to be a bordism invariant. The reason for considering only the torsion subgroup of $\Omega_{SO,D}(BG)$ is that homomorphisms from the free part of  $\Omega_{SO,D}(BG)$ to $U(1)$ depend on a continuous parameter and so do not describe disconnected phases of matter. The classification of Ref.~\onlinecite{KapustinBos} also extends to the case where $G$ contains time-reversal symmetry. In particular, SPT phases of bosons in $D$ dimensions protected by time-reversal symmetry were proposed to be classified by  ${\rm Hom}({\rm Tor}(\Omega_{O, D}(pt)), U(1))$, where $\Omega_{O, D}(pt)$ is the unoriented bordism group of a point in $D$-dimensions. This group is known in any dimension: it is pure torsion, and the Pontryagin dual ${\rm Hom}({\rm Tor}(\Omega_{O, D}(pt)), U(1))$ is generated by the Stiefel-Whitney characteristic numbers. The bordism classification of bosonic SPT phases with $T$-symmetry agrees with results obtained via less formal methods in dimensions $D \leq 4$.\cite{Chen2013,AVTS, Burnell2013} In particular, in $3+1$ dimensions,  $\Omega_{O, 4}(pt) = \bZ^2_2$, corresponding to the already mentioned two root phases eTmT and FFF, with partition functions given by Eqs.~(\ref{eq:eTmTa}) and (\ref{eq:FFFa}).

Ref.~\onlinecite{KapustinFerm} also proposed a bordism classification of certain SPT phases of fermions. In particular, it was conjectured that time-reversal protected phases of fermions with $T^2 = -1$ (so-called topological superconductors in class DIII) are classified by ${\rm Hom}\left[{\rm Tor}\left(\Omega_{{\rm Pin}_+, D}(pt)\right), U(1)\right]$, where $\Omega_{{\rm Pin}_+, D}(pt)$ is the $D$-dimensional Pin$_+$ bordism group. In low dimensions $D \leq 4$, this classification agrees with previously known/conjectured results. In particular, in $D = 3+1$, $\Omega_{{\rm Pin}_+, 4} = \bZ_{16}$ in agreement with the breakdown of the non-interacting integer classification $\bZ \to \bZ_{16}$.\cite{Kitaev_pc,FidkowskiChenAV,Wang2014,MetlitskiChenFidkowskiAV2014}

One may ask how to extend the bordism classification to topological superconductors of fermions with $u(1) \times T$ symmetry (class AIII) and topological insulators of fermions with $u(1) \rtimes T$ symmetry (class AII) considered in this paper. We discussed that non-interacting topological superconductuctors in class AIII in $3+1$D can be conveniently represented by a massive charged Dirac fermion. In order to place such a Dirac fermion on a  manifold in the presence of a background $u(1)$ gauge field, one needs to endow the manifold with a Pin$_c$ structure. We then saw that the partition function of a non-interacting phase $\nu \in \bZ$ is given by,
\beq Z^{\nu}_{{\rm TSc}} = e^{2 \pi i \nu \eta} \eeq
where $\eta$ is the spectral asymmetry of a Dirac operator (\ref{eq:etadef}). Crucially, $\eta$ is a Pin$_c$ bordism invariant. Since in $3+1$D $\eta$ is a multiple of $1/8$, we immediately see that the non-interacting phases $\nu$ and $\nu + 8$ collapse to a single phase in the presence of interactions.

The above discussion leads us to conjecture that (interacting) topological superconductors of fermions with $u(1) \times T$ symmetry in any space-time dimension $D$ are classified by  
${\rm Hom}\left[{\rm Tor}\left(\Omega_{{\rm Pin}_c, D}(pt)\right), U(1)\right]$
 with $\Omega_{{\rm Pin}_c, D}(pt)$ - the Pin$_c$ bordism group. This group has been computed in any dimension $D$.\cite{BahriGilkey} It is trivial in any odd space-time dimension. In dimension $D  = 1+1$,  $\Omega_{{\rm Pin}_c, D}(pt) = \bZ_4$ and in dimension $D = 3+1$, $\Omega_{{\rm Pin}_c, D}(pt) = \bZ_8 \times \bZ_2$. These results are in agreement with previously known/conjectured classification. In particular, in $D = 3+1$, physical arguments for a $\bZ_8 \times \bZ_2$ classification were presented in Ref.~\onlinecite{Wang2014}: the $\bZ_8$ subgroup corresponds to the non-interacting topological superconductor phases (reduced modulo $8$), and the $\bZ_2$ subgroup corresponds to the FFF SPT phase of neutral bosons. This is in detailed agreement with the bordism classification. Indeed, the $\bZ_8$ factor of the bordism group $\Omega_{{\rm Pin}_c, D}(pt)$ is generated by the manifold $\mRP$ with a particular choice of a Pin$_c$ structure (say one with $\eta = + 1/8$). The $\bZ_2$ factor corresponds to the manifold $\mathbb{CP}^2$ with a $\pi$ flux through its single 2-cycle - the corresponding value of $\eta = 0$ (Eq.~(\ref{eq:CP2}) with $n = 0$).  The non-interacting topological superconductor phases $\nu$ have partition functions $e^{2 \pi i \nu \eta}$, and so completely detect the $\bZ_8$ bordism subgroup generated by $\mRP$ with $\eta = 1/8$. At the same time, these non-interacting phases are blind to the $\bZ_2$ bordism subgroup generated by $\mathbb{CP}^2$ (with $\eta = 0$). However, the partition function of the FFF phase (\ref{eq:FFFa}) is equal to $-1$ on $\mathbb{CP}^2$ and to $-1$ on $\mRP$. Thus, the $\nu  =1$ TSc and the FFF phase serve as generators of ${\rm Hom}\left[{\rm Tor}\left(\Omega_{{\rm Pin}_c, D}(pt)\right), U(1)\right] = \bZ_8 \times \bZ_2$. We, therefore, conclude that the bordism classification of topological superconductors in class AIII in 3+1D is in complete agreement with the classification proposed in Ref.~\onlinecite{Wang2014}. 

We can also now prove Eq.~(\ref{eq:ZeTmT}), which identifies the $\nu = 4$ non-interacting TSc in class AIII with the eTmT phase. Indeed, since $e^{2 \pi i \eta}$ is a Pin$_c$ bordism invariant and so is $\exp\left(\pi i \int_M w^4_1\right)$, it suffices to prove Eq.~(\ref{eq:ZeTmT}) on generators of the bordism group: $\mRP$ and $\mathbb{CP}^2$. As already noted, on $\mRP$, $e^{8\pi i \eta} =  \exp\left(\pi i \int_M w^4_1\right) = -1$, and on $\mathbb{CP}^2$, $e^{8\pi i \eta} =  \exp\left(\pi i \int_M w^4_1\right) = +1$, completing the proof.

We can also ask about the bordism classification of (interacting) topological insulators with $u(1) \rtimes T$ symmetry and $T^2 = (-1)^F$ (class AII). We saw that the non-interacting topological insulator in 3+1D can be represented by a massive Dirac fermion. Placing this Dirac fermion on a manifold required a Pin$_{\tilde{c}}$ structure introduced in section \ref{sec:curved}. We saw that the partition function of the Dirac fermion was given by $(-1)^{N^{\chi}_0/2}$, with $N^{\chi}_0$ - the number of zero-modes of a certain doubled Dirac operator. Again, $(-1)^{N^{\chi}_0/2}$ is a bordism invariant of the Pin$_{\tilde{c}}$ structure. Thus, one may wonder if the classification of topological insulators in any dimension is given by ${\rm Hom}\left[{\rm Tor}\left(\Omega_{{\rm Pin}_{\tilde{c}}, D}(pt)\right), U(1)\right]$, with $\Omega_{{\rm Pin}_{\tilde{c}}, D}(pt)$ - the Pin$_{\tilde{c}}$ bordism group. Such bordism group does not seem to appear in the mathematics literature. It would be interesting to compute this bordism group in low dimensions and check whether it agrees with previously known/conjectured classification of topological insulators. For instance, in $3+1$D it has been argued that interacting topological insulators have a $\bZ^3_2$ classification, with one of the $\bZ_2$ root phases being the non-interacting TI, and the other two $\bZ_2$ root phases being $T$-symmetric SPT phases of neutral bosons: eTmT and FFF.\cite{ChongScience}

\section{Acknowledgements}
I am very grateful to Edward Witten for comments on the manuscript and, in particular, for pointing out the relation to Ray-Singer torsion and Reidemeister torsion in section \ref{sec:bosonsS}. I am also very grateful to Ashvin Vishwanath for a previous collaboration on a related problem, as well as for his encouragement and comments on the manuscript. I also thank T. Senthil and D. Son for useful discussions. This work was performed in part at the Aspen Center for Physics, which is supported by National Science Foundation grant PHY-1066293, and at the Kavli Institute for Theoretical Physics, supported in part by the National Science Foundation under Grant No. NSF PHY11-25915. This work was supported in part by the U.S. Army Research Office, grant No. W911NF-14-1-0379. Research
at Perimeter Institute is supported by the Government of Canada through Industry Canada and by the Province of Ontario through the Ministry of Research and Innovation.

\appendix
\section{Forms on manifolds}
\label{app:forms}
We review some notation here. $d$  is the exterior derivative and $\delta = - * d *$. The Hodge dual operator $*$ acting on a $k$-form $\eta$ gives,
\beq (* \eta)_{\mu_1 \mu_2 \ldots \mu_{4-k}} = \frac{1}{k!} \sqrt{g} \epsilon_{\nu_1 \nu_2 \ldots \nu_k \mu_1 \mu_2 \ldots \mu_{4-k}} \eta^{\nu_1 \nu_2 \ldots \nu_k} \eeq
and $* * \eta = (-1)^k \eta$. We have $d^2 = 0$, $\delta^2 = 0$. The Laplace operator is $\Delta = d \delta + \delta d$. 

An inner product on $k$-forms $\eta$, $\xi$ is given by,
\beq \langle \eta, \xi \rangle = \frac{1}{k!} \int_M d^4 x \, \sqrt{g} \,\eta_{\mu_1, \mu_2, \ldots, \mu_k} \xi^{\mu_1, \mu_2, \ldots, \mu_k} \label{eq:innerapp}\eeq

\section{Fermion determinant}
\label{sec:Eta}
In section \ref{sec:Zftimes} we gave a somewhat cavalier derivation of Eq.~(\ref{eq:Zftimesfinal}). Here, we will derive this equation more carefully by using $\zeta$-function regularization. We follow the approach of Ref.~\onlinecite{Cognola}.

We begin with the partition function (\ref{eq:Zdettimes}) of a massive Dirac fermion,
\beq F(m) = - \log Z(m) = - \sum_{\lambda} \log (i \lambda +m) \equiv \lim_{s \to 0} \frac{d}{ds} \sum_{\lambda} (i \lambda +m)^{-s} \label{eq:Zmreg} \eeq
The last equation will be taken as the definition of the partition function for both signs of $m$. Here, $\lambda$ are eigenvalues of the Dirac operator $i {\slashed D}$ on a Pin$_c$ manifold. The $s \to 0$ limit is taken by analytically continuing from large $Re(s)$ (this limit will be assumed in all the expressions below). Here and below the branch-cut in $z^{-s}$ will always be taken to lie along the negative real axis. This choice of the branch-cut ensures that the regularized $Z(m)$ in Eq.~(\ref{eq:Zmreg}) is an analytic function of $m$ and $\lambda$. Then,
\bea F(m)  &=& -N_0 \log m + \frac{d}{ds} \sum_{\lambda > 0} e^{- \pi i s/2} (\lambda - im)^{-s} +   \frac{d}{ds} \sum_{\lambda < 0} e^{\pi i s/2} (|\lambda| + im)^{-s} \nn\\
&=& - N_0 \log m + \frac{d}{ds}\left[ \frac{e^{-\pi i s/2}}{\Gamma(s)} \int_0^{\infty} dt\, t^{s-1} \sum_{\lambda > 0} e^{-t (\lambda - im)} +\frac{e^{\pi i s/2}}{\Gamma(s)} \int_0^{\infty} dt\, t^{s-1} \sum_{\lambda < 0} e^{-t (|\lambda| + im)}\right]\nn\\   \eea
where $N_0$ is the number of zero-modes $\slashed{D}$. Thus,
\bea F(-|m|) - F(|m|) &=& -i \pi N_0 -2 \frac{d}{ds}\left[\frac{\sin (\pi s/2)}{\Gamma(s)}  \int_0^{\infty} dt\, t^{s-1} Z_+(t) \sin(|m| t) \right]  \nn\\
&-& 2i\frac{d}{ds}\left[\frac{\cos (\pi s/2)}{\Gamma(s)}  \int_0^{\infty} dt\, t^{s-1} Z_-(t) \sin(|m| t)\right] \label{eq:Fmm} \eea
where
\bea Z_+(t) &=& \sum_{\lambda \neq 0} e^{-t |\lambda|} \\
Z_-(t) &=& \sum_{\lambda \neq 0} \mathrm{sgn}(\lambda) e^{-t |\lambda|} \eea
We will show that
\beq I_+(s) = \int_0^{\infty} dt \, t^{s-1} Z_+(t) \sin(|m| t) \label{eq:Ip}\eeq
extends to a function which has no poles at $s = 0$, therefore, $Re\left[F(-|m|) - F(|m|)\right] = 0$. Furthermore, $Z_-(t)$ is smooth for $t \to 0^+$ and 
\beq \lim_{t \to 0^+} Z_-(t) = \eta(0) \label{eq:Zt0} \eeq 
where $\eta(s)$ is given by Eq.~(\ref{eq:etasdef}). Therefore, the last term in Eq.~(\ref{eq:Fmm}) can be directly analytically continued to $s = 0$,
\beq F(-|m|) - F(|m|) = - i \pi N_0  - 2i  \int_0^{\infty} dt \frac{\sin |m| t}{t} Z_-(t) \eeq
Therefore, from Eq.~(\ref{eq:Zt0}) we obtain the desired result
\beq \lim_{m \to \infty} F(-|m|) - F(|m|) = - \pi i (N_0 + \eta(0)) = -2 \pi i \eta\eeq

To derive Eq.~(\ref{eq:Zt0}), observe that
\beq \eta(s) = \sum_{\lambda \neq 0} \mathrm{sgn}(\lambda) |\lambda|^{-s} = \frac{1}{\Gamma(s)} \int_0^{\infty} dt\, t^{s-1} Z_-(t) \label{eq:etaZm} \eeq
Using an inverse Mellin transform, we may invert the above equation,
\beq Z_-(t) = \frac{1}{2 \pi i} \int_{a-i \infty}^{a+ i \infty} ds\, t^{-s} \Gamma(s) \eta(s) \label{eq:Mellin}\eeq
for a sufficiently large real $a$. To generate an asymptotic expansion of $Z_-(t)$ for $t \to 0$ we need to know analytic properties of $\eta(s)$ in the complex plane. To extract these, we may alternatively write,
\beq \eta(s) = \frac{1}{\Gamma(\frac{s+1}{2})} \int_0^{\infty}dt\, t^{(s-1)/2} \mathrm{Tr}\left[ i {\slashed D} e^{-t (i \slashed D)^2}\right] \label{eq:etaintsq} \eeq
An asymptotic power series expansion for $t \to 0$ of $C(t) = \mathrm{Tr}\left[ i {\slashed D} e^{-t (i \slashed D)^2}\right]$ can be obtained using the heat-kernel method. Moreover, since $\gamma^5$ locally anticommutes with ${\slashed D}$, all coefficients in this expansion are vanishing. Therefore, $C(t)$ vanishes exponentially as $t \to 0$. This implies that Eq.~(\ref{eq:etaintsq}) allows us to directly continue $\eta(s)$ to a function holomorphic in the entire complex plane. Therefore, the integrand in Eq.~(\ref{eq:Mellin}) has simple poles at the poles of $\Gamma(s)$, $s = -n$, $n \ge 0$, and we can obtain an asymptotic expansion of $Z_-(t)$ for $t \to 0$  by pushing the contour in Eq.~(\ref{eq:Mellin}) to the left, picking up residues at these poles:
\beq Z_-(t) \approx \sum_{n = 0}^{\infty} \frac{(-1)^n}{n!} \eta(-n) t^n\eeq
In particular, the $t \to 0$ limit is given by Eq.~(\ref{eq:Zt0}).

Similar manipulations can be used to show that $I_+(s)$ in Eq.~(\ref{eq:Ip}) extends to a function that has no pole at $s =0$. To see this, we will need an asymptotic expansion of $Z_+(t)$ for $t  \to 0$. We define,
\beq \zeta_D(s) = \sum_{\lambda \neq 0} |\lambda|^{-s}  = \frac{1}{\Gamma(s)} \int_0^{\infty} dt \, t^{s-1} Z_+(t) \eeq
and so
\beq Z_+(t) = \frac{1}{2 \pi i} \int_{a-i \infty}^{a+ i \infty} ds\, t^{-s} \Gamma(s) \zeta_D(s) \label{eq:zetaMellin}\eeq
We again need analytic properties of $\zeta_D(s)$ in the complex plane. To extract these, we write
\beq \zeta_D(s) = \frac{1}{\Gamma(s/2)} \int_0^{\infty} dt \, t^{s/2-1} \left(B(t) - N_0\right) \label{eq:zetaB}\eeq
where
\beq B(t) = \mathrm{Tr}\left[e^{-t (i \slashed D)^2}\right] \stackrel{t \to 0}{\approx} \sum_{n = 0}^{\infty} A_n t^{(n -4)/2}\eeq
Here, we've noted the heat kernel expansion of $B(t)$ with $A_n$ - the heat kernel coefficients. Crucially, all coefficients $A_n$ with $n$ - odd vanish.\cite{Vassilevich} Thus, from Eq.~(\ref{eq:zetaB}), we may analytically continue $\zeta_D(s)$ to arbitrary $s$ using,
\bea \zeta_D(s) &=& \frac{1}{\Gamma(s/2)} \int_0^{1} dt \, \sum_{j = 0}^{\infty} \tilde{A}_{2 j} t^{\frac{s}{2} + j -3} + \frac{1}{\Gamma(s/2)} \int_0^{1} dt \, t^{s/2-1} \left(B(t) -  \sum_{j = 0}^{\infty} A_{2 j} t^{j-2}\right) \nn\\ &+& \frac{1}{\Gamma(s/2)} \int_1^{\infty} dt \, t^{s/2-1} \left(B(t) - N_0\right) \label{eq:Dsexpan}\eea
where $\tilde{A}_n = A_n - N_0 \delta_{n,4}$. The last two terms in Eq.~(\ref{eq:Dsexpan}) extend to holomorphic functions of $s$ in the entire complex plane. So,
\beq \zeta_D(s) = \frac{2}{\Gamma(s/2)} \sum_{j = 0}^{\infty} \frac{\tilde{A}_{2j}}{s + 2j - 4} + \mathrm{analytic} = \frac{2 A_0}{s-4} + \frac{2 A_2}{s-2} + \mathrm{analytic}\eeq
We can now generate an asymptotic expansion of $Z_+(t)$ for $t \to 0$ from Eq.~(\ref{eq:zetaMellin}):
\beq Z_+(t) \stackrel{t \to 0}{\approx} 12 A_0 t^{-4} + 2 A_2 t^{-2} + \sum_{n = 0}^{\infty} \frac{(-1)^n}{n!} \zeta_D(-n) t^n \eeq
We can now analytically continue $I_+(s)$ in Eq.~(\ref{eq:Ip}) to the neighborhood of $s = 0$:
\bea I_+(s) &=& \int_0^{1}  dt\, t^{s-1} \left(12 A_0 t^{-4} + 2 A_2 t^{-2} + \sum_{n = 0}^{\infty} \frac{(-1)^n}{n!} \zeta_D(-n) t^n\right) \sin(|m|t)\nn\\
&+& \int_0^1 dt \, t^{s-1} \left(Z_+(t) - 12 A_0 t^{-4} - 2 A_2 t^{-2} - \sum_{n = 0}^{\infty} \frac{(-1)^n}{n!} \zeta_D(-n) t^n \right)  \sin(|m|t) \nn\\
&+& \int_1^{\infty} dt \, t^{s-1} Z_+(t) \sin(|m|t)\eea
The last two terms extend to functions of $s$ analytic in the entire complex plane, while the first term contributes simple poles at integer $s \leq 3$. The two poles in the region $Re(s) > -1$ are,
\beq I_+(s) \approx \frac{12 A_0 |m|}{s-3} - \frac{2 A_0 |m|^3}{s-1} + \frac{2 A_2 |m|}{s-1} + \mathrm{analytic}, \quad Re(s) > -1\eeq
In particular, $I_+(s)$ has no pole at $s  =0$, as claimed.

\section{Partition function on $\mRP$, bosonic matter}
\label{app:RP4}
In this appendix, we explicitly evaluate the partition functions (\ref{eq:ZbRP4}), (\ref{eq:ZbrRP4}) of $u(1)$ gauge theory with bosonic matter on $\mRP$. We will use $\zeta$-function regularization. We will need the eigenvalues of the Laplacians $\Delta^0$ and $\Delta^1$ on $S^4$, as well as the parity of the corresponding eigenfunctions under the antipodal map $\imath$: 
\beq \Delta^0:\quad \lambda = \frac{n (n + 3)}{R^2}, \quad d(n) = \frac{1}{6} (n+1)(n+2)(2n + 3), \quad P = (-1)^n, \quad n \ge 0 \label{eq:DeltaS}\eeq
\beq \Delta^1: \quad \lambda = \frac{n (n+1)}{R^2}, \quad d(n) = \frac12 (n-1)(n+2)(2n+1), \quad P = (-1)^n, \quad n \ge 2\eeq
Here $\lambda$ denotes the eigenvalue of $\Delta$, $d(n)$ - the corresponding degeneracy and $P$ - the parity. Note that we only list transverse eigenmodes of $\Delta^1$  ($\delta a = 0$). $R$ is the radius of $S^4$. Here and below all distances are measured in some fixed units, i.e. the dimensionless radius $R$ in our formulas really stands for $\mu R_{p}$ with $R_p$ - the dimensionful radius and $\mu$ - the renormalization scale.

We have,
\bea F^b_{CT}(e) &=& -\log Z^b_{CT}(e) = F^0_+ + F^1_+ + \log \pi + \frac12 \log V (\mRP) \label{eq:Fbtimes}\\
F^b_T(e) &=& - \log Z^b_T(e) = F^0_- + F^1_- + \log 2 \label{eq:Fbrtimes}\eea 
where  the volume of $\mRP$, $V(\mRP) = \frac12 V(S^4)= \frac{4\pi^2}{3}R^4$ and
\bea F^0_+ &=& -\frac{1}{2} \mathrm{Tr}'_+ \log \Delta^0 = -\frac{1}{2} \sum_{\substack{n = 2\\ n - \mathrm{even}}}^{\infty} d^0(n) \log \lambda^0(n)\\
F^0_- &=& -\frac{1}{2} \mathrm{Tr}_- \log \Delta^0 = -\frac{1}{2} \sum_{\substack{n = 1\\ n - \mathrm{odd}}}^{\infty} d^0(n) \log \lambda^0(n) \\
F^1_+ &=& \frac{1}{2} \mathrm{Tr}_{\perp,+} \log \frac{\Delta^1}{2 \pi e^2} = \frac{1}{2} \sum_{\substack{n = 2\\ n - \mathrm{even}}}^{\infty} d^1(n) \log \frac{\lambda^1(n)}{2 \pi e^2}\\
F^1_- &=& \frac{1}{2} \mathrm{Tr}_{\perp,-} \log \frac{\Delta^1}{2 \pi e^2} = \frac{1}{2} \sum_{\substack{n = 3\\ n - \mathrm{odd}}}^{\infty} d^1(n) \log \frac{\lambda^1(n)}{2 \pi e^2}\eea

Let's begin by calculating $F^0_+$,
\beq F^0_+ = \frac{1}{2} \lim_{s \to 0} \frac{d}{ds}  \sum_{\substack{n = 2\\ n - \mathrm{even}}}^{\infty} d^0(n) (\lambda^0(n))^{-s} \eeq
where $s \to 0$ limit is understood to be taken by analytically continuing from large $Re(s)>0$. We will implicitly assume this limit in all subsequent expressions. From Eq.~(\ref{eq:DeltaS}),
\bea F^0_+ &=& \frac{1}{12}  \frac{d}{ds}\left[ R^{2s} \sum_{\substack{n = 2\\ n - \mathrm{even}}}^{\infty} \frac{(n+1)(n+2)(2n+3)}{n^{s} (n+3)^{s}}\right] \\
&=&  \frac{1}{12}  \frac{d}{ds}\left[ R^{2s} \frac{\Gamma(2s)}{\Gamma(s)^2} \sum_{\substack{n = 2\\ n - \mathrm{even}}}^{\infty} \int_0^{1} dx\, x^{s-1} (1-x)^{s-1} \frac{(n+1)(n+2)(2n+3)}{(n+3x)^{2s}}\right]  \eea
Writing 
\bea (n+1) (n+2) (2n +3) &=& 2(n+3x)^2 - 9(2x-1)(n+3x)^2 + (54x^2 - 54x+13)(n+3x) \nn\\ && - 3 (2x -1)(3x-1) (3x-2)\eea
we obtain,
\beq F^0_+ = \frac{1}{12}  \frac{d}{ds}\left[ R^{2s} \frac{\Gamma(2s)}{\Gamma(s)^2} \int_0^{1} dx\, x^{s-1} (1-x)^{s-1} f(s,x) \right] \label{eq:intF}\eeq
where
\bea f(s,x) &=& 2^{-2s}\Bigg( 16 \zeta(-3 + 2s, \frac{3x}{2}) - 36 (2x-1) \zeta(-2+2s, \frac{3x}{2}) + 2 (54x^2-54x + 13) \zeta(-1+2s, \frac{3x}{2}) \nn\\
&&- 3(2x-1)(3x-1)(3x-2) \zeta(2s, \frac{3x}{2}) - 6 \left(\frac{3x}{2}\right)^{-2s}\Bigg)\eea
with $\zeta(s,x)$ - the Hurwitz-zeta function, obtained by analytically continuing 
\beq \zeta(s,x) = \sum_{n = 0}^{\infty} \frac{1}{(n+x)^s}\eeq
For $s \to 0$, the $x$-integral in Eq.~(\ref{eq:intF}) becomes singular in the regions $x \to 0$, $x \to 1$, resulting in $1/s$ poles. To isolate these poles we define
\beq \tilde{f}(s,x) = f(s,x) - x f(s,1) - (1-x) f(s,0)\eeq
and write
\beq F^0_+ = \frac{1}{12}  \frac{d}{ds}\left[ R^{2s} \frac{\Gamma(2s)}{\Gamma(s)^2} \int_0^{1} dx\, x^{s-1} (1-x)^{s-1}\tilde{f}(s,x) + \frac{1}{2} R^{2s} (f(s,0) + f(s,1)) \right] \label{eq:intF2}
\eeq
The integral involving $\tilde{f}(s,x)$ is no longer singlular for $x \to 0$, $x\to 1$ in the regime $s \to 0$. Since $\frac{\Gamma(2s)}{\Gamma(s)^2} \approx \frac s 2$ for $s \to 0$, we can now set $s  =0$ in the integral in Eq.~(\ref{eq:intF2}):
\beq  F^0_+ = \frac{1}{24} \int_0^{1} dx\, x^{-1} (1-x)^{-1} \tilde{f}(0,x)  + \frac{1}{24} \frac{d}{ds} \left(R^{2s} (f(s,0) + f(s,1)) \right)\eeq
We can simplify the above experession as follows. For $x \to 0$,
\beq \zeta(s, x)  = x^{-s} + \zeta(s) + O(x) \eeq
where $\zeta(s) = \sum_{n = 1}^{\infty} \frac{1}{n^s}$ is the regular $\zeta$-function. Therefore,
\beq f(s,0) = 2^{-2s}(16 \zeta(-3+2s) + 36\zeta(-2+2s) + 26 \zeta(-1+2s) + 6 \zeta(2s)) \eeq
Also, $\zeta(s, 3/2) = (2^s-1)\zeta(s) - 2^s$, so
\bea f(s,1) &=& 16(\frac{1}{8} -2^{-2s})\zeta(-3+2s) - 36(\frac{1}{4} - 2^{-2s}) \zeta(-2+2s) + 26(\frac{1}{2} - 2^{-2s}) \zeta(-1+2s) \nn\\
&& - 6 (1-2^{-2s}) \zeta(2s) - 6 \cdot 3^{-2s}\eea
Also, for $n \ge 0$, $n$ - integer, $\zeta(-n, x) = -\frac{1}{n+1} B_{n+1}(x)$ where $B_n(x)$ is the Bernouli polynomial,
\beq B_1(x) = x-\frac12, \quad B_2(x) = x^2 - x + \frac16, \quad B_3(x) = x^3 - \frac3 2 x^2 + \frac 1 2 x, \quad B_4(x) = x^4 - 2 x^3 + x^2 - \frac{1}{30}\eeq
from which we obtain,
\beq \tilde{f}(0,x) = -\frac{9}{4} x (1-x) (9x^2-9x+4) \eeq
Therefore,
\bea F^0_+ &=& -\frac{151}{180} \log R  + \frac{1}{12} (2 \zeta'(-3) + 63\zeta'(-2) + 13 \zeta'(-1) + 6 \zeta'(0))  -\frac{15}{64} + \frac{1}{2} (\log 2 + \log 3)\nn\\\eea
Analogous calculations give,
\bea F^0_- &=&  \frac{29}{180} \log R  + \frac{1}{12} (2 \zeta'(-3) - 63\zeta'(-2) + 13 \zeta'(-1) -6 \zeta'(0))  -\frac{15}{64} - \frac{1}{2} \log 2\nn\\
F^1_+ &=& -\frac{49}{120} \log(2 \pi e^2 R^2) - \frac{1}{4} (2 \zeta'(-3) +  21\zeta'(-2) -3 \zeta'(-1) -2 \zeta'(0))  -\frac{25}{192} + \frac{1}{2} \log 2\nn\\
F^1_- &=& \frac{11}{120} \log(2 \pi e^2 R^2) - \frac{1}{4} (2 \zeta'(-3) -  21\zeta'(-2) -3 \zeta'(-1) + 2 \zeta'(0))  -\frac{25}{192} - \frac{1}{2} \log 2\nn\\
\eea
Collecting terms in Eqs.~(\ref{eq:Fbtimes}), (\ref{eq:Fbrtimes}) and using $\zeta'(0) = -\frac{1}{2} \log (2 \pi)$,
\bea F^b_{CT} &=& \frac{31}{90} \log R - \frac{49}{120} \log (2 \pi e^2) + \frac{3}{2} \log (2\pi) -\frac{35}{96} +\frac{1}{6} (-2 \zeta'(-3) + 11 \zeta'(-1)) \nn\\
F^b_T &=& \frac{31}{90} \log R + \frac{11}{120} \log (2 \pi e^2) + \frac{1}{2} \log (2\pi) -\frac{35}{96} +\frac{1}{6} (-2 \zeta'(-3) + 11 \zeta'(-1))\nn\\
\label{eq:Ffinal}\eea
Note that the coefficient of the $\log R$ term above is consistent with the conformal anomaly,
\beq F^b \sim -(N^1 - 2 N^0) \log R =\frac{1}{90 (64 \pi^2)}  \int d^4 x \sqrt{g} \left(a_g E_4 - c_g C^2 \right) \log R\eeq 
where from table \ref{tbl:charges}, $a_g = 62$ and $c_g = 36$. Since the Weyl tensor $C$ vanishes on $S^4$ (and so on $\mRP$) and the Euler-number of $\mRP$ is $\chi =1$, we have
\beq F^b \sim \frac{a_g}{90}  \frac{\chi}{2} \log R = \frac{31}{90} \log R\eeq
in agreement with the explicit calculation, Eq.~(\ref{eq:Ffinal}).

Next, we consider the ``renormalized" partition function (\ref{eq:barZbred})
\beq \bar{F}^b(e) = - \log \bar{Z}^b(e) = F(e) + \frac{\chi}{4} \log\frac{e^2}{2 \pi} + (N^1 - N^0) \log(\sqrt{2 \pi} e) \eeq
with $N^1 - N^0$ given by Eq.~(\ref{eq:NvmNs}),
\beq N^1 - N^0 = \frac{1}{90} \left(-63 \frac{\chi}{2} + \frac{5}{64\pi^2} \int d^4 x \sqrt{g} {\cal R}^2\right) = - \frac{11}{60}\eeq
Here we have used the fact ${\cal R} = \frac{12}{R^2}$ on $S^4$ (and so on $\mRP$). Thus,
\bea \bar{F}^b_{CT}(e) = -\frac{1}{4} \log \frac{e^2}{2\pi} + c_1\nn\\
\bar{F}^b_T(e) = +\frac{1}{4} \log \frac{e^2}{2 \pi} + c_1\eea
with $c_1 = \frac{31}{90} \log R+ \frac{1}{2} \log (2\pi) -\frac{35}{96} +\frac{1}{6} (-2 \zeta'(-3) + 11 \zeta'(-1))$. So Eq.~(\ref{eq:ZbRP4final}) holds.

\section{Dirac operator on $\mRP$}
\label{app:DRP4}
In this section we compute the spectrum of the Dirac operator on $\mRP$. 

We begin with the case of ${\cal L}_{CT}$, where our objective is to calculate the $\eta$-invariant on $\mRP$. 

It will be convenient to work on the double cover of $\mRP$ - $S^4$. We will use stereographic coordinates on $S^4$:  $\vec{X} = (\frac{4 \vec{u}}{\vec{u}^2+4}, \frac{\vec{u}^2 - 4}{\vec{u}^2+4})$, with $\vec{u} \in R^4$. The metric is,
\beq g_{\mu \nu} = \frac{16}{(u^2+4)^2} \delta_{\mu \nu}\eeq
We choose the following vielbein,
\beq e^{a}_{\mu} = \frac{4}{u^2+4} \delta^{a}_{\mu} \eeq
The associated spin-connection is,
\beq \omega^{ab}_{\mu} = -\frac{2}{u^2+4}(\delta^{a}_{\mu} u^b - \delta^{b}_{\mu} u^a)\eeq 
The antipodal map on $S^4$ is 
\beq \imath:\,\, \vec{u} \to -\frac{4 \vec{u}}{u^2}\eeq

There are two Pin$_c$ structures on $\mRP$. These are actually simultaneously Pin$_+$ structures, i.e. they do not involve the $u(1)$ part of the $(u(1) \times {\rm Pin}_+)/\bZ_2$ group and can be chosen to have $a_{\mu} = 0$.  Upon lifting to $S^4$, the two Pin$_c$ structures correspond to a condition on the Dirac spinor $\psi$, 
\beq P \psi = \pm \psi \label{eq:Ppsi}\eeq
where $P$ is the parity operator,
\beq (P \psi)(u) = \frac{i \gamma^a u^a}{u}\psi\left(-\frac{4 \vec{u}}{u^2}\right) \label{eq:Ppsiexpl} \eeq
To compute the $\eta$ invariant of $\mRP$ for each Pin$_c$ structure, we will first find the spectrum of $i \slashed{D} = i e^{\mu}_a \gamma^a (\d_{\mu} + i \omega_{\mu})$ on $S^4$ and then keep the part of the spectrum satisfying Eq.~(\ref{eq:Ppsi}).  Note that $\{\gamma^5, P \} = 0$, so $\gamma^5$ maps one Pin$_c$ structure into the other. Furthermore, since $\{\gamma^5, \slashed{D}\} = 0$, the two Pin$_c$ structures have opposite $\eta$ invariants. 

The spectrum of $i \slashed{D}$ on $S^4$ can be computed as follows (see Ref.~\onlinecite{Bar} for more details). We first find a set of Killing-spinors on $S^4$, i.e. spinor fields $\psi_0$ satisfying,
\beq \tilde{\nabla}_{\mu} \psi_0 = 0 \label{eq:Killing}\eeq
where
\beq \tilde{\nabla}_{\mu}  = \d_{\mu} + i \omega - i r e^a_{\mu} \gamma^a, \quad r  =1/2 \eeq
Here, we choose $r  = 1/2$ (there also exists a different set of Killing spinors with $r  =-1/2$). An explicit expression for $\psi_0$ is,
\beq \psi_0(u) = \frac{1}{\sqrt{4+u^2} } (2+ i u^a \gamma^a) \psi_0(u = 0)\eeq
We have a four-dimensional basis for $\psi_0(u = 0)$ which translates into a four-dimensional basis for $\psi_0(u)$. Note that 
\beq P \psi_0 = + \psi_0 \eeq
Now one can show that 
\beq \left(i {\slashed D} + \frac{1}{2}\right)^2 = -\frac{1}{\sqrt{g}} \tilde{\nabla}_{\mu} \left(g^{\mu \nu} \tilde{\nabla}_{\nu}\right) + \frac{9}{4}\eeq 
From Eq.~(\ref{eq:Killing}), for $\psi(u) = \phi(u) \psi_0(u)$, with $\phi$ - a function,
\beq -\frac{1}{\sqrt{g}} \tilde{\nabla}_{\mu} \left(g^{\mu \nu} \tilde{\nabla}_{\nu}\right) (\phi \psi_0) = (\Delta^0 \phi) \psi_0\eeq
The spectrum of the Laplacian $\Delta^0$ is given by Eq.~(\ref{eq:DeltaS}). Thus, the spectrum of $(i \slashed D + 1/2)^2$ is,
\beq \left(i {\slashed D} + \frac{1}{2}\right)^2  (\phi_n \psi_0) = \left(n + \frac{3}{2}\right)^2 (\phi_n \psi_0),\quad d(n) =  \frac{2}{3} (n+1)(n+2)(2n + 3), \quad P = (-1)^n, \quad n \ge 0 \eeq
where $\phi_n$ is an eigenfunction of $\Delta^0$ with eigenvalue $n(n+3)$. The eigenvalues of $i {\slashed D}$ are then of form $\lambda_{\pm}(n) = -\frac{1}{2} \pm \left(n + \frac{3}{2}\right)$, $n \ge 0$, with parities $P_{\pm}(n) = (-1)^n$, and the corresponding degeneracies $d_\pm(n)$ must satisfy 
\beq d_+(n) + d_-(n) = \frac{2}{3} (n+1)(n+2)(2n + 3) \label{eq:dsum}\eeq
Moreover, as $\{ \gamma^5, {\slashed D} \} = 0$, all non-zero eigenvalues of ${i \slashed D}$ must come in pairs $\pm \beta$, so 
\beq d_+(n+1) = d_-(n), \quad n \ge 0 \label{eq:dpdm}\eeq
A direct calculation shows that $d_+(0) = 0$ and $d_-(0) = 4$. Then solving the recursion (\ref{eq:dsum}), (\ref{eq:dpdm}) we arrive at the final expression for the spectrum of $i {\slashed D}$,
\bea  i {\slashed D} = \left\{ \begin{array}{cccc} (n+2), & d(n) = \frac{2}{3} (n+1)(n+2)(n+3), & P = (-1)^{n+1}, & n \ge 0\\
-(n+2), & d(n) = \frac{2}{3} (n+1)(n+2)(n+3), & P = (-1)^n, &  n \ge 0 \end{array}\right. \label{eq:Diracspectfinal}\eea
Note that while the spectrum itself is symmetric about $0$, the parities of the eigenstates are not symmetric.

The $\eta$ invariant can now be calculated directly. Focusing on the Pin$_c$ structure with $P  =+1$,
\beq \eta(s) = \sum_{\lambda > 0}  \lambda^{-s} - \sum_{\lambda < 0} \lambda^{-s} = \frac{2}{3} \sum_{\substack{n = 1 \\ n - \mathrm{odd}}}^{\infty} (n+1)(n+2)^{1-s} (n+3) -  \frac{2}{3} \sum_{\substack{n = 0 \\ n - \mathrm{even}}}^{\infty} (n+1)(n+2)^{1-s} (n+3)\eeq
From this,
\beq \eta(0) = -\frac{1}{4} \eeq
and 
\beq \eta = \frac{1}{2} (\eta(0) + N_0) = -\frac{1}{8} \eeq
in agreement with Ref.~\onlinecite{GilkeyPinC}. Likewise, the Pin$_c$ structure with $P = -1$ has $\eta = +\frac{1}{8}$.

We conclude by discussing the case of ${\cal L}_T$. Here we need to find the number of zero-modes $N^{\chi}_0$ of the  ``doubled" operator ${\slashed D}^{\chi}$ in Eq.~(\ref{eq:Dchi}). Now there is only a single Pin$_{\tilde{c}}$ structure, which can again be chosen to have $a_{\mu} = 0$. We can again work on a sphere, choosing the Dirac spinor $\chi = (\chi_1, \chi_2)$ to satisfy
\beq \rho^3 P \chi = + \chi \eeq
We see that the two components $\chi_1$, $\chi_2$ of $\chi$ decouple. $\chi_1$ is a Dirac spinor with $P = +1$ and $\chi_2$ is a Dirac spinor with $P = -1$. The corresponding spectra can be obtained from Eq.~(\ref{eq:Diracspectfinal}). The spectrum of $\chi_2$ is the inverted spectrum of $\chi_1$, in accordance with the discussion above Eq.~(\ref{eq:Zfrtimesfinal}). In particular, ${\slashed D}^{\chi}$ has no zero-modes: $N^{\chi}_0 = 0$.

\section{Reidemeister torsion}
\label{app:Reidemeister}
In this appendix, we review the definition of Reidemeister torsion ($R$-torsion) of a manifold and show that it reduces to Eq.~(\ref{eq:Tc}). Our discussion follows Ref.~\onlinecite{Muller}. We consider only a special simple case of the definition in Ref.~\onlinecite{Muller} relevant for our present purposes.

Let us start with a smooth triangulation of a $D$-dimensional manifold $M$. Consider the cochain complex,
\beq C^0(M, \bR) \stackrel{d}{\to} C^1(M, \bR) \stackrel{d}{\to} \ldots \stackrel{d}{\to}  C^{q-1}(M, \bR)  \stackrel{d^{q-1}}{\to} C^{q}(M, \bR)  \stackrel{d^q}{\to} C^{q+1}(M, \bR) \stackrel{d}{\to}  \ldots  \stackrel{d}{\to} C^{D}(M, \bR) \eeq
$C^q(M, \bR)$ has a ``preferred" basis $c^q_i$ generated by the $q$-simplices in the triangulation, i.e. $c^q_i$ is a basis for the lattice $C^q(M, \bZ)$.   Let $B^q \subset C^q(M, \bR)$ be the image of $d^{q-1}$, and let $K^q \subset C^q(M, \bR)$ be the kernel of $d^q$. The cohomology group $H^q(M, \bR) \equiv K^q/B^q$. For each $q$, let us pick a basis $h^q_i$ for $H^q(M, \bR)$. Let us also pick a basis $b^q_i$ for  $B^q$. Furthermore, let $\tilde{b}^q_i \in C^q(M, \bR)$ be the pre-image of $b^{q+1}_i \in B^{q+1}$. Then $b^q_i$, $h^q_i$ and $\tilde{b}^q_i$ together form a basis for $C^q(M, \bR)$. Consider a matrix   whose columns are $b^q_i$, $h^q_i$ and $\tilde{b}^q_i$ written in the $c^q_i$ basis. We will denote the absolute value of the determinant of this matrix by $| b^q, h^q, \tilde{b}^q/c^q|$. Now define the $R$-torsion $T^c$ via
\beq \log T^c(M) = \sum_{q = 0}^D (-1)^{q+1} \log |b^q, h^q, \tilde{b}^q/c^q| \eeq
The $R$-torsion is clearly independent of the choice of basis $b^q_i$ for $B^q$, however, it does depend on the choice of basis $h^q_i$ for $H^q(M, \bR)$. A theorem proved by Cheeger\cite{Cheeger} and independently M{\"u}ller\cite{Muller} states that for a particular choice of basis $h^q_i$, which we specify below, the $R$-torsion $T^c(M)$ is equal to the analytic torsion $T^{RS}(M)$ in Eq.~(\ref{eq:TRSgenD}). To obtain this ``preferred" basis, start with the de-Rham cohomology $H^q_{dR}(M)$ on forms and choose an orthonormal basis $\tilde{h}^{q,n}_i$ with respect to the inner product (\ref{eq:innerapp}), $\langle \tilde{h}^{q,n}_i, \tilde{h}^{q,n}_j \rangle = \delta_{ij}$. Now the preferred basis $h^{q,n}_i$ on the simplicial cohomology group $H^q(M, \bR)$ is obtained by integrating $\tilde{h}^{q,n}_i$ over the simplices.

We now demonstrate that the $R$-torsion with the preferred basis for cohomology groups as defined above coincides with Eq.~(\ref{eq:Tc}) for a 4-dimensional non-orientable manifold. As a first step, it will be convenient to change bases from the ``preferred" basis on cohomology $h^{q,n}_i$ to a basis for cohomology with integer coefficients ${\rm Free}(H^q(M, \bZ))$. Choose a basis $\tilde{h}^{q, \bZ}_i$ for the lattice of $q$-forms in $H^q_{dR}(M)$ with integer periods. This basis is related to the orthonormal basis $\tilde{h}^{q,n}_i$ on $H^q_{dR}(M)$ via,
\beq \tilde{h}^{q, \bZ}_i = A^q_{i j} \tilde{h}^{q,n}_j \eeq
with $A^q$ - an invertible matrix. Following the notation in section \ref{sec:bosonsS}, let
\beq {\cal N}^{+,q}_{ij} \equiv \langle \tilde{h}^{q,\bZ}_i, \tilde{h}^{q, \bZ}_j \rangle = (A^q (A^q)^T)_{ij} \label{eq:NA}\eeq
We can now obtain a basis $h^{q,\bZ}_i$ of the simplicial cohomology $H^q(M, \bR)$ by integrating $\tilde{h}^{q, \bZ}$ over the simplices. Moreover, we can write $h^{q, \bZ}_i = \hat{h}^{q}_i + \hat{b}_i$ where $\hat{h}^{q} \in C^q(M, \bZ)$ is an integer valued closed cocycle and $\hat{b}_i \in B^q_i$. $\hat{h}^q_i$ form a basis for the lattice ${\rm Free}(H^q(M, \bZ))$. We now have,
\beq |b^q, h^q, \tilde{b}^q/c^q| = \left( \det A^q \right)^{-1} |b^q, \hat{h}^q, \tilde{b}^q/c^q| \label{eq:hhath} \eeq
It will be convenient to make the following choice of $b^q$. The image, $B^{q, \bZ} = d(C^{q-1}(M, \bZ)) \subset C^q(M, \bZ)$ is a lattice. Let's choose $b^q_i$ to be a basis of this lattice. Likewise, $\tilde{b}^q_i$ can now be choosen as elements of $C^q(M, \bZ)$. Let $K^{q, {\bZ}} = \ker d^q \subset C^q(M, \bZ)$. We have, $B^{q+1,\bZ} \sim C^{q}(M, \bZ)/K^{q, \bZ}$. If we choose a basis  $k^q_i$ for the $K^{q, \bZ}$ lattice then 
\beq |k^q, \tilde{b}^q/c^q| = 1\eeq
Therefore,
\beq |b^q, \hat{h}^q, \tilde{b}^q/c^q| =|b^q, \hat{h}^q/k^q| \label{eq:bk}\eeq
Now, $H^q(M, \bZ) = K^{q, \bZ}/B^{q, \bZ}$ and ${\rm Tor}(H^q(M, \bZ)) = H^q(M, \bZ)/{\rm Free}(H^q(M, \bZ))$. Therefore, the torsion subgroup ${\rm Tor}(H^q(M, \bZ))$ is obtained by starting with the $K^{q, \bZ}$ lattice and moding out by vectors $b^q_i$ and $\hat{h}^{q}_i$. The order of the torsion subgroup is given precisely by the volume of a unit cell generated by $b^q_i$, $\hat{h}^q_i$ in the $K^{q, \bZ}$ lattice, i.e. 
\beq |{\rm Tor}(H^q(M, \bZ))| = |b^q, \hat{h}^q/k^q| \label{eq:Tordet}\eeq
Then combining equations (\ref{eq:NA}), (\ref{eq:hhath}), (\ref{eq:bk}), (\ref{eq:Tordet}), we obtain,
\beq  \log T^c(M) = \sum_{q = 0}^D (-1)^{q+1} \left(\log |{\rm Tor} H^q(M, \bZ)| - \frac12 \log {\cal N}^{+, q}\right) \label{eq:Tcappfinal}\eeq
We see that for a non-orientable manifold this exactly coincides with Eq.~(\ref{eq:Tc}), as $H^0(M, \bZ) = \bZ$ and $H^1(M, \bZ)$ have no torsion, and there is no $\log {\cal N}^{+,4}$ term in Eq.~(\ref{eq:Tcappfinal}) as $H^4(M, \bZ) = \bZ_2$ has no free part. 

\bibliography{Refs}

\begin{thebibliography}{45}
\expandafter\ifx\csname natexlab\endcsname\relax\def\natexlab#1{#1}\fi
\expandafter\ifx\csname bibnamefont\endcsname\relax
  \def\bibnamefont#1{#1}\fi
\expandafter\ifx\csname bibfnamefont\endcsname\relax
  \def\bibfnamefont#1{#1}\fi
\expandafter\ifx\csname citenamefont\endcsname\relax
  \def\citenamefont#1{#1}\fi
\expandafter\ifx\csname url\endcsname\relax
  \def\url#1{\texttt{#1}}\fi
\expandafter\ifx\csname urlprefix\endcsname\relax\def\urlprefix{URL }\fi
\providecommand{\bibinfo}[2]{#2}
\providecommand{\eprint}[2][]{\url{#2}}

\bibitem[{\citenamefont{Cardy and Rabinovici}(1982)}]{CardyRabinovici}
\bibinfo{author}{\bibfnamefont{J.~L.} \bibnamefont{Cardy}} \bibnamefont{and}
  \bibinfo{author}{\bibfnamefont{E.}~\bibnamefont{Rabinovici}},
  \bibinfo{journal}{Nuclear Physics B} \textbf{\bibinfo{volume}{205}},
  \bibinfo{pages}{1} (\bibinfo{year}{1982}).

\bibitem[{\citenamefont{Cardy}(1982)}]{Cardy}
\bibinfo{author}{\bibfnamefont{J.}~\bibnamefont{Cardy}},
  \bibinfo{journal}{Nuclear Physics B} \textbf{\bibinfo{volume}{205}},
  \bibinfo{pages}{17} (\bibinfo{year}{1982}).

\bibitem[{\citenamefont{Witten}(1995)}]{WittenS}
\bibinfo{author}{\bibfnamefont{E.}~\bibnamefont{Witten}},
  \bibinfo{journal}{Selecta Math.} \textbf{\bibinfo{volume}{1}},
  \bibinfo{pages}{383} (\bibinfo{year}{1995}), \eprint{arXiv:hep-th/9505186}.

\bibitem[{\citenamefont{Metlitski et~al.}(2013)\citenamefont{Metlitski, Kane,
  and Fisher}}]{MetlitskibTI}
\bibinfo{author}{\bibfnamefont{M.~A.} \bibnamefont{Metlitski}},
  \bibinfo{author}{\bibfnamefont{C.~L.} \bibnamefont{Kane}}, \bibnamefont{and}
  \bibinfo{author}{\bibfnamefont{M.~P.~A.} \bibnamefont{Fisher}},
  \bibinfo{journal}{Phys. Rev. B} \textbf{\bibinfo{volume}{88}},
  \bibinfo{pages}{035131} (\bibinfo{year}{2013}), \eprint{arXiv:1302.6535}.

\bibitem[{\citenamefont{{Wang} and {Senthil}}(2015)}]{Wang2015}
\bibinfo{author}{\bibfnamefont{C.}~\bibnamefont{{Wang}}} \bibnamefont{and}
  \bibinfo{author}{\bibfnamefont{T.}~\bibnamefont{{Senthil}}},
  \bibinfo{journal}{ArXiv e-prints: 1505.03520}  (\bibinfo{year}{2015}).

\bibitem[{\citenamefont{Metlitski and Vishwanath}(2015)}]{MVDuality}
\bibinfo{author}{\bibfnamefont{M.}~\bibnamefont{Metlitski}} \bibnamefont{and}
  \bibinfo{author}{\bibfnamefont{A.}~\bibnamefont{Vishwanath}},
  \bibinfo{journal}{ArXiv e-prints: 1505.05142}  (\bibinfo{year}{2015}).

\bibitem[{\citenamefont{Kapustin}(2014)}]{KapustinBos}
\bibinfo{author}{\bibfnamefont{A.}~\bibnamefont{Kapustin}},
  \bibinfo{journal}{ArXiv e-prints: 1403.1467}  (\bibinfo{year}{2014}).

\bibitem[{\citenamefont{{Kapustin}}(2014)}]{KapustinBosTI}
\bibinfo{author}{\bibfnamefont{A.}~\bibnamefont{{Kapustin}}},
  \bibinfo{journal}{ArXiv e-prints: 1404.6659}  (\bibinfo{year}{2014}).

\bibitem[{\citenamefont{Kapustin et~al.}(2014)\citenamefont{Kapustin,
  Thorngren, Turzillo, and Wang}}]{KapustinFerm}
\bibinfo{author}{\bibfnamefont{A.}~\bibnamefont{Kapustin}},
  \bibinfo{author}{\bibfnamefont{R.}~\bibnamefont{Thorngren}},
  \bibinfo{author}{\bibfnamefont{A.}~\bibnamefont{Turzillo}}, \bibnamefont{and}
  \bibinfo{author}{\bibfnamefont{Z.}~\bibnamefont{Wang}},
  \bibinfo{journal}{ArXiv eprints: 1406.7329}  (\bibinfo{year}{2014}).

\bibitem[{\citenamefont{Witten}(2015)}]{WittenSPT}
\bibinfo{author}{\bibfnamefont{E.}~\bibnamefont{Witten}},
  \bibinfo{journal}{ArXiv e-prints: 1508.04715}  (\bibinfo{year}{2015}).

\bibitem[{\citenamefont{Gilkey}(1985)}]{GilkeyPinC}
\bibinfo{author}{\bibfnamefont{P.~B.} \bibnamefont{Gilkey}},
  \bibinfo{journal}{Adv. in Math.} \textbf{\bibinfo{volume}{58}},
  \bibinfo{pages}{243} (\bibinfo{year}{1985}).

\bibitem[{\citenamefont{Cheeger}(1977)}]{Cheeger}
\bibinfo{author}{\bibfnamefont{J.}~\bibnamefont{Cheeger}},
  \bibinfo{journal}{Proc. Natl. Acad. Sci. USA} \textbf{\bibinfo{volume}{74}},
  \bibinfo{pages}{2651} (\bibinfo{year}{1977}).

\bibitem[{\citenamefont{Muller}(1978)}]{Muller}
\bibinfo{author}{\bibfnamefont{W.}~\bibnamefont{Muller}},
  \bibinfo{journal}{Adv. in Math.} \textbf{\bibinfo{volume}{28}},
  \bibinfo{pages}{233} (\bibinfo{year}{1978}).

\bibitem[{\citenamefont{Wang and Senthil}(2015)}]{ChongSenthilDirac}
\bibinfo{author}{\bibfnamefont{C.}~\bibnamefont{Wang}} \bibnamefont{and}
  \bibinfo{author}{\bibfnamefont{T.}~\bibnamefont{Senthil}},
  \bibinfo{journal}{ArXiv e-prints: 1505.05141}  (\bibinfo{year}{2015}).

\bibitem[{\citenamefont{Chen et~al.}(2014)\citenamefont{Chen, Fidkowski, and
  Vishwanath}}]{Chen2014PRB}
\bibinfo{author}{\bibfnamefont{X.}~\bibnamefont{Chen}},
  \bibinfo{author}{\bibfnamefont{L.}~\bibnamefont{Fidkowski}},
  \bibnamefont{and}
  \bibinfo{author}{\bibfnamefont{A.}~\bibnamefont{Vishwanath}},
  \bibinfo{journal}{Phys. Rev. B} \textbf{\bibinfo{volume}{89}},
  \bibinfo{pages}{165132} (\bibinfo{year}{2014}), \eprint{arXiv:1306.3250}.

\bibitem[{\citenamefont{Bonderson et~al.}(2013)\citenamefont{Bonderson, Nayak,
  and Qi}}]{Bonderson2013}
\bibinfo{author}{\bibfnamefont{P.}~\bibnamefont{Bonderson}},
  \bibinfo{author}{\bibfnamefont{C.}~\bibnamefont{Nayak}}, \bibnamefont{and}
  \bibinfo{author}{\bibfnamefont{X.-L.} \bibnamefont{Qi}},
  \bibinfo{journal}{Journal of Statistical Mechanics: Theory and Experiment}
  \textbf{\bibinfo{volume}{2013}}, \bibinfo{pages}{P09016}
  (\bibinfo{year}{2013}), \eprint{arXiv:1306.3230}.

\bibitem[{\citenamefont{Fidkowski et~al.}(2013)\citenamefont{Fidkowski, Chen,
  and Vishwanath}}]{FidkowskiChenAV}
\bibinfo{author}{\bibfnamefont{L.}~\bibnamefont{Fidkowski}},
  \bibinfo{author}{\bibfnamefont{X.}~\bibnamefont{Chen}}, \bibnamefont{and}
  \bibinfo{author}{\bibfnamefont{A.}~\bibnamefont{Vishwanath}},
  \bibinfo{journal}{Phys. Rev. X} \textbf{\bibinfo{volume}{3}},
  \bibinfo{pages}{041016} (\bibinfo{year}{2013}), \eprint{arXiv:1305.5851}.

\bibitem[{\citenamefont{Chen et~al.}(2013)\citenamefont{Chen, Gu, Liu, and
  Wen}}]{Chen2013}
\bibinfo{author}{\bibfnamefont{X.}~\bibnamefont{Chen}},
  \bibinfo{author}{\bibfnamefont{Z.-C.} \bibnamefont{Gu}},
  \bibinfo{author}{\bibfnamefont{Z.-X.} \bibnamefont{Liu}}, \bibnamefont{and}
  \bibinfo{author}{\bibfnamefont{X.-G.} \bibnamefont{Wen}},
  \bibinfo{journal}{Phys. Rev. B} \textbf{\bibinfo{volume}{87}},
  \bibinfo{pages}{155114} (\bibinfo{year}{2013}), \eprint{arXiv:1106.4772}.

\bibitem[{\citenamefont{Levin and Gu}(2012)}]{LevinGu}
\bibinfo{author}{\bibfnamefont{M.}~\bibnamefont{Levin}} \bibnamefont{and}
  \bibinfo{author}{\bibfnamefont{Z.-C.} \bibnamefont{Gu}},
  \bibinfo{journal}{Phys. Rev. B} \textbf{\bibinfo{volume}{86}},
  \bibinfo{pages}{115109} (\bibinfo{year}{2012}), \eprint{arXiv:1202.3120}.

\bibitem[{\citenamefont{Fu et~al.}(2007)\citenamefont{Fu, Kane, and
  Mele}}]{FuKaneMele}
\bibinfo{author}{\bibfnamefont{L.}~\bibnamefont{Fu}},
  \bibinfo{author}{\bibfnamefont{C.~L.} \bibnamefont{Kane}}, \bibnamefont{and}
  \bibinfo{author}{\bibfnamefont{E.~J.} \bibnamefont{Mele}},
  \bibinfo{journal}{Phys. Rev. Lett.} \textbf{\bibinfo{volume}{98}},
  \bibinfo{pages}{106803} (\bibinfo{year}{2007}),
  \eprint{arXiv:cond-mat/0607699}.

\bibitem[{\citenamefont{{Kitaev}}(2009)}]{KitaevNI}
\bibinfo{author}{\bibfnamefont{A.}~\bibnamefont{{Kitaev}}}, in
  \emph{\bibinfo{booktitle}{American Institute of Physics Conference Series}},
  edited by \bibinfo{editor}{\bibfnamefont{V.}~\bibnamefont{{Lebedev}}}
  \bibnamefont{and}
  \bibinfo{editor}{\bibfnamefont{M.}~\bibnamefont{{Feigel'man}}}
  (\bibinfo{year}{2009}), vol. \bibinfo{volume}{1134} of
  \emph{\bibinfo{series}{American Institute of Physics Conference Series}}, pp.
  \bibinfo{pages}{22--30}, \eprint{arXiv:0901.2686}.

\bibitem[{\citenamefont{Schnyder et~al.}(2008)\citenamefont{Schnyder, Ryu,
  Furusaki, and Ludwig}}]{LudwigNI}
\bibinfo{author}{\bibfnamefont{A.~P.} \bibnamefont{Schnyder}},
  \bibinfo{author}{\bibfnamefont{S.}~\bibnamefont{Ryu}},
  \bibinfo{author}{\bibfnamefont{A.}~\bibnamefont{Furusaki}}, \bibnamefont{and}
  \bibinfo{author}{\bibfnamefont{A.~W.~W.} \bibnamefont{Ludwig}},
  \bibinfo{journal}{Phys. Rev. B} \textbf{\bibinfo{volume}{78}},
  \bibinfo{pages}{195125} (\bibinfo{year}{2008}), \eprint{arXiv:0803.2786}.

\bibitem[{\citenamefont{Wang and Senthil}(2014)}]{Wang2014}
\bibinfo{author}{\bibfnamefont{C.}~\bibnamefont{Wang}} \bibnamefont{and}
  \bibinfo{author}{\bibfnamefont{T.}~\bibnamefont{Senthil}},
  \bibinfo{journal}{Phys. Rev. B} \textbf{\bibinfo{volume}{89}},
  \bibinfo{pages}{195124} (\bibinfo{year}{2014}), \eprint{arXiv:1401.1142}.

\bibitem[{\citenamefont{{Metlitski} et~al.}(2014)\citenamefont{{Metlitski},
  {Fidkowski}, {Chen}, and {Vishwanath}}}]{MetlitskiChenFidkowskiAV2014}
\bibinfo{author}{\bibfnamefont{M.~A.} \bibnamefont{{Metlitski}}},
  \bibinfo{author}{\bibfnamefont{L.}~\bibnamefont{{Fidkowski}}},
  \bibinfo{author}{\bibfnamefont{X.}~\bibnamefont{{Chen}}}, \bibnamefont{and}
  \bibinfo{author}{\bibfnamefont{A.}~\bibnamefont{{Vishwanath}}},
  \bibinfo{journal}{ArXiv e-prints: 1406.3032}  (\bibinfo{year}{2014}).

\bibitem[{\citenamefont{{Witten}}(1979)}]{WittenEffect}
\bibinfo{author}{\bibfnamefont{E.}~\bibnamefont{{Witten}}},
  \bibinfo{journal}{Physics Letters B} \textbf{\bibinfo{volume}{86}},
  \bibinfo{pages}{283} (\bibinfo{year}{1979}).

\bibitem[{\citenamefont{Goldhaber}(1976)}]{Goldhaber}
\bibinfo{author}{\bibfnamefont{A.~S.} \bibnamefont{Goldhaber}},
  \bibinfo{journal}{Phys. Rev. Lett.} \textbf{\bibinfo{volume}{36}},
  \bibinfo{pages}{1122} (\bibinfo{year}{1976}).

\bibitem[{\citenamefont{Metlitski et~al.}(2015)\citenamefont{Metlitski, Kane,
  and Fisher}}]{Metlitski2013}
\bibinfo{author}{\bibfnamefont{M.~A.} \bibnamefont{Metlitski}},
  \bibinfo{author}{\bibfnamefont{C.~L.} \bibnamefont{Kane}}, \bibnamefont{and}
  \bibinfo{author}{\bibfnamefont{M.~P.~A.} \bibnamefont{Fisher}},
  \bibinfo{journal}{Phys. Rev. B} \textbf{\bibinfo{volume}{92}},
  \bibinfo{pages}{125111} (\bibinfo{year}{2015}), \eprint{arXiv:1306.3286}.

\bibitem[{\citenamefont{Wang et~al.}(2013)\citenamefont{Wang, Potter, and
  Senthil}}]{Wang2013a}
\bibinfo{author}{\bibfnamefont{C.}~\bibnamefont{Wang}},
  \bibinfo{author}{\bibfnamefont{A.~C.} \bibnamefont{Potter}},
  \bibnamefont{and} \bibinfo{author}{\bibfnamefont{T.}~\bibnamefont{Senthil}},
  \bibinfo{journal}{Phys. Rev. B} \textbf{\bibinfo{volume}{88}},
  \bibinfo{pages}{115137} (\bibinfo{year}{2013}), \eprint{arXiv:1306.3223}.

\bibitem[{\citenamefont{Vishwanath and Senthil}(2013)}]{AVTS}
\bibinfo{author}{\bibfnamefont{A.}~\bibnamefont{Vishwanath}} \bibnamefont{and}
  \bibinfo{author}{\bibfnamefont{T.}~\bibnamefont{Senthil}},
  \bibinfo{journal}{Phys. Rev. X} \textbf{\bibinfo{volume}{3}},
  \bibinfo{pages}{011016} (\bibinfo{year}{2013}), \eprint{arXiv:1209.3058}.

\bibitem[{\citenamefont{{Burnell} et~al.}(2014)\citenamefont{{Burnell}, {Chen},
  {Fidkowski}, and {Vishwanath}}}]{Burnell2013}
\bibinfo{author}{\bibfnamefont{F.~J.} \bibnamefont{{Burnell}}},
  \bibinfo{author}{\bibfnamefont{X.}~\bibnamefont{{Chen}}},
  \bibinfo{author}{\bibfnamefont{L.}~\bibnamefont{{Fidkowski}}},
  \bibnamefont{and}
  \bibinfo{author}{\bibfnamefont{A.}~\bibnamefont{{Vishwanath}}},
  \bibinfo{journal}{Phys. Rev. B} \textbf{\bibinfo{volume}{90}},
  \bibinfo{pages}{245122} (\bibinfo{year}{2014}), \eprint{arXiv:1302.7072}.

\bibitem[{\citenamefont{Hirzebruch and Hopf}(1958)}]{HH}
\bibinfo{author}{\bibfnamefont{F.}~\bibnamefont{Hirzebruch}} \bibnamefont{and}
  \bibinfo{author}{\bibfnamefont{H.}~\bibnamefont{Hopf}},
  \bibinfo{journal}{Math. Annalen} \textbf{\bibinfo{volume}{136}},
  \bibinfo{pages}{156} (\bibinfo{year}{1958}).

\bibitem[{\citenamefont{Bahri and Gilkey}(1987)}]{BahriGilkey}
\bibinfo{author}{\bibfnamefont{A.}~\bibnamefont{Bahri}} \bibnamefont{and}
  \bibinfo{author}{\bibfnamefont{P.}~\bibnamefont{Gilkey}},
  \bibinfo{journal}{Pacific J. of Math.} \textbf{\bibinfo{volume}{128}},
  \bibinfo{pages}{1} (\bibinfo{year}{1987}).

\bibitem[{\citenamefont{Bobienski and Trautman}(2002)}]{RP2RP2}
\bibinfo{author}{\bibfnamefont{M.}~\bibnamefont{Bobienski}} \bibnamefont{and}
  \bibinfo{author}{\bibfnamefont{A.}~\bibnamefont{Trautman}},
  \bibinfo{journal}{Annals of Global Analysis and Geometry}
  \textbf{\bibinfo{volume}{22}}, \bibinfo{pages}{291} (\bibinfo{year}{2002}),
  \eprint{arXiv:math/0507340}.

\bibitem[{\citenamefont{Atiyah et~al.}(1973)\citenamefont{Atiyah, Bott, and
  Patodi}}]{Atiyah}
\bibinfo{author}{\bibfnamefont{M.}~\bibnamefont{Atiyah}},
  \bibinfo{author}{\bibfnamefont{R.}~\bibnamefont{Bott}}, \bibnamefont{and}
  \bibinfo{author}{\bibfnamefont{V.}~\bibnamefont{Patodi}},
  \bibinfo{journal}{Inv. Math.} \textbf{\bibinfo{volume}{19}},
  \bibinfo{pages}{279} (\bibinfo{year}{1973}).

\bibitem[{\citenamefont{Barkeshli et~al.}(2014)\citenamefont{Barkeshli,
  Bonderson, Cheng, and Wang}}]{Maissam}
\bibinfo{author}{\bibfnamefont{M.}~\bibnamefont{Barkeshli}},
  \bibinfo{author}{\bibfnamefont{P.}~\bibnamefont{Bonderson}},
  \bibinfo{author}{\bibfnamefont{M.}~\bibnamefont{Cheng}}, \bibnamefont{and}
  \bibinfo{author}{\bibfnamefont{Z.}~\bibnamefont{Wang}},
  \bibinfo{journal}{ArXiv e-prints: 1410.4540}  (\bibinfo{year}{2014}).

\bibitem[{\citenamefont{Atiyah et~al.}(1975)\citenamefont{Atiyah, Patodi, and
  Singer}}]{APS}
\bibinfo{author}{\bibfnamefont{M.}~\bibnamefont{Atiyah}},
  \bibinfo{author}{\bibfnamefont{V.~K.} \bibnamefont{Patodi}},
  \bibnamefont{and} \bibinfo{author}{\bibfnamefont{I.~M.}
  \bibnamefont{Singer}}, \bibinfo{journal}{Math. Proc. Camb. Phil. Soc.}
  \textbf{\bibinfo{volume}{77}}, \bibinfo{pages}{43} (\bibinfo{year}{1975}).

\bibitem[{\citenamefont{Stolz}(1988)}]{Stolz}
\bibinfo{author}{\bibfnamefont{S.}~\bibnamefont{Stolz}}, \bibinfo{journal}{Inv.
  Math.} \textbf{\bibinfo{volume}{94}}, \bibinfo{pages}{147}
  (\bibinfo{year}{1988}).

\bibitem[{\citenamefont{Klebanov et~al.}(2012)\citenamefont{Klebanov, Pufu,
  Sachdev, and Safdi}}]{KlebanovF}
\bibinfo{author}{\bibfnamefont{I.~R.} \bibnamefont{Klebanov}},
  \bibinfo{author}{\bibfnamefont{S.~S.} \bibnamefont{Pufu}},
  \bibinfo{author}{\bibfnamefont{S.}~\bibnamefont{Sachdev}}, \bibnamefont{and}
  \bibinfo{author}{\bibfnamefont{B.~R.} \bibnamefont{Safdi}},
  \bibinfo{journal}{JHEP} \textbf{\bibinfo{volume}{1205}}, \bibinfo{pages}{036}
  (\bibinfo{year}{2012}), \eprint{arXiv:1112.5342}.

\bibitem[{\citenamefont{Hatcher}(2002)}]{Hatcher}
\bibinfo{author}{\bibfnamefont{A.}~\bibnamefont{Hatcher}},
  \emph{\bibinfo{title}{``Algebraic topology"}} (\bibinfo{publisher}{Cambridge
  University Press}, \bibinfo{year}{2002}).

\bibitem[{\citenamefont{Vassilevich}(2003)}]{Vassilevich}
\bibinfo{author}{\bibfnamefont{D.~V.} \bibnamefont{Vassilevich}},
  \bibinfo{journal}{Phys. Rept.} \textbf{\bibinfo{volume}{388}},
  \bibinfo{pages}{279} (\bibinfo{year}{2003}), \eprint{arXiv:hep-th/0306138}.

\bibitem[{\citenamefont{Ray and Singer}(1971)}]{RaySinger}
\bibinfo{author}{\bibfnamefont{D.~B.} \bibnamefont{Ray}} \bibnamefont{and}
  \bibinfo{author}{\bibfnamefont{I.~M.} \bibnamefont{Singer}},
  \bibinfo{journal}{Adv. in Math.} \textbf{\bibinfo{volume}{7}},
  \bibinfo{pages}{145} (\bibinfo{year}{1971}).

\bibitem[{\citenamefont{Kitaev}()}]{Kitaev_pc}
\bibinfo{author}{\bibfnamefont{A.}~\bibnamefont{Kitaev}},
  \bibinfo{howpublished}{unpublished}.

\bibitem[{\citenamefont{{Wang} et~al.}(2014)\citenamefont{{Wang}, {Potter}, and
  {Senthil}}}]{ChongScience}
\bibinfo{author}{\bibfnamefont{C.}~\bibnamefont{{Wang}}},
  \bibinfo{author}{\bibfnamefont{A.~C.} \bibnamefont{{Potter}}},
  \bibnamefont{and}
  \bibinfo{author}{\bibfnamefont{T.}~\bibnamefont{{Senthil}}},
  \bibinfo{journal}{Science} \textbf{\bibinfo{volume}{343}},
  \bibinfo{pages}{629} (\bibinfo{year}{2014}), \eprint{arXiv:1306.3238}.

\bibitem[{\citenamefont{Cognola et~al.}(2003)\citenamefont{Cognola, Elizalde,
  and Zerbini}}]{Cognola}
\bibinfo{author}{\bibfnamefont{G.}~\bibnamefont{Cognola}},
  \bibinfo{author}{\bibfnamefont{E.}~\bibnamefont{Elizalde}}, \bibnamefont{and}
  \bibinfo{author}{\bibfnamefont{S.}~\bibnamefont{Zerbini}},
  \bibinfo{journal}{Commun. Math. Phys.} \textbf{\bibinfo{volume}{237}},
  \bibinfo{pages}{507} (\bibinfo{year}{2003}), \eprint{arXiv:hep-th/9910038}.

\bibitem[{\citenamefont{Bar}(1996)}]{Bar}
\bibinfo{author}{\bibfnamefont{C.}~\bibnamefont{Bar}}, \bibinfo{journal}{J.
  Math. Soc. Japan} \textbf{\bibinfo{volume}{48}}, \bibinfo{pages}{69}
  (\bibinfo{year}{1996}).

\end{thebibliography}


\end{document}